\documentstyle[aaspp4]{article}
\def\etal{\emph{et al.\ }}
\def\qmin{$Q_{\rm min}\ $}
\def\mrat{$M_D/M_*$}
\def\td{$T_{D}$}
\def\lsun{$L_\odot$}
\def\msun{$M_\odot$}
\def\rsun{$R_\odot$}
\def\pone{Paper~I\ }

\begin{document}

\title{Dynamics of Circumstellar Disks~II: Heating and Cooling}

\author{Andrew F. Nelson}
\affil{Department of Physics, The University of Arizona, Tucson AZ 85721\ and \\
Max Planck Institut f\"ur Astronomie, K\"onigstuhl 17, D-69117 Heidelberg,
Germany}
\author{Willy Benz}
\affil{Physikalisches Institut, Universit\"at Bern, Sidlerstrasse 5, CH-3012 Bern,
Switzerland\ and \\
Steward Observatory, The University of Arizona, Tucson AZ 85721}
\author{Tamara V. Ruzmaikina}
\affil{Lunar and Planetary Laboratory, The University of Arizona,
Tucson AZ 85721}

\begin{abstract}
We present a series of 2-dimensional ($r,\phi$) hydrodynamic simulations 
of marginally self gravitating (\mrat=0.2, with $M_*=0.5M_\odot$ and 
with disk radius $R_D=50$ and 100~AU) disks around protostars using a 
Smoothed Particle Hydrodynamic (SPH) code. We implement simple and 
approximate prescriptions for heating via dynamical processes in 
the disk.  Cooling is implemented with a simple radiative cooling 
prescription which does not assume that local heat dissipation
exactly balances local heat generation. Instead, we compute 
the local vertical ($z$) temperature and density structure of the 
disk and obtain `photosphere temperature', which is then used to
cool that location as a black body. We synthesize spectral energy
distributions (SEDs) for our simulations and compare them to fiducial
SEDs derived from observed systems, in order to understand the 
contribution of dynamical evolution to the observable character of 
a system.

We find that these simulations produce less distinct spiral structure
than isothermally evolved systems, especially in approximately the 
inner radial third of the disk. Pattern amplitudes are similar
to isothermally evolved systems further from the star but do not 
collapse into condensed objects. We attribute the differences in 
morphology to increased efficiency for converting kinetic energy into
thermal energy in our current simulations. Our simulations produce 
temperatures in the outer part of the disk which are very low ($\sim 10$~K). 
The radial temperature distribution of the disk photosphere is well 
fit to a power law with index $q\sim1.1$.  Far from the star, 
corresponding to colder parts of the disk and long wavelength 
radiation, known internal heating processes ($PdV$ work and shocks)
are not responsible for generating a large fraction of the thermal
energy contained in the disk matter. Therefore gravitational torques
responsible for such shocks cannot transport mass and angular momentum
efficiently in the outer disk. 

Within $\sim$5--10~AU of the star, rapid break up and reformation 
of spiral structure causes shocks, which provide sufficient dissipation
to power a larger fraction of the near infrared radiated energy output.
In this region, the spatial and size distribution of grains can have
marked consequences on the observed near infrared SED of a given disk,
and can lead to increased emission and variability on $\lesssim 10$ year
time scales. The inner disk heats to the destruction temperature of dust
grains. Further temperature increases are prevented by efficient cooling
when the hot disk midplane is exposed. When grains are vaporized in the
midplane of a hot region of the disk, we show that they do not reform
into a size distribution similar to that from which most opacity 
calculations are based. With rapid grain reformation into the original 
size distribution, the disk does not emit near infrared photons. With a
plausible modification of the opacity, it contributes much more.

\end{abstract}


\section{Introduction}\label{coolintro}

In the early stages of the formation of a star (see the review paper of 
Shu, Adams \& Lizano 1987), a cloud of gas and dust collapses and forms
a protostar with a disk surrounding it.  Later on, while the accretion
from the cloud continues, the star/disk system also begins to eject matter
into outflows whose strength varies in time. Finally, accretion and outflow
cease and over the next million or so years the star loses its disk and 
evolves onto the main sequence. A major refinement of this paradigm over 
the past decade has been to account for the formation of multiple objects 
from a single collapse. While this picture provides a good qualitative 
picture of the star formation process, many important issues remain poorly 
understood.

Once a well developed star/disk system evolves, whether as a single star 
or a multiple star system, the dynamics of the disk itself as well as its 
interaction with the star or a possible binary companion become
important in determining the system's final configuration. Depending
on the mass and temperature of a disk, one may expect spiral density
waves and viscous effects to develop and play roles of varying
importance. Each may be capable of processing matter through the disk
as well as influencing how the disk eventually decays away as the star
evolves onto the main sequence.

Until recently the primary observational evidence for circumstellar 
accretion disks has been the existence of sources with strong infrared 
excesses which extend from the near infrared to submillimeter and 
millimeter wavelengths. A number of papers (Adams, Lada \& Shu 1987, 
1988, Adams, Emerson \& Fuller 1990, Beckwith \etal 1990--hereafter 
BSCG, Osterloh \& Beckwith 1995) have successfully modeled these 
excesses assuming a geometrically thin accretion disk with or without
additional circumstellar material. Other recent observations (Roddier
\etal 1996, Close \etal 1997) have used adaptive optics to image the
disks of several young star systems. Other disk systems (so called
`proplyds') have been observed in the Orion molecular cloud with the 
Hubble Space Telescope (O'Dell \& Wen 1994, McCaughrean \& O'Dell 1996)
in silhouette against the bright cloud background or through interactions
with winds from nearby massive stars. 

With these direct and indirect observations it has become clear that
disk systems are quite common around young stars. Many efforts to
model the dynamical processes involved in their formation (Laughlin \&
R\'o\.zyczka 1996, Bonnell \& Bate 1997) and evolution (Nelson \etal 1998,
hereafter \pone, Boss 1997, Artymowicz \& Lubow 1996, Pickett, Durisen \& 
Link 1997) have so far resulted only in a summary of what is possible 
rather than strong limits on what types of evolution are impossible.
Many gaps remain in the understanding of the physical processes 
important in different regimes and even in the configurations of
systems at various points in their history. 

Other efforts have been applied to modeling the spectral energy 
distributions (SEDs) of young stellar systems. The SEDs of passive 
disks (i.e. those disks which only reprocess radiation from the 
central star) have been successfully modeled in recent work by Chiang 
\& Goldreich (1997). Axisymmetric models of a disk and a mixing 
length approximation for the vertical structure (Bell \& Lin 1994-hereafter
BL94, Bell \etal 1995, Bell \etal 1997-hereafter BCKH) have been used to 
model the most dynamic properties of disks seen in FU~Orionis stars. Time 
dependent radiative transport calculations (Simonelli, Pollack \& McKay 
1997, Chick, Pollack \& Cassen 1996) have also been incorporated into 
calculations of the structure of infalling gas and dust. They model 
the destruction of grains in material falling onto the star/disk 
system from the surrounding circumstellar cloud and find that under
many conditions grains can be partially or totally destroyed prior
to their accretion into the star/disk system. Heating mechanisms in
the cloud and infalling envelope are of comparable effectiveness 
in heating the grains as in the accretion shock itself.

\pone showed that in the limit of a disk modeled with a locally isothermal
equation of state, spiral arm formation and later collapse into clumps 
totaling at least a few percent of the disk mass was prevalent in all 
disks whose minimum initial Toomre stability was \qmin$\lesssim 1.5-1.7$.
Boss (1997) and Pickett \etal (1998) have concluded that a locally adiabatic
equation of state will also produce spiral arm collapse as instabilities grow. 
Each of these works are limited in the sense that a predefined temperature law 
is assumed: the gas is locally isothermal or locally adiabatic, but is not
globally isothermal or globally adiabatic. In this approximation, any radial 
motion of gas within the disk causes the parcel of gas to heat or cool, even if 
no other processes occur to change its state. Compression and shock events are
likewise artificially managed.  Heating and cooling are instantaneous,
but only act when the state of the gas deviates from a predefined `steady
state' value. 

We present a series of two dimensional (radius and azimuth) numerical 
simulations using Smoothed Particle Hydrodynamics (SPH) modeled under 
the assumption that the disk is able to heat or cool depending only on
local conditions within the disk.  Our goal for this work is to understand
the dynamical growth characteristics of instabilities in systems with 
heating and cooling incorporated into the models and to understand which 
heating and cooling mechanisms are likely to be responsible for which 
features in the spectral energy distributions of observed systems. In
section \ref{coolphys}, we summarize the initial conditions adopted for
the disks studied. In section \ref{energy}, we outline heating and cooling
mechanisms included in our study and the numerical method used to determine 
their magnitude at each point and time in the disk. We describe the dynamical
results obtained from our simulations in section \ref{simulations} and the
physical origin of the emitted spectra in section \ref{sed-dynam}. In 
section \ref{cmparother}, we compare our results to work in the 
literature and summarize their significance in the context of the
evolution of stars and star systems.

\section{Physical Assumptions}\label{coolphys}

\subsection{Initial Conditions} \label{coolinit}

The initial conditions used in this work are quite similar to those
used in \pone. We refer the reader to that work for a more complete 
discussion and only summarize them here. At time zero we set equal mass
particles on a series of concentric rings extending from the innermost 
ring at a radius of $0.5$~AU to either 50 or 100 AU depending upon
the simulation (see table \ref{cool-params} below). With the number
of particles used, smoothing lengths (i.e. the resolution) are less 
than a few tenths of one AU in the inner portion of the disk and up
to $\sim$1~AU in the outer disk. The star is modeled as a point mass
free to move in response to gravitational forces from the surrounding
disk. The gravitational force due to the star is softened with a softening
radius of $0.4$~AU and particles whose trajectories pass through this 
radius are absorbed by the star. Magnetic fields are neglected in our
simulations.

The disk mass is initially distributed according to a power law:
\begin{equation}\label{cooldenslaw}
\Sigma(r) = \Sigma_0 \left[ 1 + \left({r\over r_c}\right)^2\right]^
{-{p\over{2}}},
\end{equation}
while the disk midplane temperature is given according to a similar law:
\begin{equation}\label{cooltemplaw}
T(r) = T_0\left[1 + \left({r\over r_c}\right)^2\right]^{-{q\over{2}}},
\end{equation}
where the exponents $p$ and $q$ are $3/2$ and $1/2$, respectively, and
$\Sigma_0$ and T$_0$ are determined from the disk mass and a choice 
of the minimum value of Toomre's stability parameter $Q$ over
the disk. $Q$ is defined as:
\begin{equation}\label{toomre-q}
Q = {{\kappa c_s}\over{ \pi G \Sigma}},
\end{equation}
where $\kappa$ is the local epicyclic frequency and $c_s$ is the sound 
speed.  The core radius $r_c$ for the power laws is set to $r_c$=1AU.

Matter is set up on initially circular orbits assuming rotational 
equilibrium in the disk. Radial velocities are set to zero. Gravitational
and pressure forces are balanced by centrifugal forces by setting
\begin{equation}\label{coolrotlaw}
\Omega^2(r) = { {GM_*\over{r^3}} + {1\over{r}}{
       {\partial\Psi_D}\over{\partial{r}}} + {1\over{r}}{
       {{\bf\nabla}{P}}\over{\Sigma} } },
\end{equation}
where $\Psi_D$ is the gravitational potential of the disk and the
other symbols have their usual meanings. The magnitudes of the
pressure and self gravitational forces are small compared to the stellar
term, therefore the disk is nearly Keplerian in character.

We use an accretion radius of 0.4~AU as a compromise between the
numerical requirement that the integration time step not be so small
that long period evolution cannot be followed and the desire to
model as large a radial extent of the disk as possible. The exact size
of the boundary region between the stellar surface and the inner disk
edge is not well constrained, but is probably substantially smaller 
than 0.4~AU. For example, Shu \etal (1994) estimate that the inner
edge of the disk should be between two and ten times the stellar 
radius. Coupling this with a stellar radius computed from pre-main
sequence evolution (see e.g. D'Antona \& Mazzitelli 1994, 1997) of
$R_*\approx 2-5$ \rsun, depending on age, gives an inner disk edge 
as small as 0.02~AU or as large as 0.2~AU. We will also perform
a small number of simulations with an accretion radius of 0.2~AU
in order to investigate what effects the inner boundary may have
on the evolution.

\subsection{The Equation of State}

The hydrodynamic equations are solved assuming a vertically integrated 
gas pressure and a single component, ideal gas equation of state given by:
\begin{equation}\label{ideal-eos}
P=(\gamma - 1)\Sigma u
\end{equation}
where $\gamma$ is the ratio of specific heats, $P$ is the vertically
integrated pressure and $u$ is the specific internal energy of the gas.
Since we limit the motion of our particles to two dimensions ($r,\phi$), 
the effective value of $\gamma$ is different from that derived from a true 
three dimensional calculation (see e.g. the discussion of the equations of
motion derived for a vertically integrated torus in Goldreich, Goodman \&
Narayan 1986). For the systems we study, we have taken a simpler 
approach by assuming that only two translational degrees of freedom 
exist for each molecule. Helium is included as a monatomic ideal gas 
and metals are neglected. Coupled with the assumption that the gas is of 
solar composition, this means the effective value of $\gamma$ is no longer
the well known $\gamma=5/3$, but rather 
\begin{equation}
\gamma\approx 1.53.
\end{equation}
This value includes the contribution of hydrogen with its rotational degrees 
of freedom active ($\gtrsim 100$~K) but its vibrational degrees of freedom 
inactive ($\lesssim 800-1000$~K). This value will be most representative of
moderate temperature regions of the disk. In three dimensions,
$\gamma\approx1.42$.

\section{Thermal Energy Generation and Dissipation}\label{energy}

In this work we relax the common practice (see e.g. \pone, Pickett \etal 
1998, Boss 1997) of predefining the temperature or adiabatic 
constant, $K$, at each location in the disk. Instead, we allow thermal
energy to be generated by internal processes and we allow the disk to
cool radiatively at a temperature solely dependent upon local conditions
at a given time. Thermal energy may be generated in one location in 
the disk but be dissipated somewhere else if matter moves there, or the
disk may heat up or cool down over time in a single location. The disk 
may therefore equilibrate to the internal energy state that the physical 
evolution of the system requires.

In our simulations, we only require that the disk be in instantaneous 
vertically constant entropy equilibrium and instantaneous vertical thermal
balance in order to determine its structure. We do not require it to be in
long term vertical thermal balance. With the latter assumption, the radiative
cooling rate at each point is defined to be equal to the local heating rate
from internal processes (see e.g. Frank, King \& Raine 1992 section 5.4).
In some cases, the assumption also includes energy flux radiating onto the 
disk from outside, so that the radiative cooling rate includes terms due to 
both internally generated energy and passive reprocessing. Accurate 
quantification of the relative contributions of each of these terms is 
critical because by working backwards from observed spectral characteristics 
of the disk an observer can derive an evolutionary picture of the mass and 
angular momentum transfer through the system. For example, if radiation 
emitted by a disk comes entirely from passive reradiation, then no mass or 
angular momentum transport can occur, since such transport is due to internal 
dissipation of kinetic energy in the disk. Therefore, if the contributions 
due to one or more sources of the emitted radiation are incorrectly determined, 
an evolutionary picture derived from them will be flawed. In the present
work, we will neglect passive heating and reradiation in favor of accurately
quantifying internal heating mechanisms in the disk. Here we outline the
physical basis for the heating and cooling processes incorporated into 
our simulations.

\subsection{Thermal Energy Generation}\label{energy-gen}

Thermal energy in the disk is generated in our simulations from bulk 
mechanical energy via viscous processes and shocks. We model the energy
generation using an artificial viscosity common in many implementations
of hydrodynamic codes. Because the balance between thermal energy
generation and dissipation are important for both the observed character
of the systems as well as their morphology and dynamics, we outline our
implementation here. In the following section (section \ref{phys-artvisc}),
we also outline the physical interpretation of its effects on our 
simulations and the basis for conclusions about physical processes that 
can be drawn from these effects. We refer the reader to one of the many 
discussions already in the literature (e.g. Benz 1990, Monaghan 1992) for 
more complete treatment. 

The standard form of artificial viscosity used in numerical hydrodynamic 
codes takes the form of an added artificial viscous pressure to the
energy equation. In the SPH formalism the energy dissipation due 
to this additional term is the sum of contributions from all neighbor 
particles, $j$, on particle $i$ as
\begin{equation}\label{art-visc}
{{du_i}\over{dt}} = 
      {{1}\over{2}}\sum_j m_j\Pi_{ij}
	      \left({\bf v_i} - {\bf v_j}\right)\cdot {\nabla_i W_{ij}},
\end{equation}
where ${\bf v_i}$ and ${\bf v_j}$ are the velocities of each particle,
$m_i$ is the mass of the $i$th particle and $W_{ij}$ is the value of the 
SPH kernel calculated between the two particles. The factor 1/2 in eq. 
\ref{art-visc} accounts for half of the kinetic energy dissipation being
added to each particle. The term $\Pi_{ij}$ is the viscous pressure
and the other terms together compose the volume element, $dV$, in
the viscous work $\Pi_{ij}dV$. An analogous term enters in the momentum
equation, so that momentum and energy may be properly coupled.

The viscous pressure, $\Pi_{ij}$, is given by a sum of two terms
which are linear and quadratic in the velocity divergence. The form of
these terms is determined by the desire to model correctly strong
compressions while reducing numerical oscillations in the flow (see 
below). Mathematically, $\Pi_{ij}$ is
\begin{equation}\label{Piij}
\Pi_{ij} = \cases{
	   {{-\bar\alpha c_{ij}\mu_{ij} + \beta\mu_{ij}^2}\over\Sigma_{ij}}
        &if (${\bf v_i} - {\bf v_j})\cdot({\bf r_i} - {\bf r_j}) \leq 0$;\cr
      0 &otherwise,\cr}
\end{equation}
where ${\bf r_i}$ and ${\bf r_j}$ are the positions of each particle,
$c_{ij}$ is the sound speed and $\Sigma_{ij}$ is the mean surface 
density.  These two terms are the so called `bulk' or `$\bar\alpha$'
\footnote{Note that we have used the symbol $\bar\alpha$ to denote the
bulk component of artificial viscosity in order to distinguish it from
the Shakura \& Sunyaev (1973) turbulent viscosity parameter, 
$\alpha_{SS}$.} viscosity and the `von~Neumann-Richtmyer' or `$\beta$' 
viscosity. The velocity divergence $\mu_{ij}$ is defined by
\begin{equation}\label{diverg}
\mu_{ij} = {{h_{ij}({\bf v_i} - {\bf v_j})\cdot({\bf r_i} - {\bf r_j})}\over
           {\vert {\bf r_i} - {\bf r_j} \vert^2 + \epsilon h_{ij}^2}}
           {{ \left({{f_i + f_j}\over{2}}\right)}}
\end{equation}
where $\epsilon$ is a small value to prevent numerically infinite 
divergence as particles come very close and $h_{ij}$ is the average of
particle $i$ and particle $j$'s smoothing lengths. 
The $f_i$ and $f_j$ terms are due to Balsara (1995) and are defined by
\begin{equation}\label{Balsara-f}
f_i = {{ | (\nabla \cdot {\bf v_i}) | } \over 
       { | (\nabla \cdot {\bf v_i}) | + | (\nabla \times {\bf v_i} ) |
               + 0.0001c_i/h_i}}.
\end{equation}

Equations \ref{art-visc}--\ref{diverg} are little more than a restatement
of the standard form of artificial viscosity for SPH as discussed in 
Benz (1990). The standard form, while adequate for many problems,
is well known to produce a large and unphysical shear dissipation as a 
side effect in disk simulations like those in our study. Such 
dissipation arises because in order to more closely model the physical 
effects of shocks with the code, the divergence is calculated pairwise 
between particles rather than as an average over the nearest neighbors 
(Gingold \& Monaghan 1983). In a Keplerian disk, particles at 
different radii have nearly parallel trajectories but different 
velocities, which produces the numerical result that they appear as 
if they are approaching (false compression) each other or receding 
(false expansion) from each other. Another side effect is that the
standard formulation produces dissipation nearly everywhere in the 
fluid rather than only in regions in which it is required. Numerical
modeling of an adiabatic compression is impossible with artificial
viscous terms included for example.

The specific form of the terms in eqn. \ref{art-visc}--\ref{diverg}
are quite arbitrary. Any improvements to them which decrease the 
importance of unphysical side effects, while retaining the required
stability and dissipative effects of the code, are desirable. We 
incorporate two adaptations which improve the performance of the 
code by reducing these side effects.

First, we modify the computation of the velocity divergence from its 
usual form by including the factors $f_i$ and $f_j$ in eq. \ref{diverg}.
These factors act to reduce shear viscosity. They are near 
unity when the flow is strongly compressive, but near zero in shear
flows. For the simulations we have performed we find that typically
the reduction due to this term is a factor of three or better. The
second improvement is due to Morris and Monaghan (1997). They implement
a time dependence to the coefficient $\bar\alpha$ which allows grow
in regions where it is physically appropriate (strong compressions) and 
decay in quiescent regions where it is inappropriate. The decay takes place 
over at distance of a few SPH smoothing lengths, after which the coefficient
stabilizes to a constant, quiescent value. In our formulation, which includes
both the $\bar\alpha$ and $\beta$ terms, we define the ratio 
$\bar\alpha/\beta\equiv 0.5$, but allow their magnitudes to vary in time and
space according to the Morris \& Monaghan formulation. Thus, except in strongly
compressing regions (shocks) where it is required to stabilize the flow, 
artificial viscous dissipation is minimized.

\subsubsection{The origin and physical meaning of artificial 
          viscosity}\label{phys-artvisc}

The physical reason for including an artificial viscosity into a
numerical code is that in real systems, dissipation of kinetic
energy into heat occurs on scales which are much smaller than the
numerically resolvable scale of the system. Shock dissipation for
example occurs over a few mean free path lengths of the molecules
in the fluid, a size at which even the fluid assumptions of the
hydrodynamic equations break down. Without the addition of additional,
artificial terms to the finite difference equations, such dissipation
would not be included. The results obtained from such an omission 
are in general numerically unstable. 

The solution of choice has been to include macroscopic dissipation
terms (the viscous pressure $\Pi_{ij}$ above) to model the microscopic
dissipation. The form of the two terms making up $\Pi_{ij}$ is motivated
first by the fact that dissipation must be introduced to the system in 
order to reproduce the hydrodynamic quantities in shocked regions. The
$\beta$ term in eq. \ref{Piij} provides a functional dependence of the
viscous coefficient (usually denoted $\nu$) itself on the velocity 
divergence present in the flow. In this way, the magnitude of the 
dissipation becomes dependent upon a low order approximation of the
discontinuity present in a shock.

Without additional correction unphysical oscillations can still develop 
in the flow, due to the low order finite differencing in the numerical 
solution of the hydrodynamic equations. To damp out such oscillations, 
a second term (the $\bar\alpha$ term in eq. \ref{Piij}) is introduced
which acts diffusively.  With this in mind we note that, while the 
von~Neumann-Richtmyer term may have some approximate physical basis, its
counterpart bulk term can only be considered a necessary nuisance.

Although we may properly regard both terms as nuisances, since they
do not directly model any physical process, with caution we can turn 
artificial viscosity into a useful nuisance. We have already identified
the von~Neumann-Richtmyer term as a low order representative of shock
dissipation. We can make a similar identification of the bulk viscosity 
as a `black box' source of dissipation in the system like the {\it ad hoc}
`$\alpha_{SS}$' model of Shakura \& Sunyaev, which uses a single parameter,
$\alpha_{SS}$, to model the effects of turbulence. A correspondence (Murray
1995, 1996) exists between the standard $\alpha_{SS}$ form of dissipation
and the bulk artificial viscosity implemented in our simulations. In two 
dimensions the correspondence can be expressed for particle $j$ as
\begin{equation}\label{alpha-eq}
\alpha^j_{SS} = {{ f_j\bar\alpha_j h_j \Omega_j} \over {8c_j}}.
\end{equation}
where the index $j$ refers to the $j$th particle, $f_j$ is defined by
eq. \ref{Balsara-f}, $\bar\alpha_j$ is its bulk viscosity coefficient,
$\Omega_j$ is the orbit frequency, $c_j$ is the sound speed of the particle
and $h_j$ is its smoothing length. 

The conversion to an `$\alpha_{SS}$' in our work is different from 
the standard Shakura \& Sunyaev form in several ways. It is both time
and space dependent, in contrast to the usual form of a constant 
$\alpha_{SS}$ everywhere or (in a few cases), variation between an 
`on' and `off' state. The conversion neglects the contribution due to the 
von~Neumann-Richtmyer term and so by itself represents only an 
incomplete approximation of the magnitude of the total dissipation present.
Finally, it also neglects the Morris \& Monaghan time dependence of the
viscous coefficients noted above. In the context of attempting to identify
the source of the dissipation, this means that if a region experiences 
repeated compression events or shocks on short time scales, the dissipation
would be accounted for as a turbulent process rather than as a shock 
process. In this case, a short time scale means that the gas travels only
a few SPH smoothing lengths before experiencing another compression. 
Nevertheless, we find that it proves useful as an illustration of where and 
to what extent thermal energy generation processes are active.

\subsection{Thermal Energy Dissipation and the Vertical Structure 
of Accretion Disks} \label{energy-diss}

The cooling experienced by a given particle is determined first by 
calculating the approximate vertical density and temperature ($\rho, T)$
structure of the disk. Then, using these quantities we determine the 
altitude of the disk photosphere and cool the particle as a blackbody 
whose temperature is the calculated photosphere temperature.

In order to calculate $\rho(z)$ and $T(z)$ without a full three dimensional
hydrodynamic calculation we make two assumptions about the disk structure.
We assume that at each location the disk has some degree of turbulence or 
convection so that it becomes very nearly adiabatic in the $z$ direction
(i.e. that $p=K\rho^\gamma$ with $K$ and $\gamma$ constant). We also assume
that it is locally plane parallel.  In this limit, Fukue \& Sakamoto (1992)
have shown that $\rho(z)$ and $T(z)$ at a known distance from the star can
be determined from the solution of the second order, ordinary differential 
equation
\begin{equation}\label{struc-diffeq}
{{d}\over{dz}} \left({{1}\over{\rho}} {{d}\over{dz}}\left(K\rho^\gamma\right)
                +   {{GM_*z}\over{(r^2 + z^2)^{3/2}}}\right)
		=   -4\pi G\rho.
\end{equation}
A known midplane density, $\rho_{mid}$, the distance from the star, $r$,
the adiabatic constant, $K$, and the ratio of specific heats,
$\gamma$, define the conditions which completely specify the solution
in the absence of external heating of the disk surface.

In our two dimensional simulations, each SPH particle is uniquely defined 
at some time by a particular value of internal energy, surface density 
and distance from the star. These three quantities correspond to the 
three conditions $\rho_{mid}$, $K$ and $r$ which specify the structure
in $z$. The distance from the star is, of course, the same for both
the SPH and Fukue \& Sakamoto specifications. Derivation of the
quantities $\rho_{mid}$ and $K$ from the surface density and internal
energy must be done by iteration to convergence.

\begin{figure}
\plotfiddle{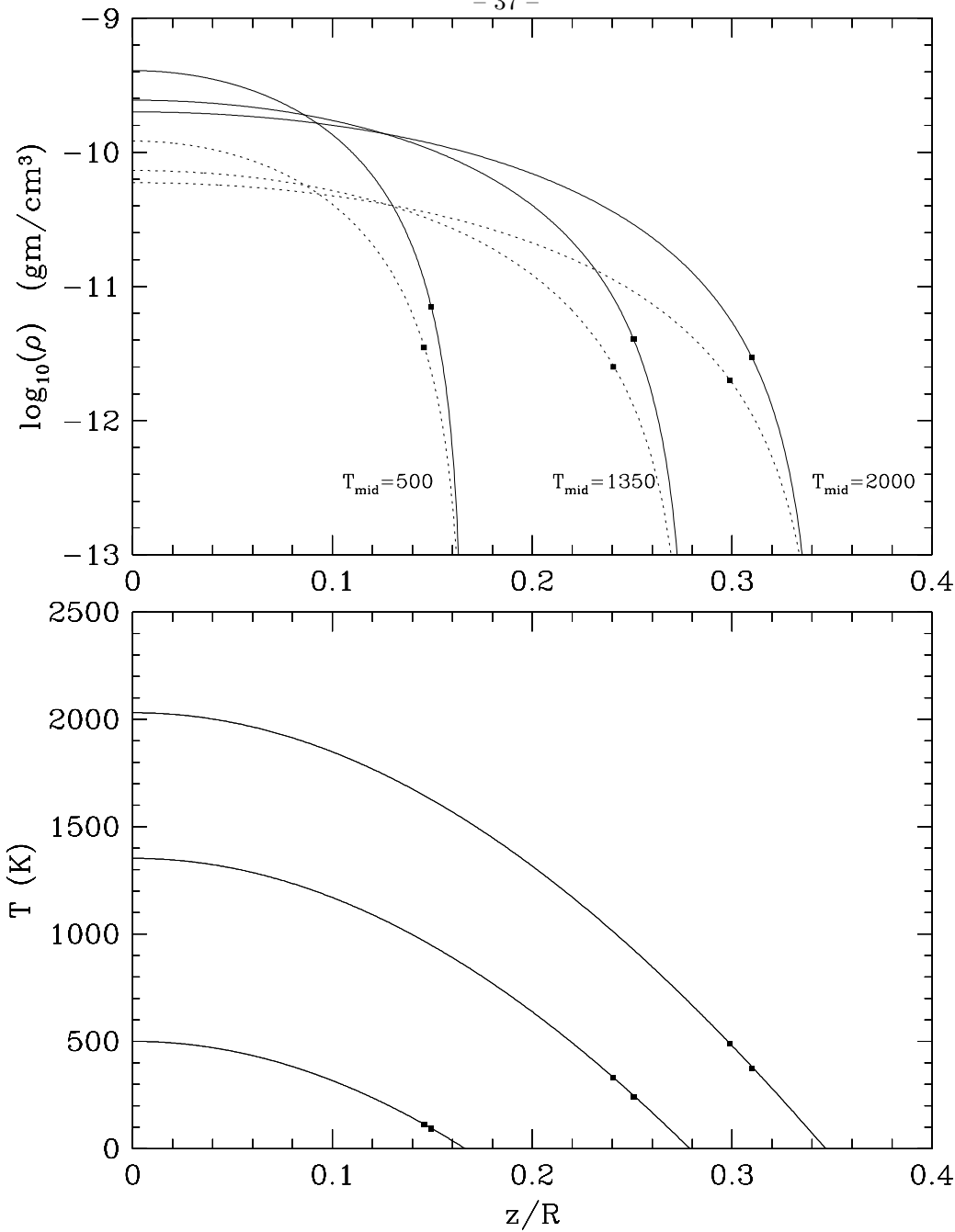}{6.41in}{0}{72}{72}{-230}{-35}
\caption[Density and temperature structure for several conditions
typical of the disks studies in our simulations]
{\label{z-strucplot}
Density and temperature structure as a function of altitude above the
midplane for conditions typical of our disk simulations at a 1~AU distance
from the star. The location of the calculated disk photosphere of
each disk is marked with a solid square attached to each curve. The solid
curves on the upper frame are typical of the density derived from our
simulations of 50~AU disks (1000 g/cm$^2$), while the dotted curves represent
the density structure typical of our 100~AU disks (300 g/cm$^2$).  Each of
the three pairs of curves in the plot show the density structure for an
assumed midplane temperature of $\sim 2000$~K, $\sim 1350$~K and $\sim 500$~K
as noted. The temperatures are plotted in the lower frame and are 
represent the models with midplane temperatures well below, approximately
equal to and well above the grain destruction temperature in the disk
midplane.}
\end{figure}

We supply an initial guess for $\rho_{mid}$ and $K$ and solve equation
\ref{struc-diffeq} numerically for $\rho(z)$. The $z$ coordinate
is discretized with 500 zones and the differential equation is solved 
numerically to second order accuracy. Once a tentative solution is 
reached, we integrate the density $\rho$ over $z$ using the trapezoid 
rule to derive the surface density of matter defined by the solution.
Specific internal energy is obtained by a similar integration over the 
vertical extent of the disk. The guesses of $\rho_{mid}$ and $K$ are
then revised using the downhill simplex method to converge to a self
consistent solution. Plots of the density and temperature structure
as a function of the altitude, $z$, are shown for several conditions 
typical of the disks in our simulations are shown in fig. \ref{z-strucplot}.

Implicit in our calculation is the assumption that the gas is locally
(and instantaneously) adiabatic as a function of $z$. In an adiabatic
medium, the gas pressure and density are related by $p=K\rho^\gamma$ and
the heat capacity of the gas, $C_V$, is a constant (by extension, also
the ratio of specific heats, $\gamma$). In fact, this will not be the 
case in general because, in various temperature regimes, molecular 
hydrogen will have active rotational or vibrational modes, it may
dissociate into atomic form or it may become ionized. As a matter of
expediency and in order to retain our prescription for the structure
calculation, we have assumed that the rotational states of hydrogen are
active, but that the vibrational states are not. Under this assumption 
and including the contribution due to helium, the effective value for
the three dimensional adiabatic exponent of the gas is $\gamma\approx1.42$. 

From the now known $(\rho, T)$ structure we derive the temperature of the 
disk photosphere by a numerical integration of the optical depth, $\tau$,
from $z=\infty$ to the altitude at which the optical depth becomes
$\tau=2/3$
\begin{equation}\label{tau-a}
\tau = 2/3 = \int_\infty^{z_{phot}}\rho(z)\kappa(\rho,T)dz.
\end{equation}
In optically thin regions, for which $\tau<2/3$ at the midplane, we 
assume the photosphere temperature is that of the midplane. The 
photosphere temperature is then tabulated as a function of the three
input variables radius, surface density and specific internal energy.
At each time we determine the photosphere temperature for each particle
from this table and cool the particle as a blackbody at that 
temperature. The cooling of any particular particle proceeds as 
\begin{equation}\label{dudt}
{{du_i}\over{dt}} = {{ -2\sigma_R T_{eff}^4 }\over{\Sigma_j} }
\end{equation}
where $\sigma_R$ is the Stefan-Boltzmann constant, $u_i$ and $\Sigma_i$
are the specific internal energy and surface density of particle $i$
and $T_{eff}$ is its photospheric temperature. The factor of two 
accounts for the two surfaces of the disk. On every particle, we
enforce the condition that the temperature (both midplane and 
photosphere) never falls below the 3~K cosmic background temperature.
Finally, we investigate the effect of optically thin emission in some
simulations by multiplying the flux by the optical depth, $\tau$,
in regions where the optical depth to the midplane is $\tau<1$). 

We use Rosseland mean opacities from tables of Pollack, McKay
\& Christofferson (1985 hereafter PMC). Opacities for packets of 
matter above the grain destruction temperature are taken from 
Alexander \& Ferguson (1994). We have chosen not to use the updated
opacity models of Pollack \etal (1994) in this work. In part this 
is due to the fact that opacity tables including both $\rho$ and $T$
variation based on this work do not exist (D. Hollenbach, personal 
communication). As the authors note however, the opacity is only a 
weak function of density (entering primarily through the change in 
vaporization temperatures of various volatiles at different densities)
and they produce a figure comparing the new opacity with that of 
the old for a single value of the density. The functional form 
derived by Henning \& Stognienko (1996) to reproduce the Pollack
\etal opacity also do not include full $\rho$ and $T$ dependence.

As we shall note in the sections ahead, it is exactly this 
vaporization of grains which we find to be an important factor 
in determining the character of the SED. We have therefore chosen
to implement the old version of the opacities until such time as 
new tabulated values become available. In any case, the opacities 
derived from the new and old works (see Pollack \etal, fig. 6) are 
similar except in the temperature region between 200 and 450~K, where 
the Pollack \etal (1994) derivation exceeds the PMC value by a factor 
of about three. The effect on our calculations would be to slightly
reduce the photosphere temperature in regions where the difference
between the two versions becomes important.

\subsubsection{Where this prescription fails, and an improvement}

We will find in section \ref{morph} that SEDs synthesized from
our simulations using the cooling prescription above fail to 
reproduce the short wavelength spectrum (near IR) of observed 
circumstellar disks. This wavelength regime corresponds to the
portion of the disk in which the disk midplane temperatures are
warm enough to sublimate grains. Our cooling prescription on the 
other hand, assumes both that the opacity source (grains) is evenly
mixed with the gaseous disk material and that the grain size 
distribution everywhere is not substantially different from that of
the interstellar medium distribution used to calculate the Rosseland
opacities. Is the failure of our simulations to correctly reproduce
observed disk SEDs due to the failure of these physical assumptions
about the grain physics? 

To address the first assumption we note that Weidenschilling (1984) 
has shown that grains smaller than $\sim$0.1--1 cm will be largely 
entrained in the gas in a turbulent disk and will therefore not settle 
to the midplane. Because the smaller grains provide the largest 
contribution to the opacity, we may assume that for the purposes of 
our model the grains are well mixed. 

\begin{figure}
\plotfiddle{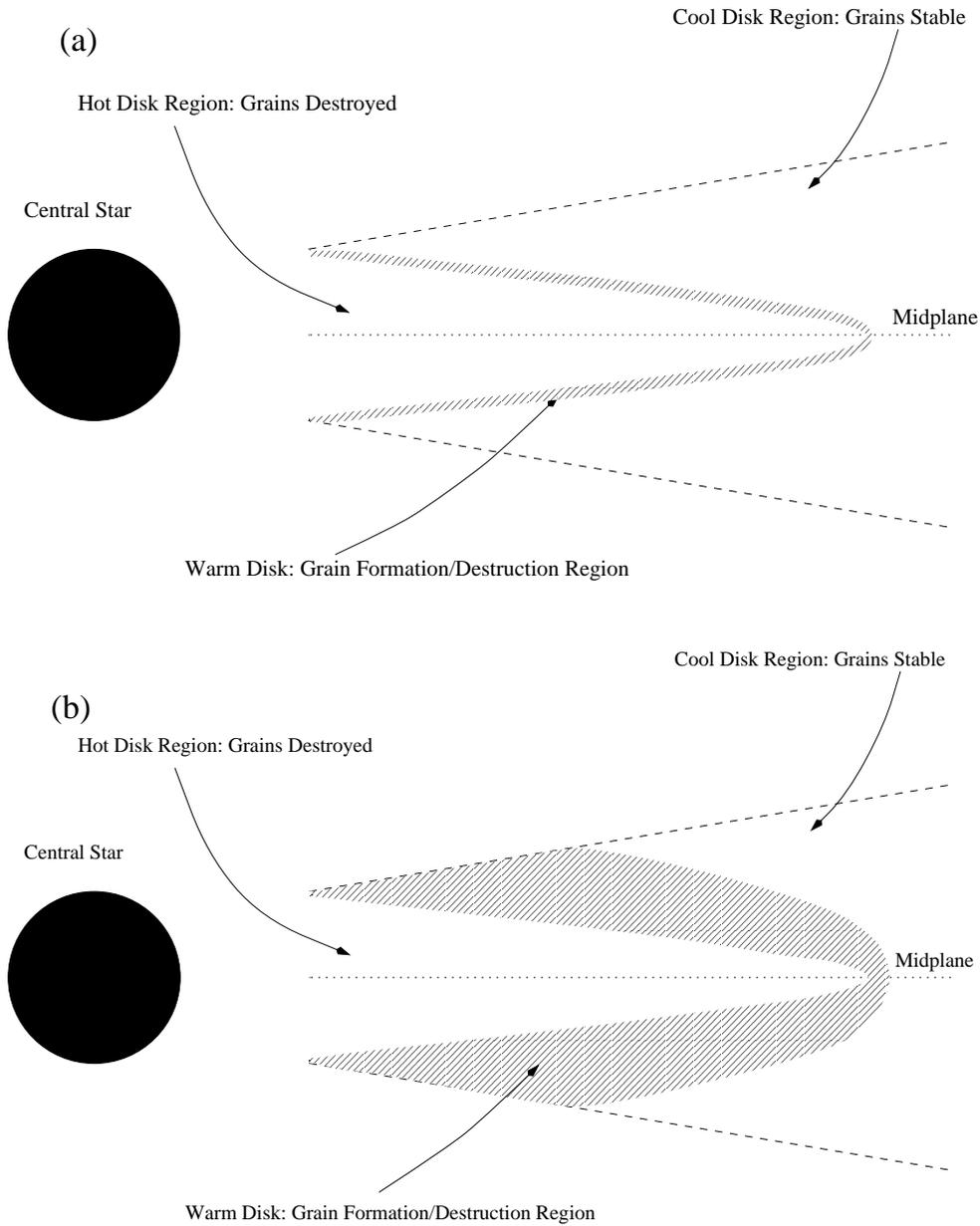}{6.61in}{0}{72}{72}{-230}{-35}
\caption[Cartoon of conditions in the inner disk where dust may be 
destroyed]
{\label{dust-convect}
(a) A cartoon of the physical conditions
of cooling prescription `A' and implemented for the simulation in figures
\ref{disk-1} and \ref{sed-1}.  Under this assumption, even if the midplane
temperature lies well above the grain destruction temperature, grains 
embedded in high altitude, cool gas block radiation from the hot midplane 
matter. (b) the modified condition (cooling prescription `B') used for 
the simulation shown in figures \ref{disk-2} and \ref{sed-2} below. Under 
this modified assumption, grains are destroyed in the midplane if the
temperature is hot enough but reform into their original distribution
only slowly at high altitudes. This allows a particular column of gas to
become less opaque so that it radiates at a higher effective temperature 
and cools more quickly.}
\end{figure}

The second assumption (the size distribution of grains) proves much 
more difficult to address. Contained within our assumption of a 
vertically adiabatic disk structure is the fact that the adiabatic 
condition arises out of a convective or turbulent medium, which acts
to smooth any entropy gradients that develop. In such a case, grains
entrained in the gas should be processed through the midplane fairly
frequently and, if the midplane temperature is hot enough, destroyed. 
As refractory materials are brought to higher altitudes where 
temperatures are lower, they will begin to reform into grains. If 
they reform quickly (compared to their vertical motion) into a similar
size distribution to their original distribution, a narrow boundary
region in which grains reform will delineate a hollowed out region 
within the disk as shown in figure \ref{dust-convect}a. On the 
other hand, if grain reformation is slow compared to speed 
of vertical motion then the region in which the grain size is 
modified from its original distribution becomes much wider (fig.
\ref{dust-convect}b). In this case, calculations assuming one 
distribution of grains may not determine the true opacity.

These differences are important because of the common use of grain
distributions similar to those given in Mathis, Rumpl \& Nordsieck 
(1977 hereafter MRN) or Kim, Martin \& Hendrys (1994) and used
in many opacity calculations. Since the opacity is a function of the 
size distribution, modifying the distribution from some canonical value 
will result in differences in the calculated opacity. For example, PMC 
have shown (see their fig. 5) that a narrow size distribution with an 
average grain size of 0.1~$\mu$m will produce Rosseland mean opacities 
which are reduced from the values obtained from their standard size 
distribution values by a factor of $\sim$~10 at temperatures above
about 150~K and an additional factor of ten above $\sim$600~K for
grains of size 0.01~$\mu$m. The differences comes because the grain
size to wavelength ratio changes as the grain size changes. As the
ratio changes from small to large, the calculations move from a 
Rayleigh scattering regime, to a Mie scattering regime and finally
to a geometric optics regime. In each, the opacities produced from the
calculations change in magnitude as different scattering and absorption
processes vary in importance.

In appendix \ref{app-grains} we model the growth of grains under the 
assumption that the midplane temperature is hot enough that the grains are 
destroyed there. The result of that calculation is that the grain distribution 
does not return to a distribution similar to that used in most grain opacity
calculations. Instead, at high altitudes grains are unable to grow
beyond a few hundredths of a micron in size and at moderate altitudes
to a few tenths of a micron on dynamical time scales of the inner 
disk. Power law distributions do not reform at any altitude.

A reduced opacity implies for our cooling prescription that the 
photosphere of the disk will be found at a lower altitude and therefore
an increased effective radiating temperature will be obtained. Using 
the tabulated opacities to obtain the location and temperature 
of the photosphere will then underestimate the cooling which actually 
takes place. Furthermore, due to the higher effective radiating
temperatures, the SED will be modified from its previous form as more
radiation is emitted at shorter wavelengths.

We investigate the effect of a modified grain opacity by 
adapting our cooling prescription to include an additional assumption.
In regions of the disk where the midplane is hot enough to vaporize
grains, we assume that the grain opacity is temporarily reduced from 
its tabulated value by a constant multiplicative factor, $R$, over the 
entire vertical column of disk matter above and below the midplane. In 
other regions of the disk we assume the opacity remains unaffected, so 
that the effective opacity is
\begin{equation}\label{mod-opac}
\kappa_{eff}(\rho,T) =
	    \cases{ R\kappa(\rho,T) & if $T_{mid}> T_{crit}$,\cr
              \phm{R}\kappa(\rho,T) & otherwise \cr  }
\end{equation}
where $T_{crit}$ is the grain destruction temperature and $T_{mid}$ 
is the disk midplane temperature. The disk photosphere temperature 
and altitude are calculated in the same manner as before, with the 
modified opacity $\kappa_{eff}$ replacing $\kappa(\rho,T)$ in eq.
\ref{tau-a} above. The disk photosphere temperature therefore will 
increase in regions where the midplane temperature is hot enough to 
destroy grains and remain unaffected elsewhere.

In eq. \ref{mod-opac}, the factor $R$ is only applied when the midplane
temperature is above the grain destruction temperature. Thus, the
duration of the opacity modification is the time span for which the 
midplane temperature remains above the dust destruction
temperature. Further, the modification affects the whole vertical 
column of material in the same way. The possibility does exist for 
further revisions of this prescription. For example, one could 
implement a buffer region of some finite vertical thickness over 
which the modification was applied or a modification factor, $R$,
which varied as a function of midplane temperature or of distance
from the star. Ultimately, we believe that such additional parameters
are of limited value in the present study because of our large 
ignorance of their exact values.

\subsection{Synthesizing Observations}\label{sed-synth}

In order to connect the physical properties of our simulations of 
accretion disks to observable quantities in real systems we synthesize 
spectral energy distribution's (SEDs) from our simulations. We calculate 
the SED using the derived black-body temperature of each SPH particle at
a particular time. We assume that the disk is viewed pole on and then 
determine the luminosity of each particle at each frequency as 
\begin{equation}\label{L-particle}
L_\nu^j = {{m_j}\over{\Sigma_j}} \pi B_\nu(T_{eff})
\end{equation}
and of the disk by summing the contributions of all of the particles.
In eq. \ref{L-particle}, $m_j$ is the mass of particle $j$, $\Sigma_j$ 
is its surface density and $B_\nu$ is the Planck function. 
The area factor is given as $m_j/\Sigma_j$ in order to avoid ambiguity
in the surface area (i.e. the smoothing length and its overlap with
other particles) defined for each particle. 

Although we neglect the luminosity of the star as a source of energy
input during the calculation, we include it in the post processed SED
calculation. We assume the star contributes to the SED as a 1~\lsun\ 
blackbody with temperature $T_{eff}=$4000~K, both values are 
typical of observed T~Tauri stars (see e.g. Osterloh \& Beckwith 1995,
BSCG, Adams \etal 1990). The star's contribution is included primarily
to make the visual comparison of our synthetic SEDs to observed systems 
simpler and to provide a constant physically meaningful calibration
of the disk emission on the plot. We neglect the contribution to
the luminosity which would be present due to the accretion of mass 
onto the star, and instead merely remove particles whose trajectory
takes them inward beyond the defined accretion boundary. We expect these
two sources of luminosity to contribute primarily to the optical and UV
spectrum, while the disk will contribute primarily at longer 
wavelengths. Therefore, for our purposes, the disk luminosity will be
well separated in frequency from both the stellar and star/disk boundary
accretion luminosities.

Comparison of the results of our simulations with observed systems 
is an important step towards interpreting those results. However,
the characteristics of SEDs from observed sources are quite diverse
and a comparison our synthesized SEDs with any specific source
would suffer from failures to duplicate features in an SED which
may only be idiosyncrasies of a single object. Therefore, in order to make
a comparison to observational data, we will compare with a `fiducial' 
SED which we derive from observed systems in the literature. We construct 
these SEDs assuming that the radial dimensions of the disk are the same 
as the initial radial dimensions of our simulations. The temperature 
profile of the disk is given by a power law with exponent $q=0.6$ (i.e. 
$T\sim r^{-q}$) and the total disk luminosity is $L_D=0.5$\lsun. We derive
the SED by summing the blackbody contribution of each radial ring,
$2\times \pi \nu B_\nu(T) 2\pi rdr$ at each frequency. Finally, as in the
synthesized SEDs above, we add a 1~\lsun, 4000~K stellar black body to
the SED.

The total disk and stellar luminosities are obtained from the results of 
BSCG, who quote luminosities between about 0.5 and 2 \lsun for the star 
and 0.1 and 1.0~\lsun for the disk. Disk masses inferred from their
observations are typically $<0.1$\msun, similar to the disks in our study.
We use a temperature exponent, $q=0.6$, derived from the BSCG study
and from the Osterloh \& Beckwith (1995) study by using the approximate
median of their fitted exponent values.

\begin{figure}
\plotfiddle{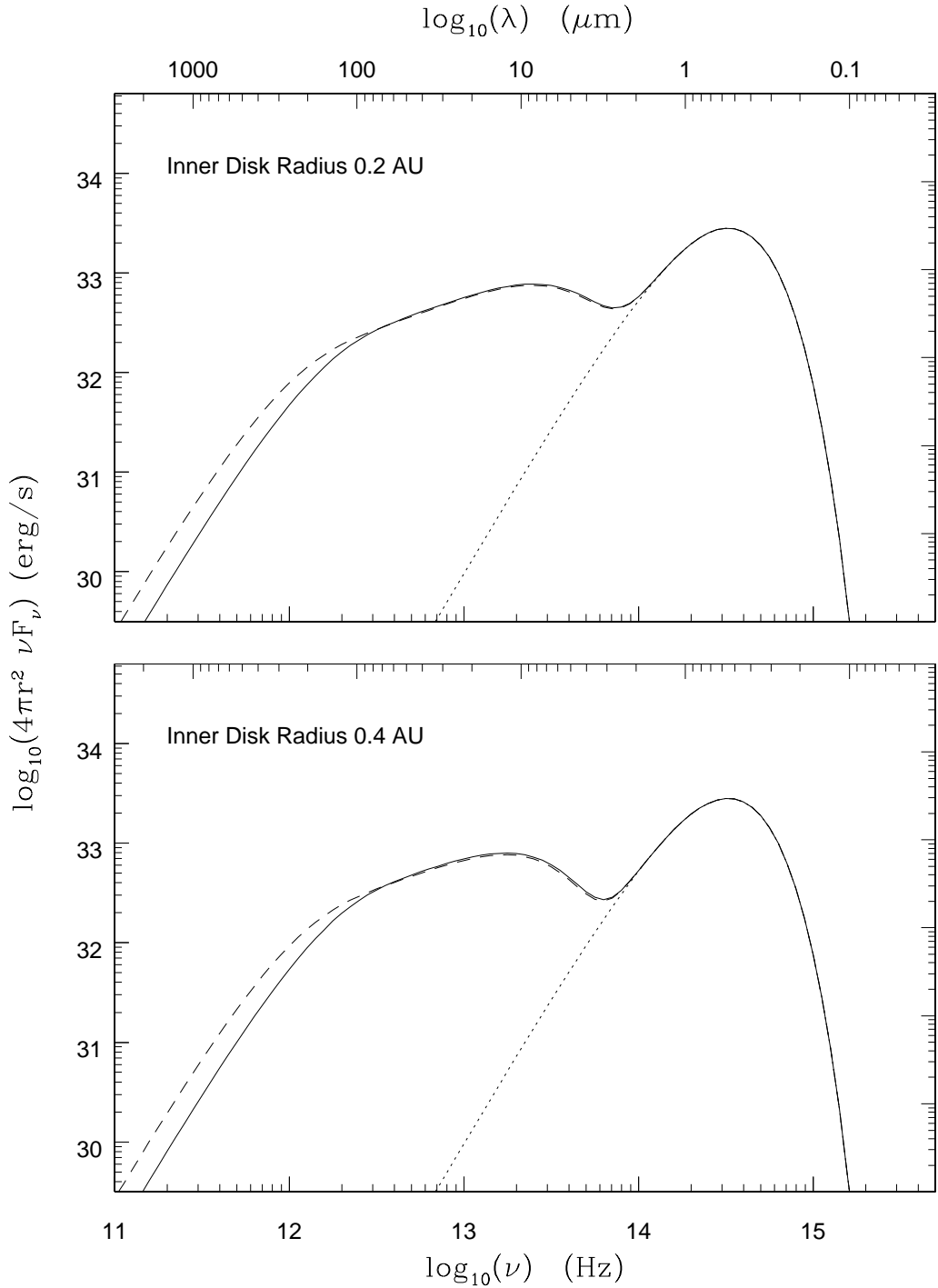}{6.61in}{0}{72}{72}{-230}{-35}
\caption[Fiducial SEDs synthesized for disks with morphologies
like those in our study.]
{\label{fiducial}
Fiducial SEDs synthesized for disks with morphologies like those in our
study. Each SED is produced by a 0.5\lsun disk with temperature power
law index $q=0.6$.  The solid line of each frame shows a disk with
outer radius, $R_D=$50~AU and the dashed line shows that produced from
a disk with $R_D=100$~AU. The top frame shows SEDs produced with an
inner disk radius of $R_I=$0.2~AU and the bottom with $R_I=0.4$~AU.
The stellar contribution is shown with a dotted line and is due to
an assumed 1\lsun, 4000~K blackbody.}
\end{figure}

In figure \ref{fiducial}, we show fiducial SEDs as described above
assuming a temperature power law index of $q=0.6$ and disk
radii of $R_D=$50~AU and $R_D=$100~AU. We also show SEDs for similar
systems with a smaller ($R_I=0.2$~AU) inner disk edge, rather than
the $R_I=0.4$~AU usually assumed in our simulations.  Both sets 
of SEDs show a gap in their near infrared emission corresponding to
material assumed to be missing between the inner disk edge and the
stellar surface, and so appear somewhat dissimilar to many observed
SEDs. Since our purpose for these fiducial SEDs is to compare the flux
from our simulated disks with the portion of flux which comes from the
same spatial part of the disk as in observed system, this discrepancy
poses no serious problem. The SEDs with the larger inner disk radius 
exhibits a gap between 1 and 10$\mu$m and which is deepest at about
3$\mu$m. The smaller inner disk radius exhibits emission which is enhanced
by a factor two over that produced by the disks with $R_I$=0.4~AU. At 
long wavelengths, the 100~AU fiducial SEDs exhibit emission further 
into the far infrared than their 50~AU counterparts, but are otherwise
similar to the 50~AU systems.

\subsection{Units: The Physical Scale of the System}

The introduction of a cooling mechanism requires an introduction of
a physical scale to the simulations. We shall assume quantities with
values typical of the early stages of protostellar evolution. The star
mass will be assumed $M_*$ = $0.5$\msun and the disk radius of either
$R_D=50$~AU or $R_D=100$ AU as noted in table \ref{cool-params}. Time 
units are given in either years or the disk orbit period defined by 
\td=$2\pi\over{\sqrt{GM_*/R_D^3}}$ which, with the stellar mass and
disk radii given above is equal to about 500 or 1400 years for disk
radii of $R_D=50$ or $100$~AU respectively.

\section{The Simulations}\label{simulations}

\pone showed that the character of disk evolution undergoes a 
marked change between disk masses of \mrat~$=0.2$ and \mrat~$=0.4$.
In this paper, we will concentrate on studies of a disk at the lower edge
of this mass boundary. In the following discussion, we simulate the
evolution of a disk with a mass ratio of \mrat~$=0.2$ and with an assumed
initial minimum Toomre stability of \qmin~$=1.5$, under varying physical
assumptions. Initial parameters of our simulations are tabulated in table
\ref{cool-params}. The first column of the table represents the name of
the simulation for identification. The second column defines the 
resolution (in number of particles) and column 3 the disk radius. The
assumed opacity modification factor (see sections \ref{energy-diss}
and \ref{cool-improve}) is defined in column 4 and the total simulation
time of each simulation in the remaining column. We examine the qualitative
nature of the simulations first, then examine in detail the structures 
which form and their characteristics.  

We have run three series of simulations under different assumptions 
about the heating and cooling mechanisms important for the disk. The first
set of simulations proceed under the assumption that gas and grains exist
in equilibrium everywhere in the disk and that the grain size distribution
is well modeled by the distributions used in opacity calculations in the
literature (e.g. PMC, Alexander \& Ferguson 1994). Vaporized material, 
upon entering a region cool enough for it to form grains, does so instantly
and in such a way as to reproduce its original grain size distribution, as
defined by PMC. These simulations are denoted by a leading `a' (100~AU disks)
or `A' (50~AU disks) in the simulation name in table \ref{cool-params}.

The second set of simulations relaxes the assumption that refractory
materials reform into their original size distribution quickly. Instead, 
we assume that they reform their original distribution slowly so that
their size distribution and therefore their opacities may be modified 
from their original form. These simulations are denoted with a leading `B' 
in table \ref{cool-params}.

We have also run models of disks under the same `isothermal evolution'
assumption used in \pone. In these simulations we assume a fixed 
temperature profile as a function of radius and a ratio of specific
heats, $\gamma$, equal to unity. These simulations are denoted with a 
leading `I' in Table \ref{cool-params}. In each case, a capital 
letter refers to a disk with outer edge at 50 AU while a lower case letter
refers to a disk with outer edge at 100 AU. In each case `lo', `me'
and `hi' refers to a simulation with low, medium or high resolution as
defined in column 2, and with the `B' simulations, the number 1--5
corresponds to an assumed opacity modification.

In the following sections we first describe the morphology of the systems
as they evolve and the energy output they produce. We then describe
the azimuth averaged thermodynamic structure produced by the evolution.
Finally we close the section with a discussion of the origin of the
differences between the results produced by isothermally evolved 
simulations as opposed to the `A' and `B' runs in our current study.

\subsection{Morphology and Spectral Energy Distributions}\label{morph}

\begin{figure}
\plotfiddle{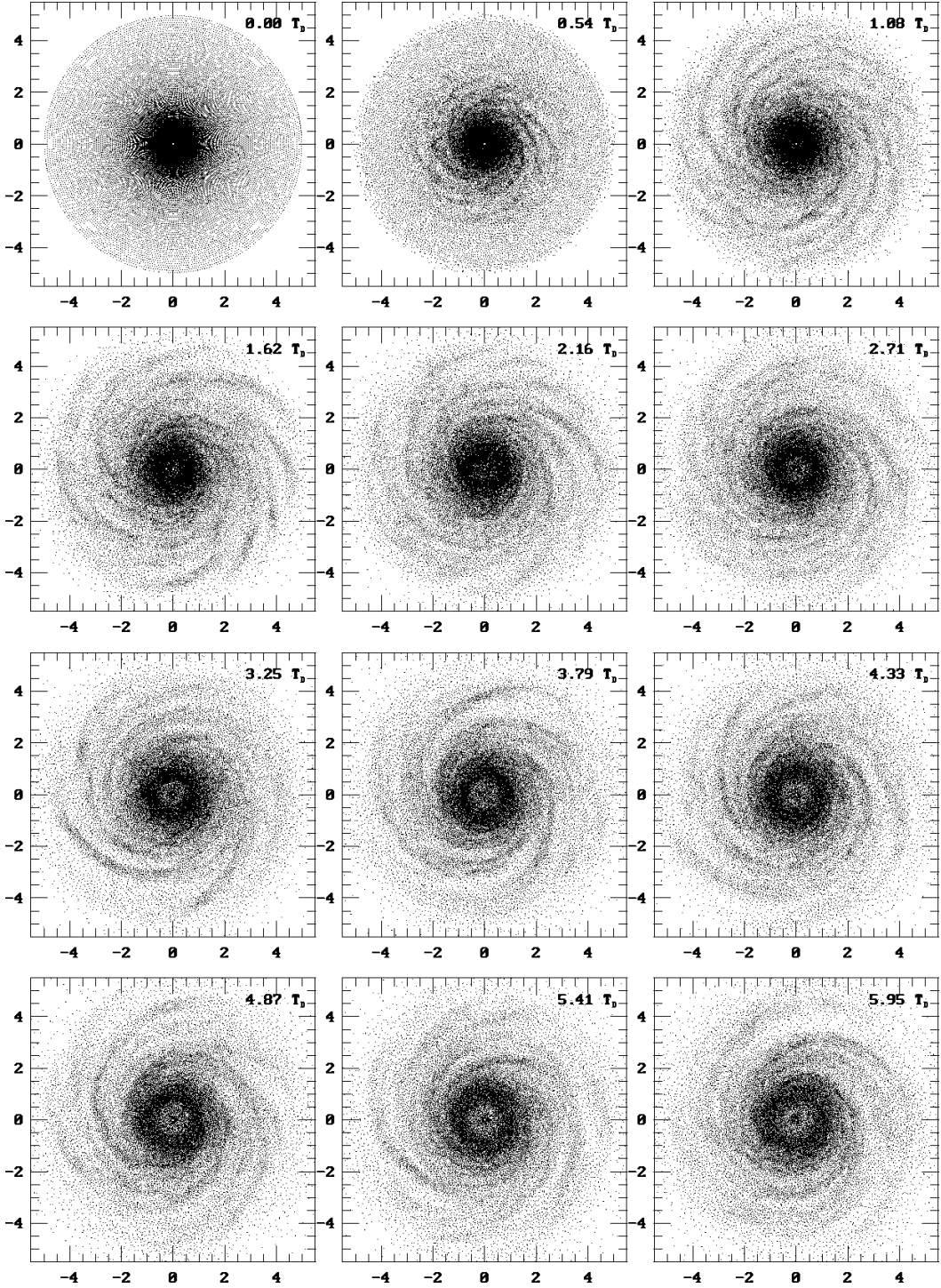}{7.33in}{0}{75}{75}{-235}{-20}
\caption[SPH simulation of a low mass disk with our `A' cooling
prescription]
{\label{disk-1}
A time series of SPH particle positions for a disk of mass $M_D/M_*$=0.2
and initial minimum Q$_{min}$=1.5 (simulation {\it A2me}). Spiral structure
varies strongly over time. Length units are defined as 1=10AU and time in
units of the disk orbit period \td= $2\pi\sqrt{ R_D^3/GM_*}$. With the
assumed mass of the star of 0.5~$M_\odot$ and the radius of the disk of 
50~AU, \td$\approx$~500~yr.}
\end{figure}

\begin{figure}
\plotfiddle{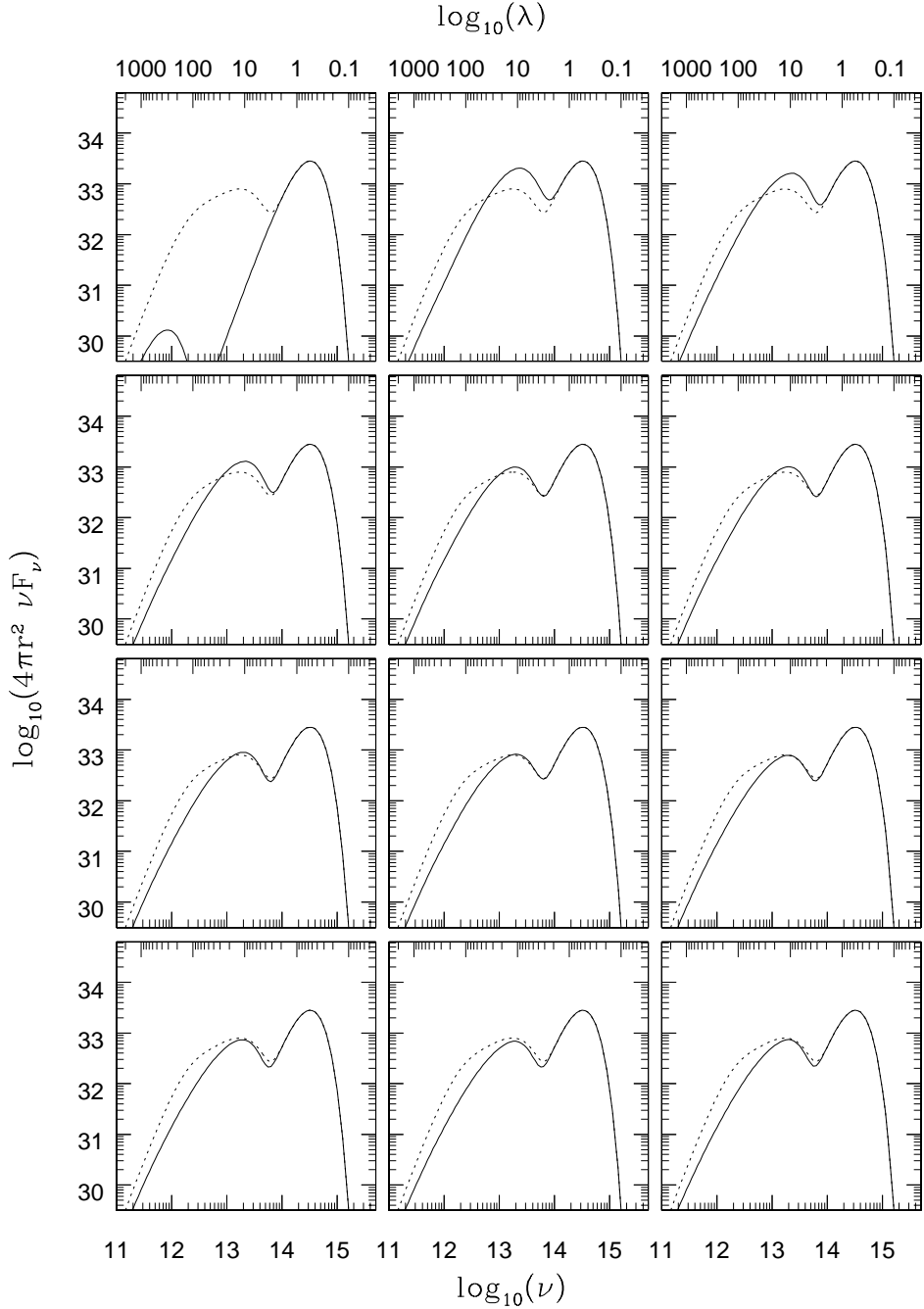}{6.57in}{0}{68}{68}{-220}{-35}
\caption[Spectral energy distribution's for the simulation shown in
figure \ref{disk-1}]
{\label{sed-1}
Spectral energy distribution's for the disk shown in figure \ref{disk-1}
(simulation {\it A2me}).  Each panel in this mosaic corresponds to the 
analogous panel in figure \ref{disk-1}.  The horizontal axes of each
panel are labeled in frequency (bottom tick marks) and in wavelength 
(top tick marks). A `fiducial' SED produced with parameters corresponding
to observed systems is shown with a dotted line. The SEDs produced from
our simulations do not reproduce the observed luminosity spectrum 
around T Tauri stars, producing instead insufficient flux at both long and 
short wavelengths.}
\end{figure}

Using the physical assumptions outlined above and cooling using the `A'
prescription we have completed a series of simulations with initial 
minimum Toomre $Q=1.5$. Snapshots of the evolution of simulation
{\it A2me} are shown in figure \ref{disk-1} and of its synthesized SED 
in figure \ref{sed-1}. As in our previous isothermally evolved simulations 
(\pone) growth of instabilities begins in the inner regions of a disk, 
engulfing the entire system over the course of about 1~\td. Initially,
spiral structures develop in the inner disk, but are later suppressed
by the heating which occurs there. The spiral structures which develop
throughout the whole disk are multi-armed and change their shape and
character over orbital time scales. At times they become somewhat 
filamentary, but in no case do they become as filamentary as in the
isothermally evolved simulations of \pone.

In order to make quantitative comparisons between systems evolved under 
different physical assumptions, we examine the amplitude of various spiral
patterns during the simulations. We derive pattern amplitudes using the 
same procedure as in \pone.  We compute the amplitude of spiral patterns
by Fourier transforming (in the azimuth coordinate) a set of annuli 
spanning the disk in radius. The amplitude of each Fourier component is
then defined as $|A_m(r)|= (|\Sigma_m(r)|/|\Sigma_0(r)|)$, where
$\Sigma_m(r)$ is
\begin{equation} \label{mode-amp}
\Sigma_m = {{1}\over{\pi}}\int_0^{2\pi}
		e^{im\phi}\Sigma(r,\phi)d\phi,
\end{equation}
for $m>0$. The $m=0$ term is defined with a normalization of $1/{2\pi}$.
With this normalization, the $\Sigma_0$ term is the mass of the disk and
the amplitudes, $A_m$, are dimensionless quantities.

As shown in \pone, SPH cannot follow low amplitude growth of structure 
in disks. In a simpler test case in which a short duration linear phase
was possible, the results produced from our SPH code and from a Piecewise 
Parabolic Method (PPM) code (which is able to follow low amplitude growth 
for longer periods) were similar. In both these test cases and disk evolution 
simulations the gross morphology of the SPH and PPM simulations were also 
similar. Therefore, apart from the inability to derive growth rates of spiral 
structures or assign linear growth modes to the system, we expect little effect 
on our results due to this characteristic of the SPH method.

\begin{figure}
\plotfiddle{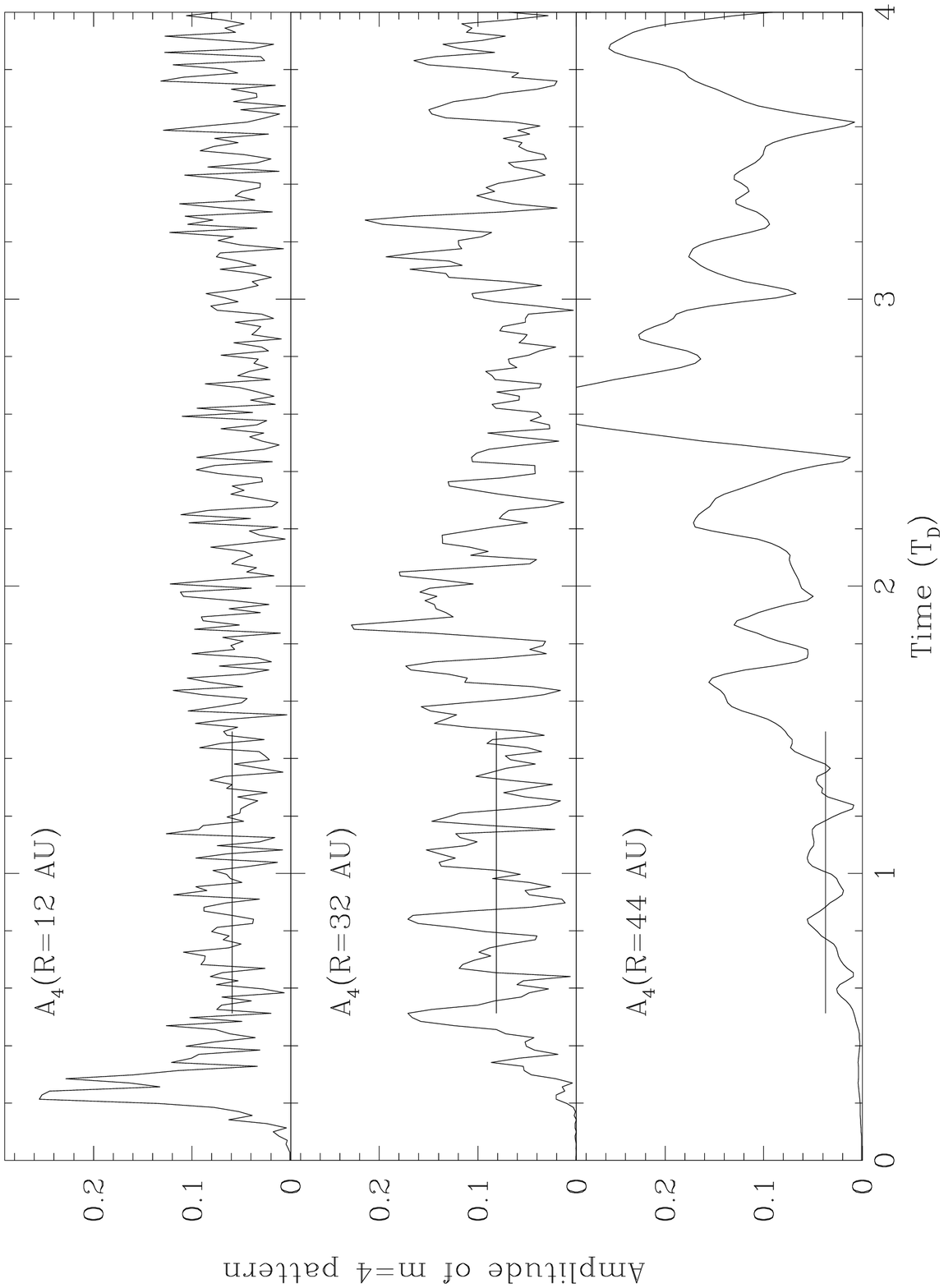}{3.02in}{-90}{45}{45}{-180}{285}
\plotfiddle{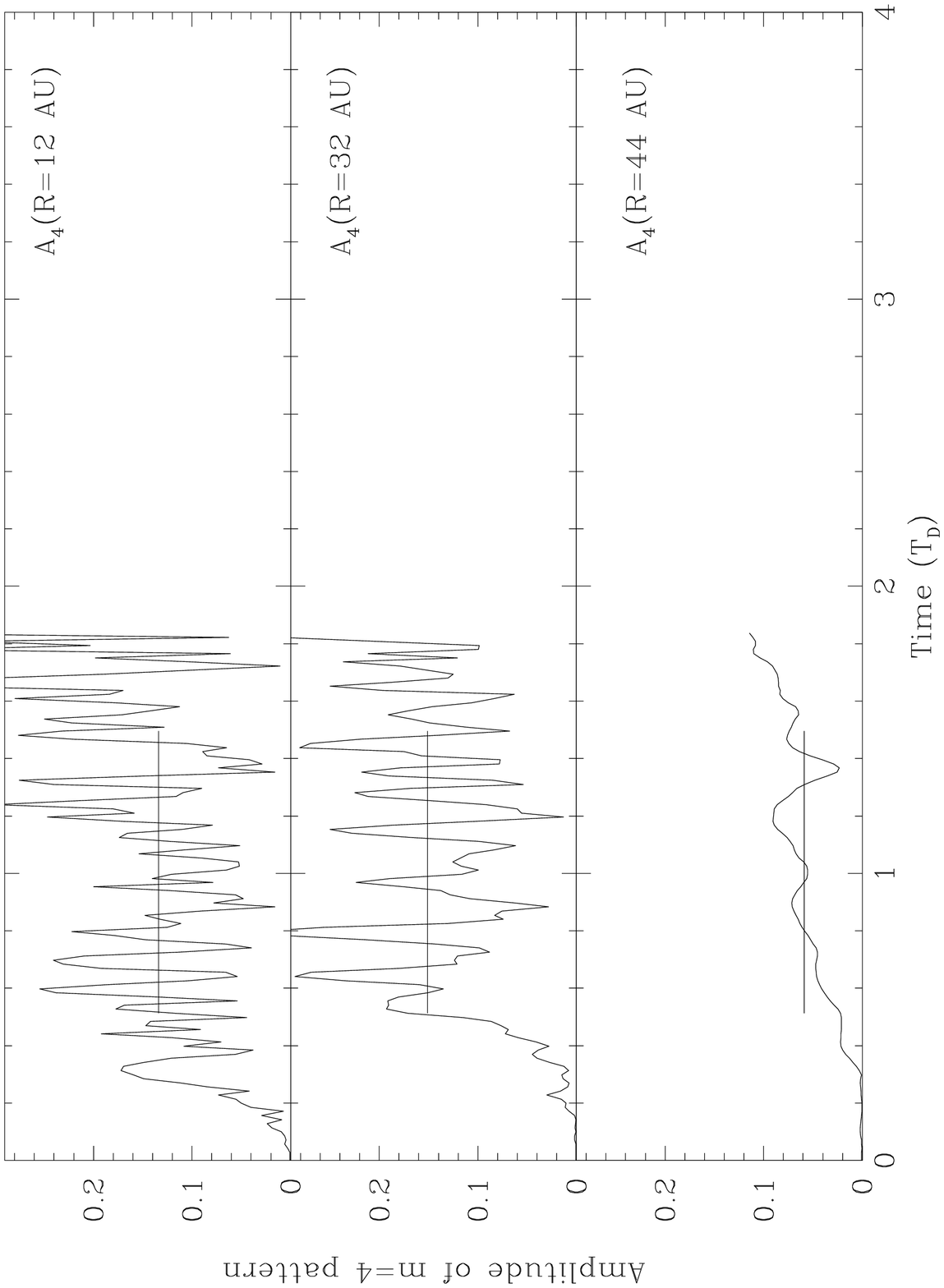}{3.02in}{-90}{45}{45}{-180}{265}
\caption[Amplitude of the $m=4$ spiral pattern at several distances from the
star from cooled and isothermally evolved simulations as a function of time.]
{\label{timeave-amp}
The amplitude of the $m=4$ spiral pattern as a function of time 
at several distances from the star. The top panel show the amplitudes
derived from the simulation shown in figure \ref{disk-1} ({\it A2me}),
while the bottom panel shows the isothermally evolved simulation 
{\it I2me}. Solid horizontal lines denote the fitted average pattern 
amplitude and extend over the time span for which the average was 
calculated.}
\end{figure}

In figure \ref{timeave-amp} we show the amplitudes at three radii
for the $m=4$ pattern produced from an isothermally evolved simulation 
and the disk shown in figure \ref{disk-1}. Short period variation occurs
in the pattern amplitudes shown, but pattern growth has saturated to the 
extent that time averages of the amplitudes can be used to characterize much
of the system. We derive the time average of a given pattern using the time
interval between \td$=0.5$ and \td$=1.5$, shown with solid horizontal
bars. These limits are used in order to ensure than most of the disk has
in fact reached its saturation amplitude (for the beginning limit), but
has not evolved long enough to form clumps (in the case of the isothermally
evolved runs). The time intervals used for both sets of runs are identical
in order to ensure that the comparison can be as close as possible. For 
example, for the outer part of the disks, it is clear that the `A' 
simulation reaches its highest amplitudes well after the isothermally
evolved simulation forms clumps and cannot be evolved further. In order
to make a fair comparison of the amplitudes we restrict the averages
to the same time window.

As we found for the isothermally evolved disks studied in \pone,
the `A' simulations (figure \ref{timeave-amp}) also produce time 
variation of the pattern amplitudes on the same time scales as the
orbital periods in the disk. The amplitude of the variation present
at each radius is about a factor two above and below the mean amplitude.
The time variation is important in a dynamical context because of the 
possible observational consequences. Dynamical variation produces 
heating from shocks as various portions of the disk collide with each 
other or break apart and reform. We will discuss the energy output 
obtained from our simulations, its origin and variation in section 
\ref{sed-dynam} below.

\begin{figure}
\plotfiddle{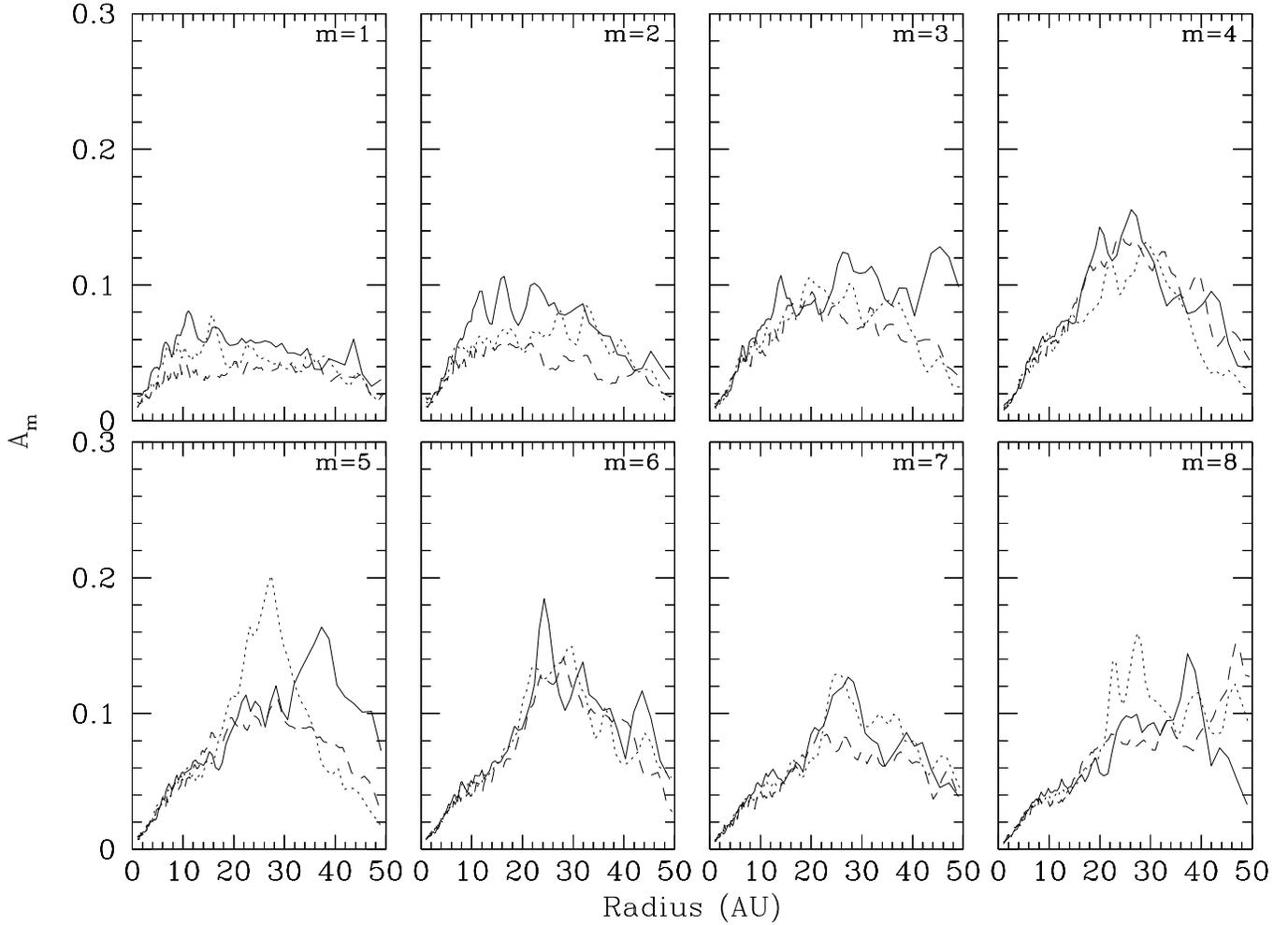}{5.5in}{-90}{70}{70}{-280}{500}
\caption[Amplitude of several spiral patterns from simulations evolved
under the `A' cooling prescription as a function of radius. ($R_D=$50~AU)]
{\label{radius-amp-a50}
The time averaged amplitude of the $m=1-8$ spiral patterns as a function 
of radius for the disk shown in figure \ref{disk-1} as well as it's high
and low resolution counterparts. The solid line denotes the low resolution
run ({\it A2lo}), while the dotted and dashed lines represent the moderate
and high resolution runs, {\it A2me} and {\it A2hi}, respectively.}
\end{figure}

\begin{figure}
\plotfiddle{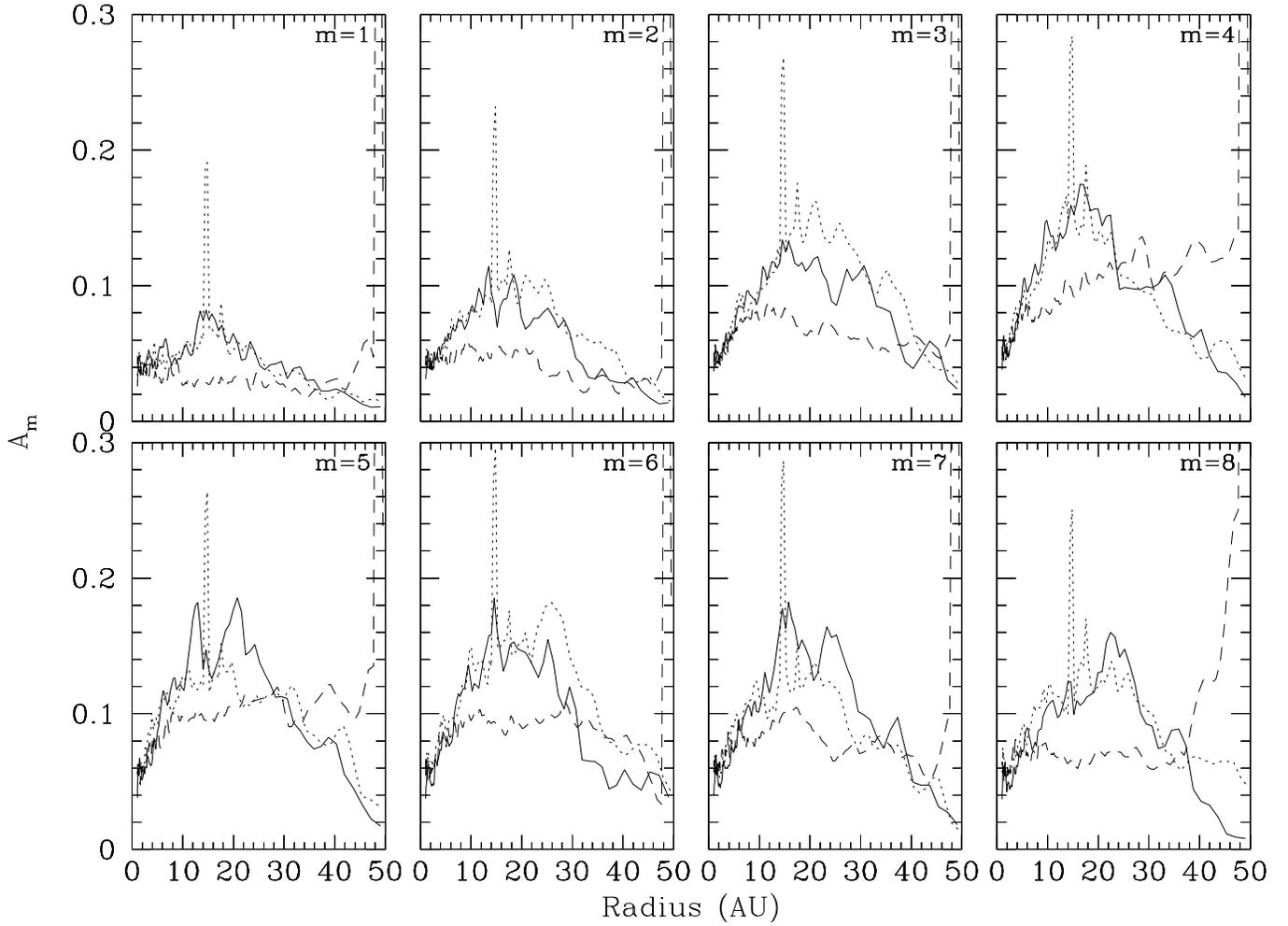}{5.5in}{-90}{70}{70}{-280}{500}
\caption[Amplitude of several spiral patterns from isothermally evolved
simulations as a function of radius. ($R_D=$50~AU)]
{\label{radius-amp-iso50}
The time averaged amplitude of the $m=1-8$ spiral patterns as a function 
of radius for isothermally evolved simulations with the same initial 
conditions as those shown in figure \ref{radius-amp-a50}. Here again, the 
solid line denotes the low resolution run ({\it I2lo}), while the dotted 
and dashed lines represent the moderate and high resolution runs ({\it I2me} 
and {\it I2hi}, respectively). Spikes appearing in the plots for the moderate
resolution run (at $\sim$12 AU) and the high resolution run (near the
outer edge) are both artifacts of clumps which formed just prior to the 
termination of the fit. They should be disregarded in comparisons with
figure \ref{radius-amp-a50}.}
\end{figure}

The time averaged amplitudes are shown as functions of radius for the same 
simulations in figures \ref{radius-amp-a50} and \ref{radius-amp-iso50}.
Approximately the inner third of the cooled disk shows suppressed
pattern amplitudes relative to the isothermally evolved simulation. In \pone
we showed that this region was the region most likely to form collapsed
objects in the isothermal evolution limit. With our new series of simulations
this conclusion must be revised. No clumps form in this region in these
simulations, as they do in the isothermally evolved runs. The time averaged
pattern amplitudes decrease as radius increases in the outer half of each
disk because the patterns have not fully saturated there during the period
over which the average is taken. Averaged over longer time baselines later
in the simulation, the averages of the `A' and `a' simulations reach values
of $\sim0.1-0.15$ throughout the outer disk. 

A similar deficit in the amplitude of spiral structure in the inner 
third of the disk is present in disks with outer radii at 100~AU. These 
simulations show the same amplitude structure as in the $R_D=$50~AU disks
above, stretched over the 100~AU extent of the disk rather than over 50~AU.
Because of the great similarity we omit a separate figure displaying
the result. Attempts to compare these results to isothermally evolved
simulations were unsuccessful, since such simulations tend to develop
clumps on the same physical time scale (750--1000 yr) as with the 
$R_D=50$~AU isothermal runs. Evolution of the 100~AU simulations had to
be terminated before the outer disk has completed even a single orbit. 

In spite of the large pattern amplitudes, only one simulation ({\it a2me},
with an assumed 100~AU disk radius) showed evidence of possible spiral 
arm collapse into a clump. A portion of a spiral arm near the outer disk
edge began to show signs of collapse at time \td$=2.3$. At this point
the cooling prescription failed to determine the vertical structure for
the particles near the forming clump and further evolution of this simulation 
becomes impossible. We cannot determine whether collapse would continue or 
dissipate once again into the background flow. No other simulations show 
indications of clump formation, including the lower and higher resolution 
counterparts of this run ({\it a2lo, a2hi}). Therefore we do not consider 
the signs of collapse in the {\it a2me} simulation to be reliable evidence 
for spiral arm collapse in our simulations. More generally, since the 
simulations in our present study do not reliably form clumps, we conclude 
that spiral arm collapse in marginally self gravitating systems like those 
we study is rare.

The SEDs synthesized from the simulation {\it A2me} are shown in figure 
\ref{sed-1}. The simulation produces a double peaked spectrum with one 
peak dominated by the assumed 4000~K stellar black body and the other 
between 10 and 20$\mu$m ($\sim$3--400~K). The long wavelength end
($>30\mu$m) portion of the SED does not match the SED well: insufficient 
radiation is emitted.  We defer a detailed discussion of the differences 
in the long wavelength behavior to section \ref{therm-gen}. The near 
infrared band (defined for our purposes as wavelengths from $\sim 1\mu$m
to 5$\mu$m) does match the fiducial SED with the same morphology as our
simulated disk, however this turns out to be a fortunate accident rather 
than a real agreement. Lower (higher) resolution simulations produce more
(less) flux, due to variations in the dissipation rate with resolution
(see appendix \ref{resol-diss}). We expect that if resolution were increased
still further, the near IR flux would continue to decrease.

The absence of near IR flux in both the fiducial SED and our simulations
is in part due to the relatively large truncation radius of our disk 
(recall that particles are accreted by the star inside 0.4~AU). The size 
of the accretion radius is motivated by numerical time step constraints
rather than a specific physical basis, but does model the physical 
existence of a 0.4~AU boundary region between the inner disk edge and 
the star if one that size exists. To test the effect of a smaller 
boundary region we ran simulations similar to {\it A2me} and {\it A2hi} 
but with a reduced accretion radius, $R_{acc}=$0.2~AU. We found that 
the total disk luminosity does increase and the near IR flux deficit is
partially filled in, but that a near IR deficit still remains. It remains
unclear the full effect that a still smaller inner truncation radius
or higher resolution will have on the SED.

In several regions, our cooling model will break down. For example, 
reprocessed stellar illumination will heat the gas and modify the assumed 
vertically adiabatic structure. Another failure will manifest 
because of the combined effect of the large, low temperature opacities
and the low temperatures implied by our model at high altitudes above the 
midplane. For any conditions at the midplane of the disk, our vertical 
structure calculation produces a region which is both cold and sufficiently
dense to make the column optically thick at high altitude. Photosphere 
temperatures higher than $\sim$500~K are not obtained in this case and the 
disk is unable to radiate efficiently in the near infrared. Such a condition
will accurately model real systems only as long as the opacity source
(dust) remains unaffected by temperature or by vertical transport. In 
regions where the gas is hot enough to vaporize embedded dust grains, 
this assumption will fail.

\subsubsection{An Attempt to Improve the Cooling Prescription}\label{cool-improve}

The dust which produces nearly all of the low temperature opacity
will be destroyed in the inner disk because the temperatures in the disk 
midplane can rise above the vaporization temperature of the dust. At 
higher altitudes temperatures are cooler and grains can reform and grow.
We have studied the growth and reformation of dust grains when the
midplane is hot and found (Appendix \ref{app-grains}) that after being
destroyed, grains require several months to grow to sizes of a few tenths
of a micron. At high altitudes where densities and the probability of 
dust-dust collisions are lower, grains grow to only $\sim .02-.03\mu$m in 
the same time. In no case do grains reform into a distribution similar to 
that assumed for most Rosseland mean opacity calculations (e.g. PMC, 
Pollack \etal 1994).

The Rosseland opacities in use in most calculations to date are calculated
with a specific grain size distribution and constitution based on models
of the dust in the interstellar medium (ISM). In the case of the disk evolution,
both densities and temperatures are much higher than the ISM and our
calculations in the appendix show that grain reprocessing occurs on
orbital time scales in the inner disk. We have not calculated the effect 
that the reprocessing discussed in appendix \ref{app-grains} will actually
have on the opacity, however other calculations for different grain sizes 
(Pollack \etal 1994, Miyake \& Nakagawa 1993) have shown significant
differences between the frequency dependent opacities calculated for 
different sized grains. Therefore, we conclude that in the inner disk
where grain size evolution occurs on short time scales, accurate opacities
require a calculation which accounts for the time dependent size and 
composition changes of the grains. Radiative transport calculations in
which such effects are not included may yield inaccurate spectra and 
thermodynamic structure of the disk. 

We have investigated the effect of a grain reprocessing on a disk system
by performing a series of simulations which begin with the same initial 
model as simulation {\it A2me}, but with an artificially modified opacity. 
In these simulations, we calculate the instantaneous midplane temperature
for each location in the disk and if it is above the grain destruction 
temperature, we multiply the opacity by a factor $R$ (see eq. 
\ref{mod-opac}). We vary $R$ between $R=0.001$ and $R=0.01$ for different
simulations. Within a single simulation $R$ is constant. The factor $R$ 
models in a crude manner the effect of a modified opacity limited to 
regions where the temperatures are hot enough that dust may be destroyed 
and reform. The photosphere temperature is determined as before, but 
will be modified because the reduced opacity allows photons closer to
the hot disk midplane to escape into space.

\begin{figure}
\plotfiddle{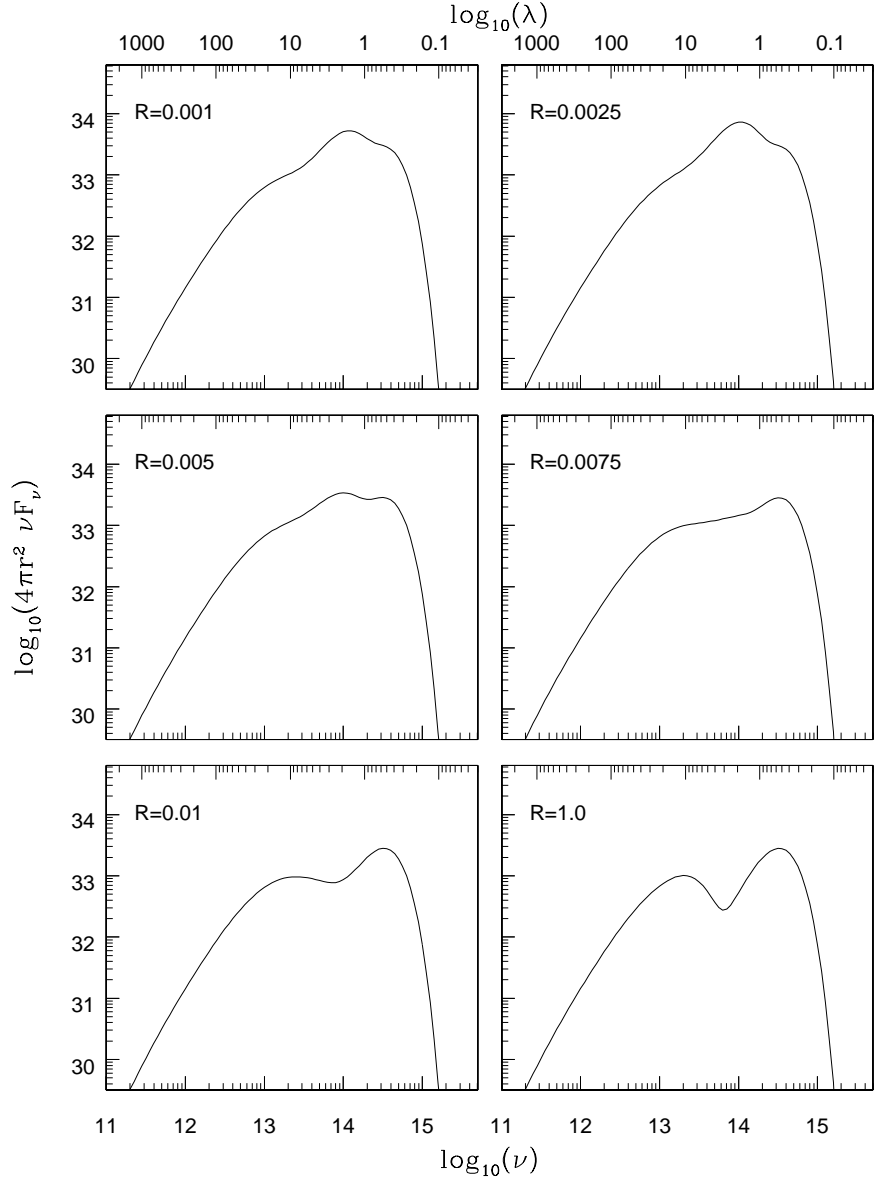}{6.0in}{0}{60}{60}{-200}{20}
\caption[Synthesized SEDs of simulations with varying modifications in
grain opacity]
{\label{varopac-sed}
Synthesized SEDs of simulations with varying modifications in grain 
opacity. The initial conditions of these simulations are identical to 
those of simulation {\it A2me}, but are each carried out under varying
physical assumptions. To remove short period time variation, we plot
a time averaged SED over the time from \td=1 to \td=5. In order of
increasing $R$, the simulations shown are {\it B2m1, B2m5, B2m3, B2m4,
B2m2} and {\it A2me}.}
\end{figure}

We show time averaged SEDs derived from these simulations in fig. 
\ref{varopac-sed}.  Depending on the assumed opacity modification, $R$,
the near infrared flux `turns off' or `turns on'. With large $R$ 
the SED appears similar to that shown above in figure \ref{sed-1} (whose 
time average is reproduced in the lower right panel of fig. \ref{varopac-sed}).
With $R$ small it appears permanently in an `outburst' phase in which the 
1--5$\mu$m band is enhanced by as much as a factor of ten over the stellar 
contribution. 

It is notable that the values of $R$ which are required to produce 
any change in the 1--5$\mu$m flux are quite small. In fact, the 
modifications in the opacity values are not inconsistent with an opacity
consisting only of a contribution from gas rather than from both gas
and grains. A `grain free' calculation (D. Alexander: personal communication)
of the opacity using the model of Alexander \& Ferguson (1994) down to
1000~K produces opacities which are similar in magnitude to our modified
values. We conclude that unless grains are completely destroyed in
a given column of disk matter no effect will be present in the
emitted spectrum.

Both a frequency dependent radiative transfer code, incorporated into
the hydrodynamic calculation at each time step, and a recalculation of 
the Rosseland opacity for each $\rho$, $T$, and grain size distribution 
accessible to our simulations are beyond the scope of the present work. 
As a parameterized factor however, we bracket the difference from the 
tabulated PMC opacities to a factor between 0.001 and 0.01 times the PMC 
values in regions where the midplane temperature rises above the grain 
vaporization temperature. Below, we shall implement a `standard'
factor of 0.005 (high resolution) or 0.0075 (medium resolution) times
the tabulated PMC values in such regions. 

Other regions of the disk may also experience the effects of grain
reprocessing and modified opacities. Except for the hottest regions,
grain size evolution occurs on a time scale of many orbits
(Weidenschilling 1997) rather than less than one, so we would expect
less short term variation in the energy output of the disk. Long
term differences may still occur as the size distribution evolves
and the opacity changes.

\subsubsection{Morphology and SEDs using modified opacities}\label{new-morph}

\begin{figure}
\plotfiddle{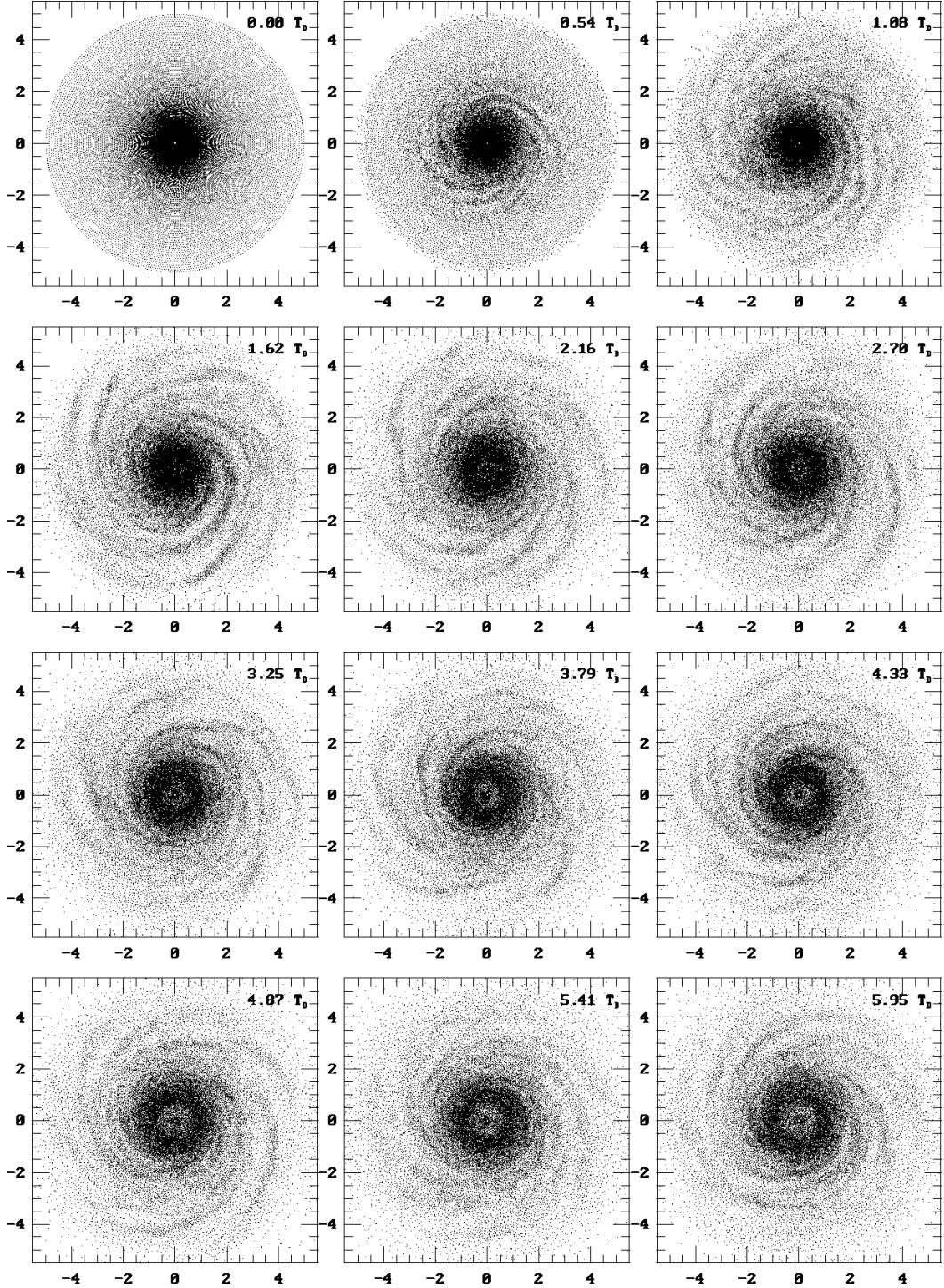}{7.15in}{0}{75}{75}{-235}{-20}
\caption[SPH simulation of a disk using our `B' cooling prescription]
{\label{disk-2}
A time series mosaic of SPH particle positions for the same initial 
condition as shown in figure \ref{disk-1}. The cooling prescription 
used in this simulation has been revised to include dust destruction 
over entire vertical columns as shown in figure \ref{dust-convect}b 
with $R=0.0075$ (simulation {\it B2m4}; see text and table
\ref{cool-params}). The gross morphology of the structures that develop
is quite similar to that shown in figure \ref{disk-1}, even though 
the cooling assumptions and gas temperatures in the hottest (inner 
disk) regions are different.}
\end{figure}

\begin{figure}
\plotfiddle{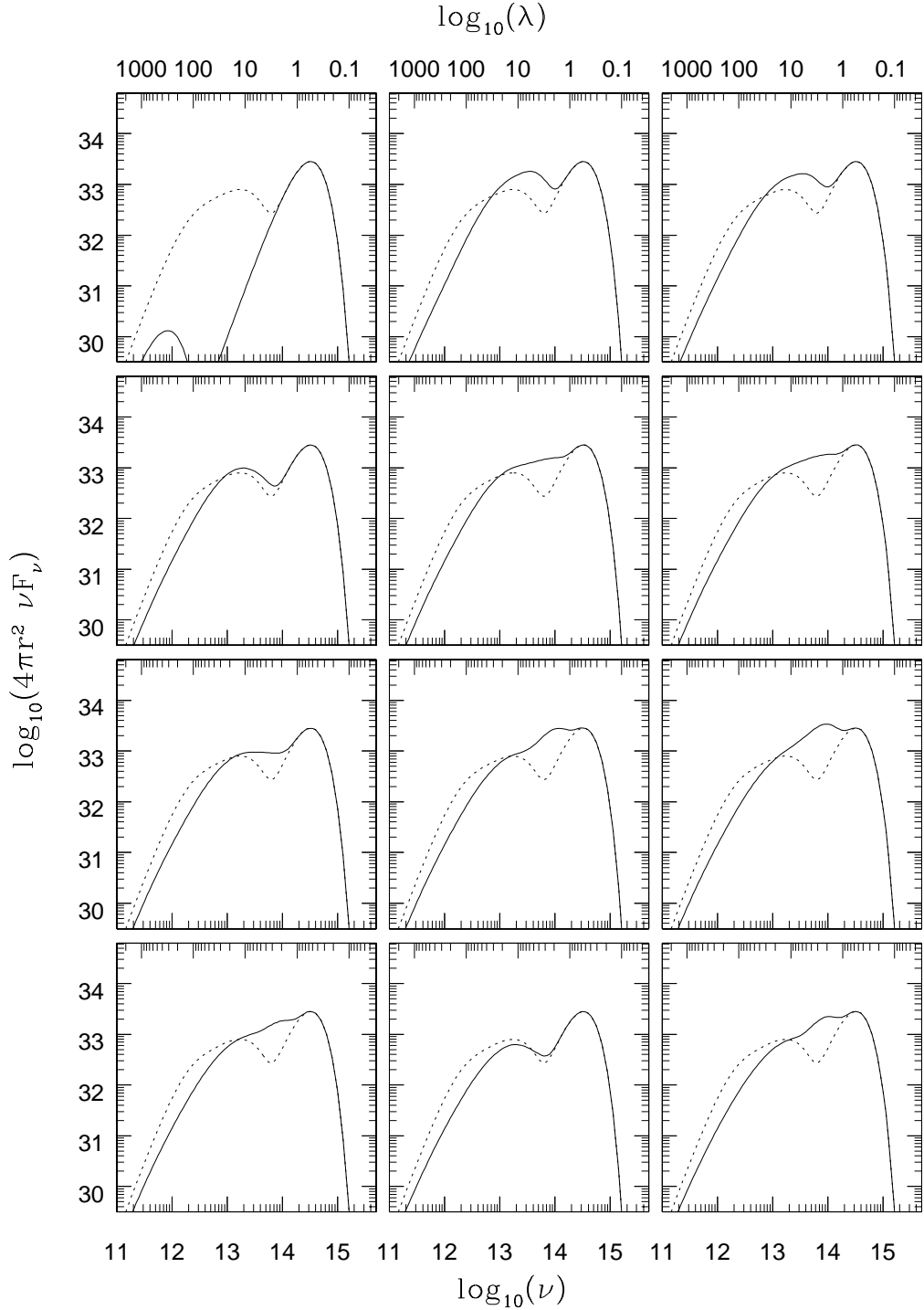}{6.97in}{0}{73}{73}{-240}{-35}
\caption[Spectral energy distribution's for the simulation shown in
figure \ref{disk-2}]
{\label{sed-2} 
SEDs for the simulation shown in figure \ref{disk-2} ({\it B2m4}). Under
the modified cooling assumption shown here, a closer correspondence to
observed systems is found at near IR wavelengths. Substantial variations
in the shape are seen over time scales of a few tens of years to a few
hundred years. At some times the contribution of the star is partially
masked by emission from the disk, while at others the star contributes
nearly all of the short wavelength flux.}
\end{figure}

\begin{figure}
\plotfiddle{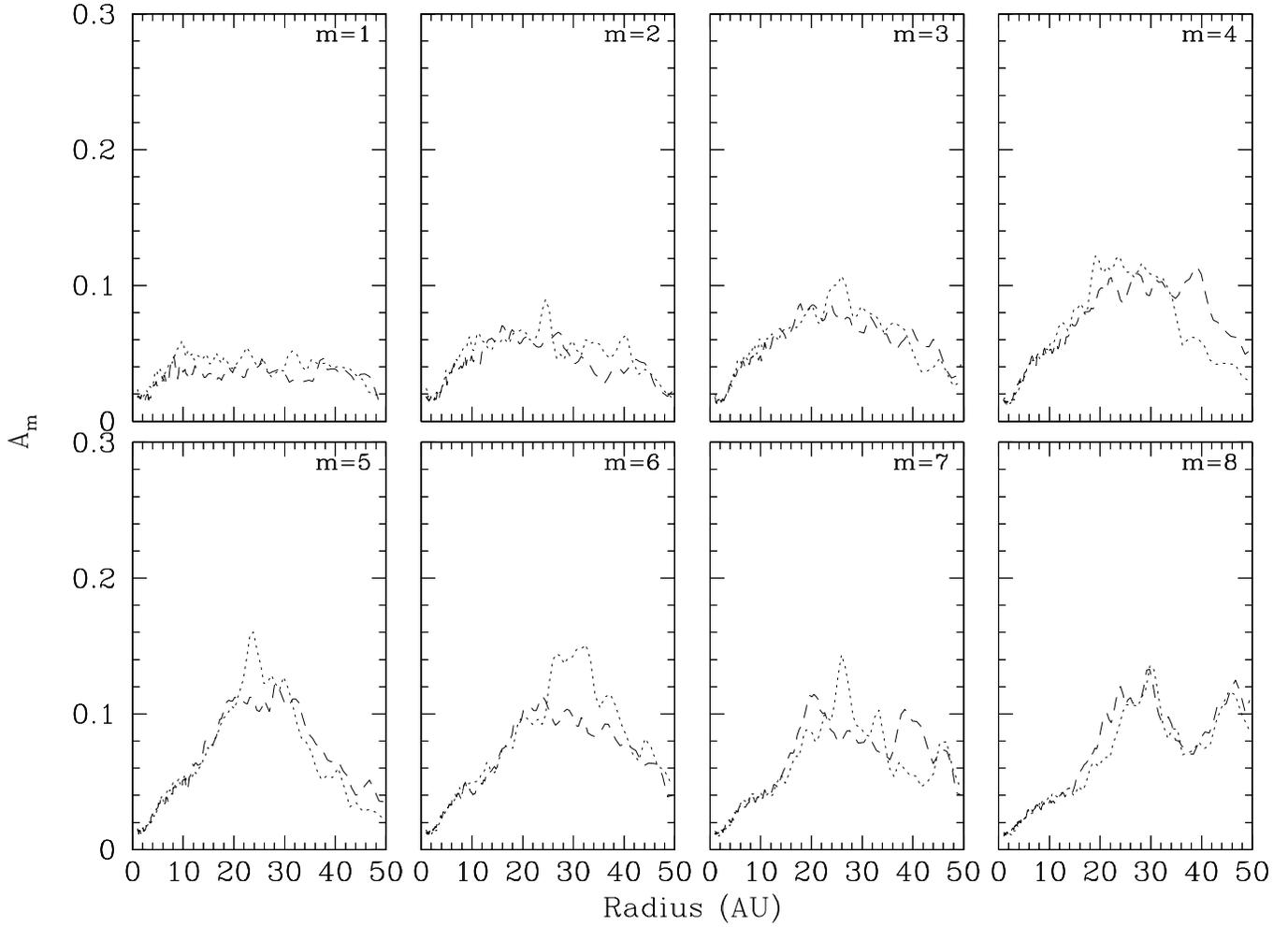}{5.5in}{-90}{70}{70}{-280}{500}
\caption[Amplitude of several spiral patterns from simulations evolved
under the `B' cooling prescription as a function of radius. ($R_D=$50~AU)]
{\label{radius-amp-b50}
The time averaged amplitude of the $m=1-8$ spiral patterns as a function 
of radius for the disk shown in figure \ref{disk-2} as well as it's high
and low resolution counterparts. As in figure \ref{radius-amp-a50},
the dotted and dashed lines represent the moderate and high resolution
runs, {\it B2m4} and {\it B2h3}, respectively.}
\end{figure}

The results of a simulation with an identical initial condition, but 
with the modified cooling prescription `B' (simulation {\it B2m4}), are 
shown in figure \ref{disk-2}. The derived SED corresponding to each 
frame is shown in figure \ref{sed-2}. The gross morphology of the system 
as evolved under this modified cooling prescription is quite similar to 
that produced with the original prescription. Instabilities begin in
the inner regions of the disk and as time progresses, engulf the entire
disk, forming filamentary, multi-armed spiral structures.  A quantitative 
measurement of the system morphology as measured by its pattern 
amplitudes (fig. \ref{radius-amp-b50}) confirms the similar behavior for
these simulations. There are no significant differences between the pattern
amplitudes derived from the `A' and `B' simulations apparent. 

The similarities are perhaps to be expected since only the inner 
regions undergo different cooling. However, if the inner regions of the
disk are responsible for the character of dynamical behavior further out, 
the modifications might create a different pattern of evolution throughout 
the entire disk. Since no such differences are evident, we may conclude 
that although the instability growth begins in the inner most regions of 
the disk, its character at large radii is not strongly dependent on the 
dynamics of the inner region, at least for the two types of cooling 
assumptions we have outlined.

In spite of only small differences in the system morphologies, the derived
SEDs show a marked difference from those produced using the original 
cooling prescription. In the present case, the SED exhibits a rising
(toward higher frequencies) spectrum between $\sim$30--50$\mu$m
($\sim10^{13}$ Hz) and $\sim1\mu$m (a few $\times 10^{14}$ Hz), then
falls off at the highest frequencies where only the star contributes
significantly to the flux. In this example, the flux in the near infrared
is variable in time, with each of the panels in the SED time mosaic 
being somewhat different from the others. The variations are concentrated 
in the near IR region of the disk, for which the calculated midplane 
temperatures are high enough to invoke the modified cooling. The flux also 
significantly overestimates that of our fiducial system, which also includes
the possibly unphysical 0.4~AU central hole region. The hot inner disk
will cause significant grain reprocessing out to much larger radii
than 0.4 AU. However, the extent to which the reprocessing actually
reduces the opacity and increases the flux, or the simple presence of 
additional matter in the hole region is responsible for producing additional
near IR flux is unclear. Models which attempt to reproduce the near IR flux
from observed systems must account for this degeneracy.

As before with our `A' cooling prescription, the disk does not emit 
sufficient flux at low frequencies ($<10^{13}$ Hz), since the modifications 
in section \ref{cool-improve} affect only the hottest portion of the disk. 
The temperature at the disk photosphere lies above the values required 
to produce the 30--100$\mu$m flux when grain destruction has begun to affect
the radiating temperature, or below them, where the matter contributes 
only minimally to the SED.

\subsection{Azimuth averaged radial structure}\label{rad-struc}

In this section we use the structure model from section \ref{energy-diss} 
to derive physical disk parameters such as the gas temperature and
the disk scale height, as well as the directly available mass density 
as functions of distance from the star. The physical and numerical 
origins of the radial structure as a function of time are addressed in 
section \ref{morph-diff} and in appendix \ref{app-numphys}.

\subsubsection{Surface Density}

\begin{figure}
\plotfiddle{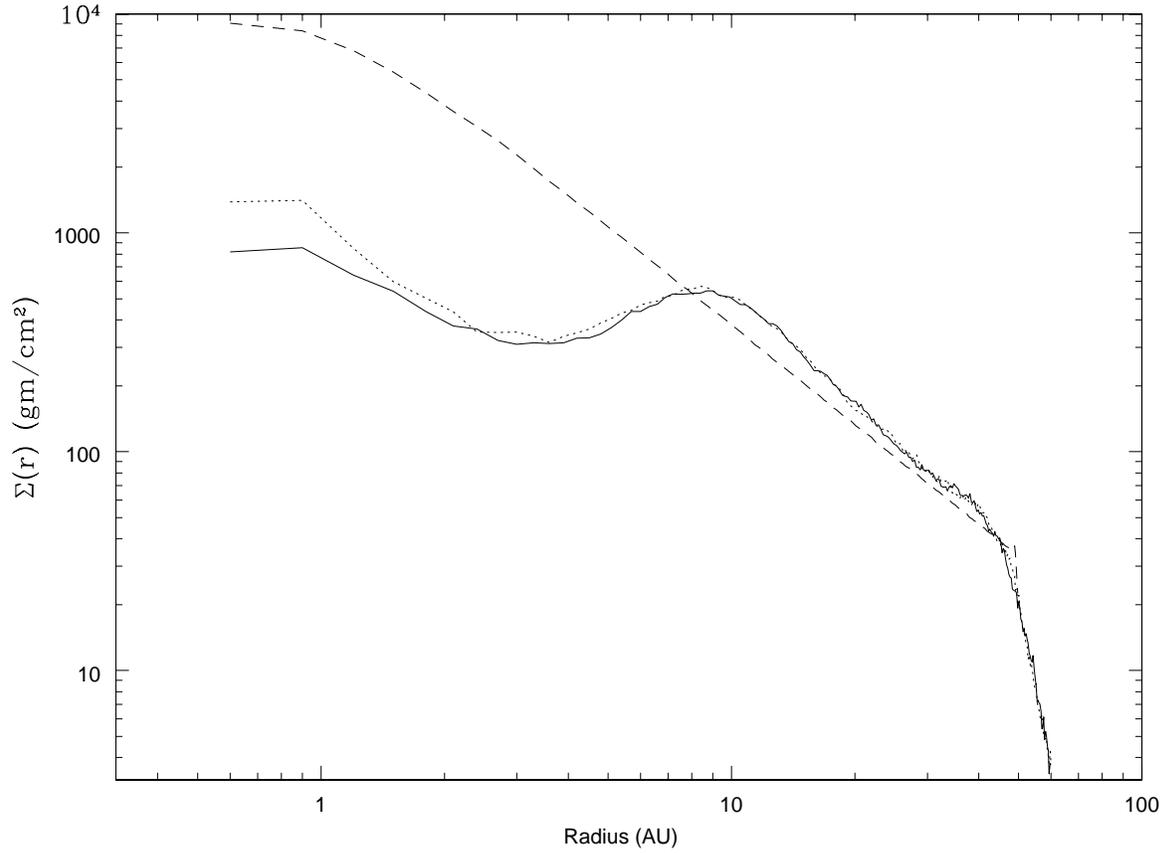}{5.5in}{-90}{60}{60}{-240}{500}
\caption[Azimuth averaged surface density of the disks.]
{\label{azavesdens}
Azimuth averaged surface density of the disks in figures \ref{disk-1}
({\it A2me}) and \ref{disk-2} ({\it B2m4}) after evolving 4\td from
their initial condition. In both the `A' (solid line) and `B' 
(dotted line) simulations, the inner disk rapidly becomes depleted of
matter with respect to the initial profile, which increases as
$r^{-3/2}$. The initial profile is shown with a dashed line.}
\end{figure}

\begin{figure}
\plotfiddle{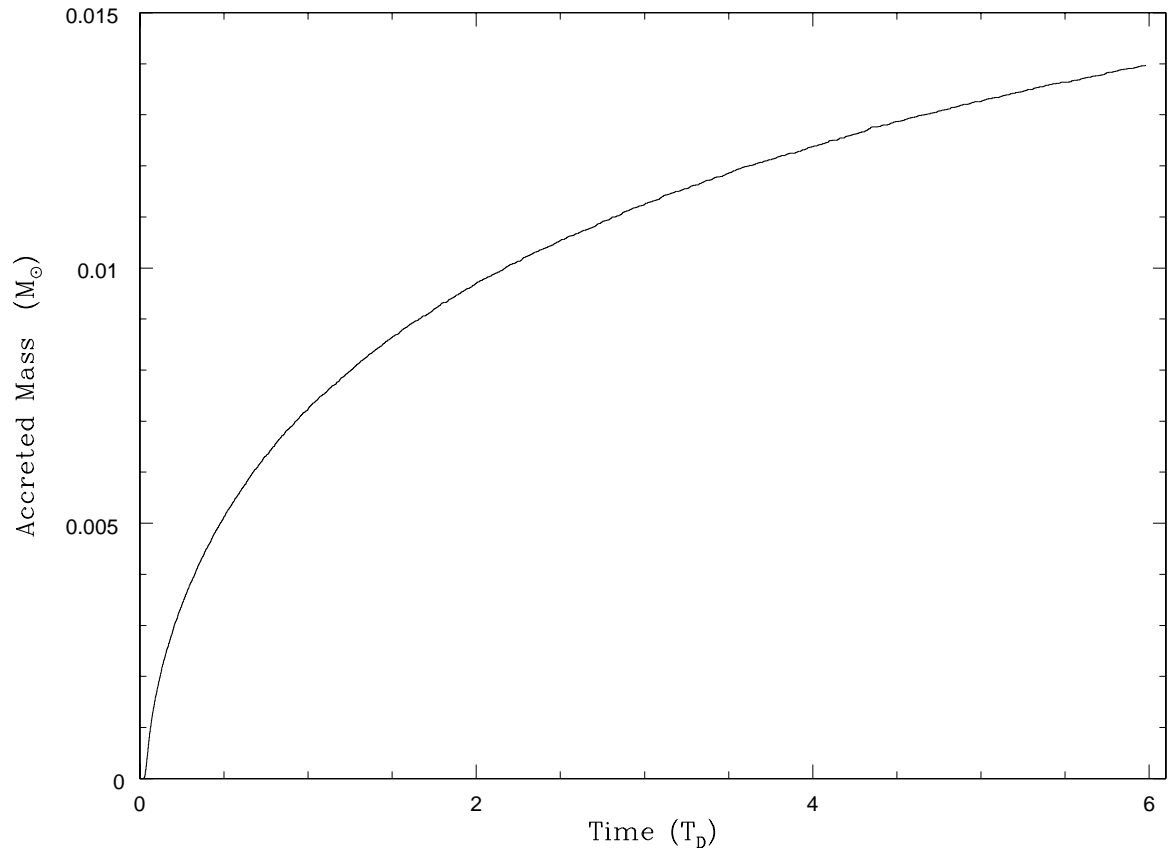}{5.5in}{-90}{60}{60}{-240}{500}
\caption[Total mass accreted by the star.]
{\label{massacc}
The total mass accreted by the star as a function of time for
the simulation {\it A2me}. Rapid mass accretion occurs for the
first $\sim$2\td, then slows as the inner disk becomes evacuated.}
\end{figure}

Dissipative processes produce radial transport of mass and momentum
through the disk. Figure \ref{azavesdens} shows the azimuth averaged
surface density profile obtained after 4\td. The density profile far
from the star does not change greatly over the period of evolution 
simulated. Further inward, we find that a portion of the mass initially
in the inner disk has been driven further outwards, increasing the 
surface density over its initial value, Another portion is accreted by
the star, as shown in fig. \ref{massacc}. The total mass accreted for 
the `A' and `B' simulations are similar. During the first two \td, the
accretion rate is much more rapid than later on. In our highest resolution
runs, the accretion begins at rate of about $6\times 10^{-6} M_\odot/$yr,
but falls by a factor of three later on. These rates compare favorably 
with accretion rates obtained for very young objects (e.g. Hartigan, 
Edwards \& Ghandour 1995), however they do depend on the resolution of
the simulation, and a further study of the details of the accretion
will be required to completely specify the accretion. In the example 
above, the total mass accreted by the star over 6\td was approximately 
$M_{acc}=0.012$\msun, or 23\% of the initial disk mass. In lower
resolution simulations the total increases to $M_{acc}=0.15$\msun, or 
30\%. 

This time dependent mass accretion pattern is present in every 
simulation we study. Over the duration of a simulation, the inner disk 
becomes depleted of much of its initial complement of mass (fig. 
\ref{azavesdens}) as the orbital energy of the gas is dissipated first 
as heat and then as radiation. Effectively, the surface density profile 
develops a much larger core radius ($r_c$ in eq. \ref{cooldenslaw}) than 
it initially has. The new core radius is approximately $r_c\approx10$~AU 
and the surface density in the core region is $\sim 500-2000$~gm/cm$^2$.
More exact values are not possible to quote because they continue to
evolve for the full duration of our simulations. 

The initial fast accretion period and its decrease later on, suggest
that the initial density profile we assume is somewhat unrealistic.
The late time, slower and more constant accretion rate suggests that
a `steady state' system will have a flatter density profile than we
assume initially. Since we begin our evolution at a somewhat arbitrary 
point in time, we conclude that a more physical initial condition for the
radial density structure for circumstellar disks will be shallower
than the $r^{-3/2}$ power law initially assumed here to extend all the 
way to the inner disk edge. We defer additional discussion of the 
redistribution of mass and angular momentum within the disk to a 
future study.

\subsubsection{Temperature}

\begin{figure}
\plotfiddle{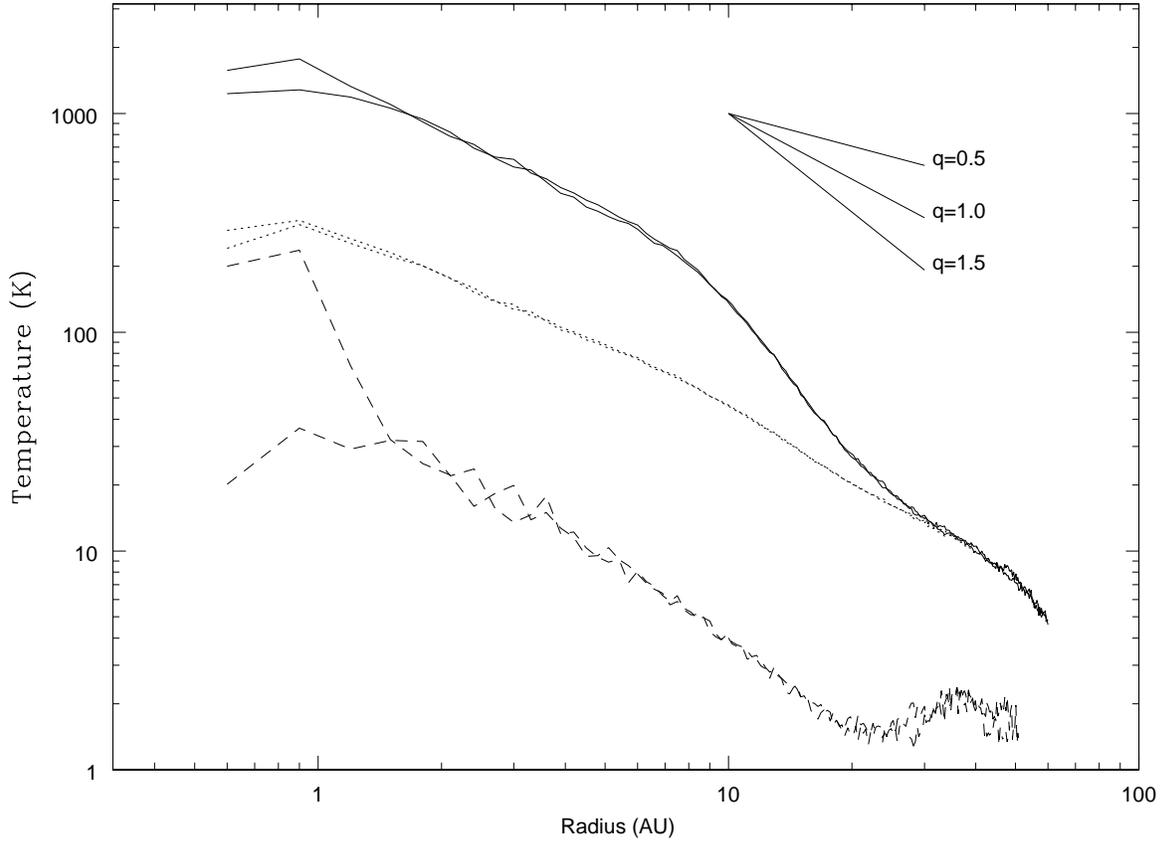}{5.5in}{-90}{60}{60}{-240}{450}
\caption[Azimuth averaged temperature structure of the disks shown in
figures \ref{disk-1} and \ref{disk-2}]
{\label{t-struct}
The azimuth averaged temperature structure of the disks shown in figures 
\ref{disk-1} ({\it A2me}) and \ref{disk-2} ({\it B2m4}). The photosphere 
temperature (dotted), the midplane temperature (solid) and the rms variation 
of the photosphere temperature in each radial ring (dashed) are shown. 
Throughout most of the system, the two simulations show near identical 
temperature structure. In the inner disk, the midplane temperatures for
the `A' simulation differs from that of the `B' run by about 300~K (`A' is 
higher than `B'). The azimuth averaged photosphere temperatures are also 
quite similar everywhere, however the variation in azimuth in regions where
the opacities were modified is of the same magnitude as the temperature 
itself, suggesting that disk matter becomes transparent intermittently on
time scales shorter than a single orbit and as local conditions dictate. The
lines drawn in the upper right of the figure represent a power law with 
index $q=$0.5, 1.0 and 1.5 respectively.}
\end{figure}

The azimuth averaged temperature profiles late in the evolution
of the disks shown in figures \ref{disk-1} and \ref{disk-2} are
shown in figure \ref{t-struct}. The midplane temperature
shows several distinct regions which individually appear 
as power laws for suitably small radius ranges, but a single
power law for the whole range of radii does not provide a good fit
to the temperature structure. Between 1 and 10~AU the midplane
temperatures of both simulations decrease roughly according to
an $r^{-1}$ power law. Further out (between 10 and 20~AU) the 
decrease steepens as the disk transitions from an optically thick
to optically thin regime. Beyond 20~AU the disk becomes optically
thin to its own radiation so that the midplane and photosphere 
temperatures are the same.

In the innermost portion ($\lesssim$~1 AU) of both sets of simulations,
the radial photospheric temperature structure is nearly a flat 
function of radius. Due to the increased efficiency of cooling when
the opacity is modified, the `B' disk midplane temperature profile 
also flattens out in the inner $\sim$1~AU, while the `A' simulation
temperatures increase to 2-400~K above those of the `B' temperatures. 
The difference is because of the changed opacity assumptions when
dust is destroyed. The hottest `B' temperatures are similar to the 
dust destruction temperatures obtained from the PMC opacity tables. 

The photosphere temperature follows a single power law much more 
closely over its radial range. Only in the inner few AU does
the temperature again becomes flatter than in rest of the 
disk. The `A' and `B' cooling prescriptions have profiles which are 
quite similar to each other.  The differences which do exist are in 
the azimuthal variation of the temperature. In regions where the midplane 
is hot and the modified cooling procedure comes into play, the rms 
photosphere temperature is nearly as large as the temperature itself, 
indicating many short lived opacity holes for which hot material near 
the midplane becomes temporarily more visible to the surrounding space.

In the `B' runs, a sensitive balance between the dynamical heating
and cooling in the inner disk produces wide variation in the opacity
because dust grains responsible for the opacity are destroyed at high
temperatures. When dynamical activity heats a vertical column 
of disk matter so that grains are destroyed, the hot disk midplane is
exposed and the column cools efficiently. As it cools below the grain
reformation temperature, the optical depth to the midplane increases
and the cooling slows as the effective temperature reflects colder,
high altitude regions.

\begin{figure}
\plotfiddle{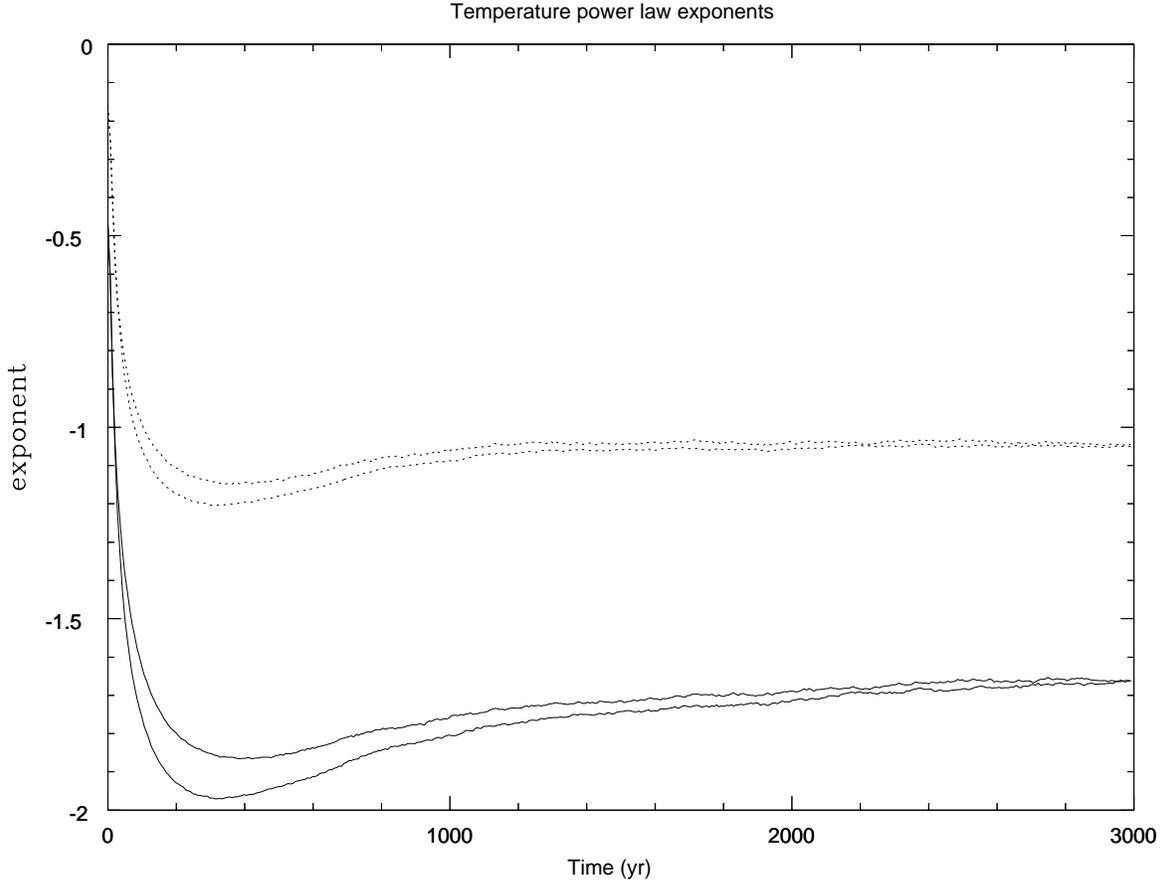}{5.5in}{-90}{60}{60}{-240}{400}
\caption[Temperature power law index at the midplane and the disk
photosphere of the simulations shown in \ref{disk-1} and \ref{disk-2}]  
{\label{t-index}
The value of the temperature power law index for the simulations shown 
in figures \ref{disk-1} ({\it A2me}) and \ref{disk-2} ({\it B2m4}) as
a function of time. Indices for both the midplane (solid) and the 
photosphere (dotted) of the disk are shown. Apart from a small 
difference in the initial behavior the fitted exponents
for each of the two simulations are identical. The indices for both
the midplane and photosphere are far larger than the values 
($0.5 \lesssim q \lesssim 0.75$) observed in proto-stellar systems.
We believe that the fitted values of the power law index have such 
large negative values because of temperatures in the outer part of
the disk which are too low, rather than by temperatures in the 
inner part of the disk which are too high.  In fact, temperatures 
in the outer part of our disk simulations are lower than those 
derived from models of observations (Adams \etal 1990).}
\end{figure}

We have fit both the midplane and the photosphere temperature profiles
to power laws whose exponents are a free parameters. The resulting 
exponents are plotted as a function of time in figure \ref{t-index}. 
Fits are made over the range of radius between the assumed accretion 
radius and the outer edge of the system, where no particles exist. 
Because the midplane temperature cannot be fit to a single power law
we do not attach significant meaning to the derived exponent.
We include it in order to highlight the differences between the value
obtained by an external observer and that relevant for the dynamically 
important disk midplane.

After an initial transient of $\sim$200~yr (0.4\td) as the disk becomes 
active, the exponents reach minimum values of $q_{\rm mid}>1.8$ and 
$q_{\rm phot}=1.2$, then settle to $q_{\rm mid}\approx 1.7$ and
$q_{\rm phot}\approx 1.1$ over the next the next 1000~yr (2\td).
This time corresponds to the end of the fast mass accretion period,
after which the dynamical evolution and the strong internal 
heating which accompanies it slows. Each of the power law exponents 
are steeper than those derived from the models of BSCG, who fit observed 
SEDs to a model of a disk with a power law temperature profile. In 
their work, exponents in the range $0.5 < q < 0.75$ were determined.
The origin of the differences are discussed in section \ref{therm-gen}.

To study the effect of using a Rosseland opacity in an optically thin region
on the cooling prescription, in a separate simulation, we multiplied the 
flux emitted from such from regions by the optical depth, $\tau$, which
will more accurately model the flux when $\tau<1$. This treatment will
produce reasonable accuracy as long as the Rosseland and Planck mean
opacities are comparable. In this case, the temperatures in the optically
thin outer disk increased to about 10~K at the outer disk edge from the
$\sim$5~K temperature seen in fig. \ref{t-struct}. The fitted exponents
changed to 1.0 and 1.6 for the photosphere and midplane respectively.
No change in the dynamical character of the simulation was observed.

\subsubsection{Toomre $Q$}

\begin{figure}
\plotfiddle{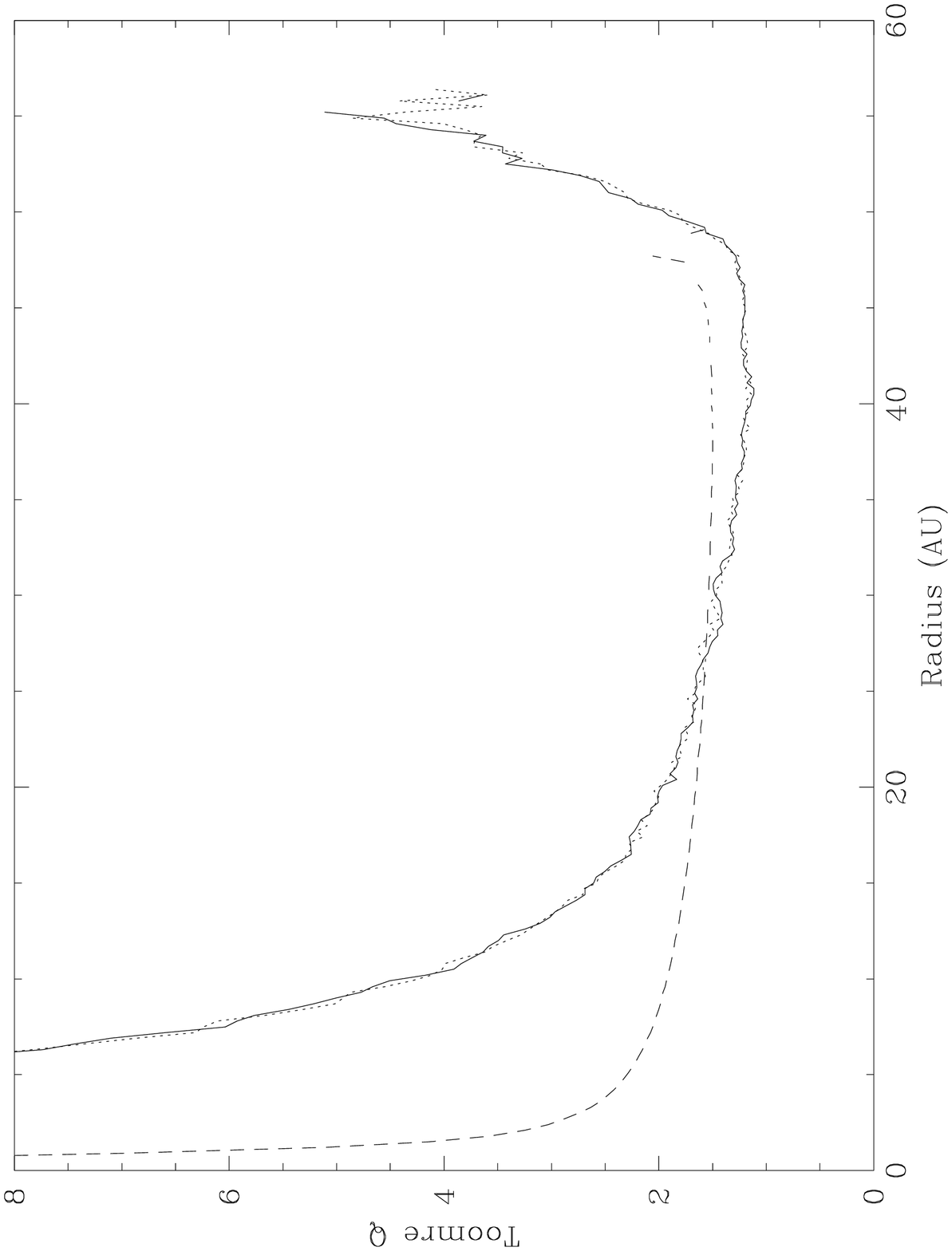}{3.5in}{-90}{50}{50}{-200}{300}
\plotfiddle{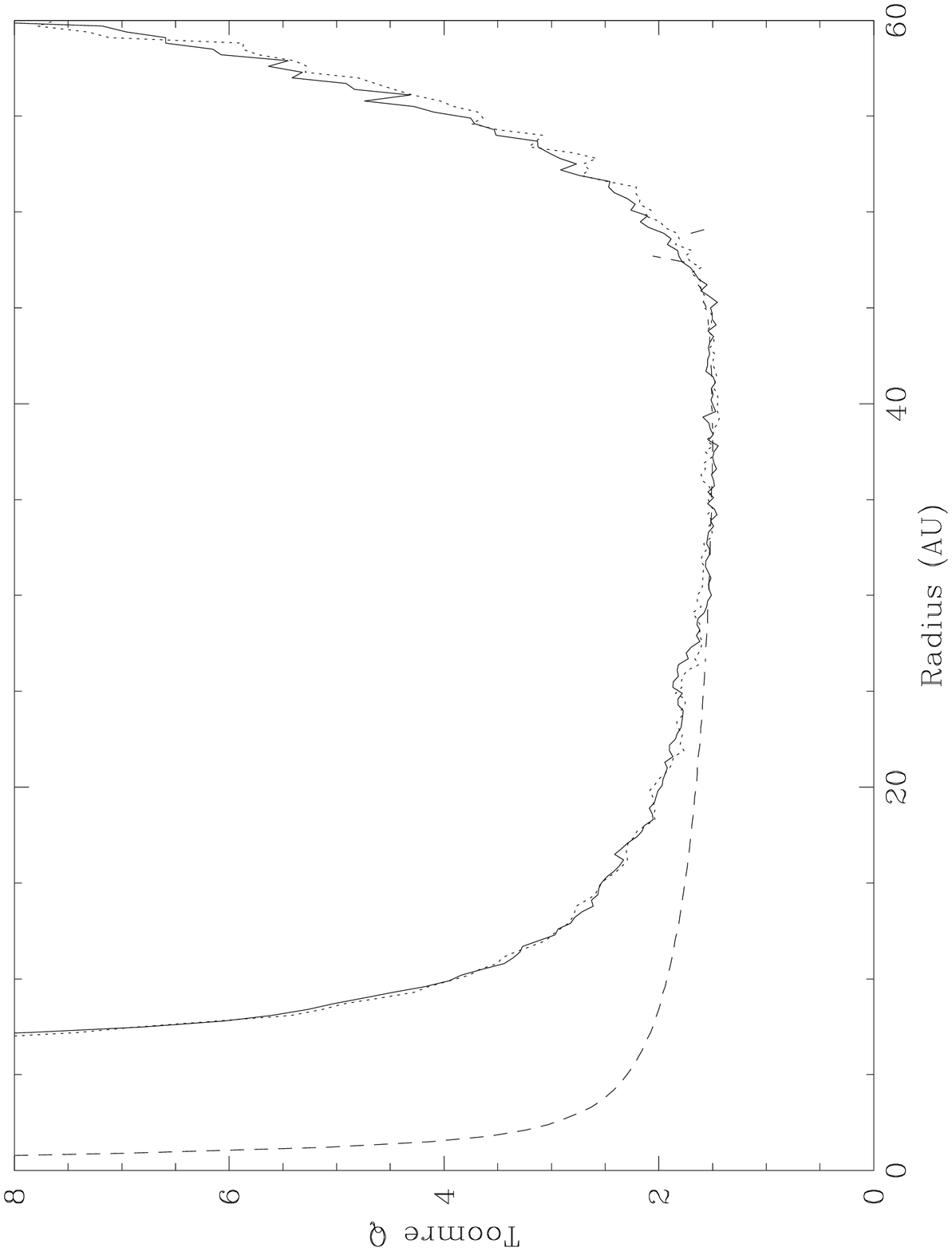}{3.5in}{-90}{50}{50}{-200}{300}
\caption[Azimuth averaged Toomre $Q$ profile of the disk]
{\label{qprofile}
Azimuth averaged Toomre $Q$ profile of the disk.  Top panel: Time
of 1\td after the beginning of the simulation. Bottom panel: Time
of 4\td after the beginning of the simulation. In each plot, the
initial profile is shown with a dashed line, and the profiles
of simulations {\it A2me} and {\it B2m4} are shown with a solid line
or a dotted line, respectively. }
\end{figure}

In figure \ref{qprofile}, we show a plot of the azimuth averaged Toomre
$Q$ values for the initial system and at 1\td, the middle of the period
over which the pattern amplitudes are averaged in section \ref{morph}.
In the inner 10~AU of the disk, the $Q$ value is far larger than 
initially, while between 10 and 30 AU a smaller increase is present.
Beyond 30~AU, $Q$ decreases below its initial value after 1\td, but 
recovers its initial profile later in time, as the spiral structure
becomes fully active and internal heating turns on throughout the disk.

The $Q$ increase at small radii is driven primarily by the decreased
surface density there, which decreases the denominator in eq.
\ref{toomre-q}. Secondarily $Q$ increases because of the increased
temperatures relative to the isothermal runs, which increase the numerator
of eq. \ref{toomre-q}, through the temperature dependence of the sound 
speed. The consequence of the increased $Q$ values are the lower 
amplitude spiral structure and the lack of clumping in the A and B 
simulations because of the increased stability to perturbations.
Note that increased Toomre stability does not necessarily imply stability
against all perturbations however, since as we shall see shock dissipation 
is largest in the inner disk.

\subsubsection{Scale Height}

\begin{figure}
\plotfiddle{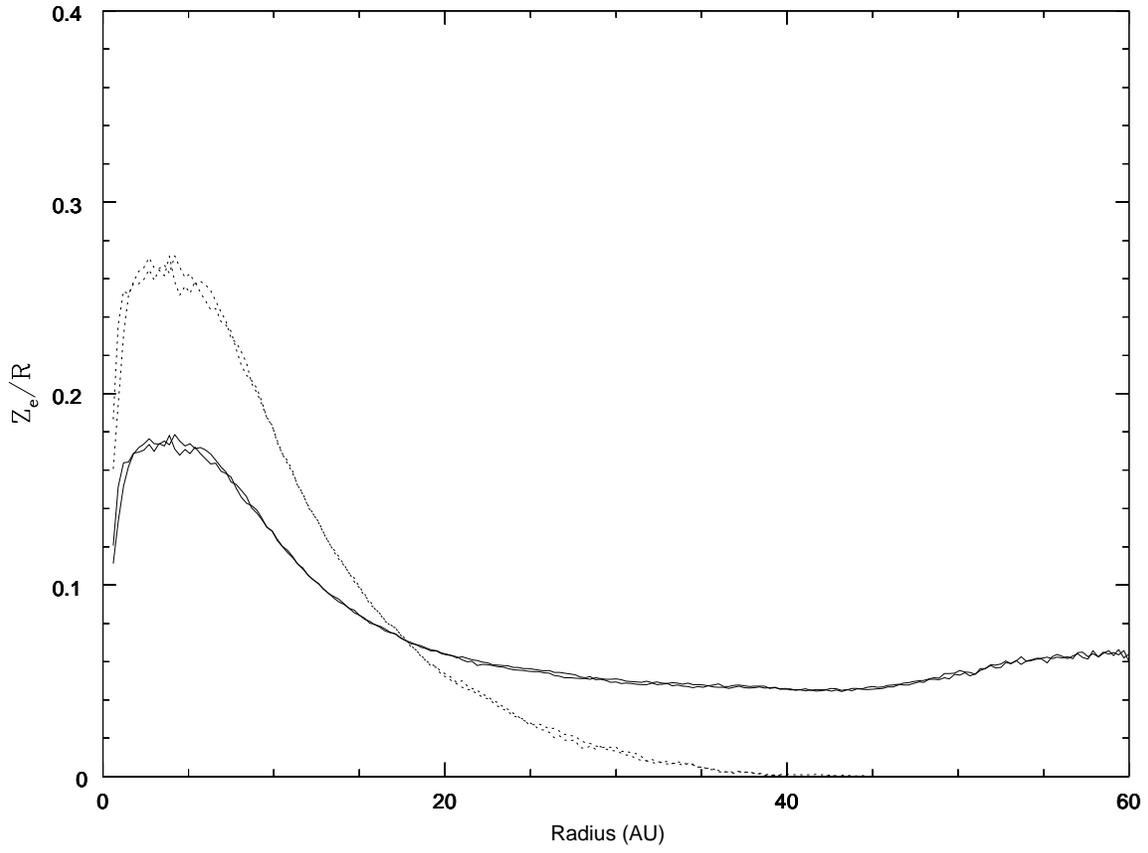}{5.5in}{-90}{60}{60}{-240}{500}
\caption[Azimuth averaged scale height of the disk]
{\label{scaleheight}
Azimuth averaged scale height, $Z_e/R$ (solid), and photosphere
altitude, $Z_{phot}/R$ (dotted), of the simulations {\it A2me} and
{\it B2m4} shown in figures \ref{disk-1} and \ref{disk-2}. }
\end{figure}

The vertical structure model outlined in section \ref{energy-diss} can 
also be used to calculate the vertical scale height at each point in the 
disk. For the purposes of this work we define the disk scale height,
$Z_e$, at some point in the disk as the altitude above the midplane at which 
the mass density, $\rho$, decreases by a factor of $1/e$ from its 
midplane value. We find (fig. \ref{scaleheight}) that as our simulations 
evolve, the disk becomes quite thick at small distances from the star. 
No differences are apparent between the scale height produced from the 
different cooling prescriptions. The altitude of the photosphere shows
an even more pronounced rise at small radii, extending vertically to
$Z_{phot}/R\approx0.27$ near 5~AU. At larger radii, the photosphere
surface drops to the midplane as the disk becomes optically thin
to its own radiation. In the same region, the scale height, $Z_e/R$,
stays nearly constant with no flaring present except at the outer 
boundary of the disk.

No doubt the vertical structure produced will change when a full description 
of the heating mechanisms is included. However, heating by absorption of
stellar photons requires that the outer disk not be in the shadow
of material further in, as the maxima in the photosphere and exponential
scale heights suggest could be the case. In order to fully address the 
question, the three dimensional distribution of the opacity sources must
be known, both for the long wavelength, ambient disk radiation and short
wavelength stellar radiation. Even though the disk may be transparent to
its own radiation, it may still absorb stellar photons efficiently at
high altitudes above the photosphere (see e.g. D'Alessio \etal 1998).

\subsection{The origin of the differences between isothermally
and non-isothermally evolved simulations}\label{morph-diff}

The differences in the pattern morphology in the `I' and `A/B' simulations 
have their origin in the way thermal energy is retained by the gas or radiated
to space.  The conversion of kinetic energy via shocks into radiated 
energy is more efficient in isothermally evolved simulations than our current
series. The conversion of kinetic energy into thermal energy is more
efficient in our current series than in isothermally evolved simulations. 
The conversion efficiencies are different because an isothermal evolution 
assumption implies three restrictive statements about the physical 
mechanisms involved in the heating and cooling of the disks, which are not 
otherwise present. Namely,

$\bullet$ Isothermal evolution assumes that the effects of heating and
cooling exactly balance each other according to a predefined temperature 
law.

$\bullet$ Isothermal evolution assumes that passive cooling and
dissipative heating processes within the disk are very rapid. Deviations 
from the predefined temperature law are restored instantaneously to 
the original profile. 

$\bullet$ Isothermal evolution as modeled assumes zero efficiency for
reversibly converting kinetic energy into thermal energy of the gas
(i.e. $PdV$ work) because of the assumption that the ratio of specific
heats, $\gamma$, is equal to unity and the temperature is constant.

Each of these statements actually also highlight subtle differences 
between the model of an isothermal system and the actual physical system.
Consider the case of an isolated hydrogen gas ($\gamma=7/5$). In this 
physical system, passage through a shock or some through some 
viscous disturbance would increase the temperature and pressure of the 
gas. If we assume the cooling is efficient so that the evolution is 
`isothermal', we assume that the energy driving the temperature increase
across the shock interface is radiated away to an external reservoir. 
The first two statements above express this balance between the rapid
dissipative heating and rapid radiative cooling.

In our disk models such dissipative heating may occur as gas passes
though a spiral arm structure. The first two statements are falsified in 
our new simulations because substantial heating of the disk occurs before
a new equilibration is reached (section \ref{rad-struc}) and because
temperature variations which occur as a result of grain growth and
destruction, since these temperature variations do not immediately
return to their `steady state' values.  The temperature profile at any
given instant in time is not the same as the predefined profile: the
gas retains thermal energy better than we assume in the isothermally 
evolved case. The changed thermodynamic state in turn changes the 
stability of the disk to perturbations. The decreased pattern 
amplitudes in the inner third of our disks are a consequence of this
stability.

Again in the hydrogen gas example, a compression which does not lead
to a shock in a hydrogen gas will only be isothermal if energy going
into raising its temperature is radiated away to an external 
reservoir. In an isothermally evolved model, rather than including a
cooling process which produces an effective $\gamma$ value of unity,
we define $\gamma=1$ and a different assumption is made instead 
(expressed in the third statement above). In this case, {\it no}
energy is used to raise the temperature of the gas in the compression.
Instead, all of the $PdV$ work used to compress the gas goes only 
into increasing its pressure and density. A given amount of $PdV$
work can therefore compress the gas to higher densities and, in our
specific example of disk evolution, spiral structure will achieve
higher amplitudes and appear more filamentary.

\section{The contribution of dynamical processes to emitted
                   radiation}\label{sed-dynam}

In addition to direct conclusions regarding the dynamical evolution of
the simulations, in this section we investigate the contribution that 
dynamical mechanisms make to the energy output by the system. Because we 
have not included all sources of heating available to the disk in our 
simulations (passive reprocessing of radiation for example), we cannot 
expect the SEDs synthesized from those simulations to accurately reproduce
observations. We can however draw conclusions about the whole by examining
the differences between our synthesized results and observed systems.
For example, to what extent do dynamical processes power observed accretion
disk SEDs (and in what wavelength regimes?) as compared to passive sources?
Given that dynamical processes do not reproduce the SED completely, what 
can we infer about the relative importance of other processes in the system?

\subsection{Variation of the SEDs with time}\label{sedtime-var}

\begin{figure}
\plotfiddle{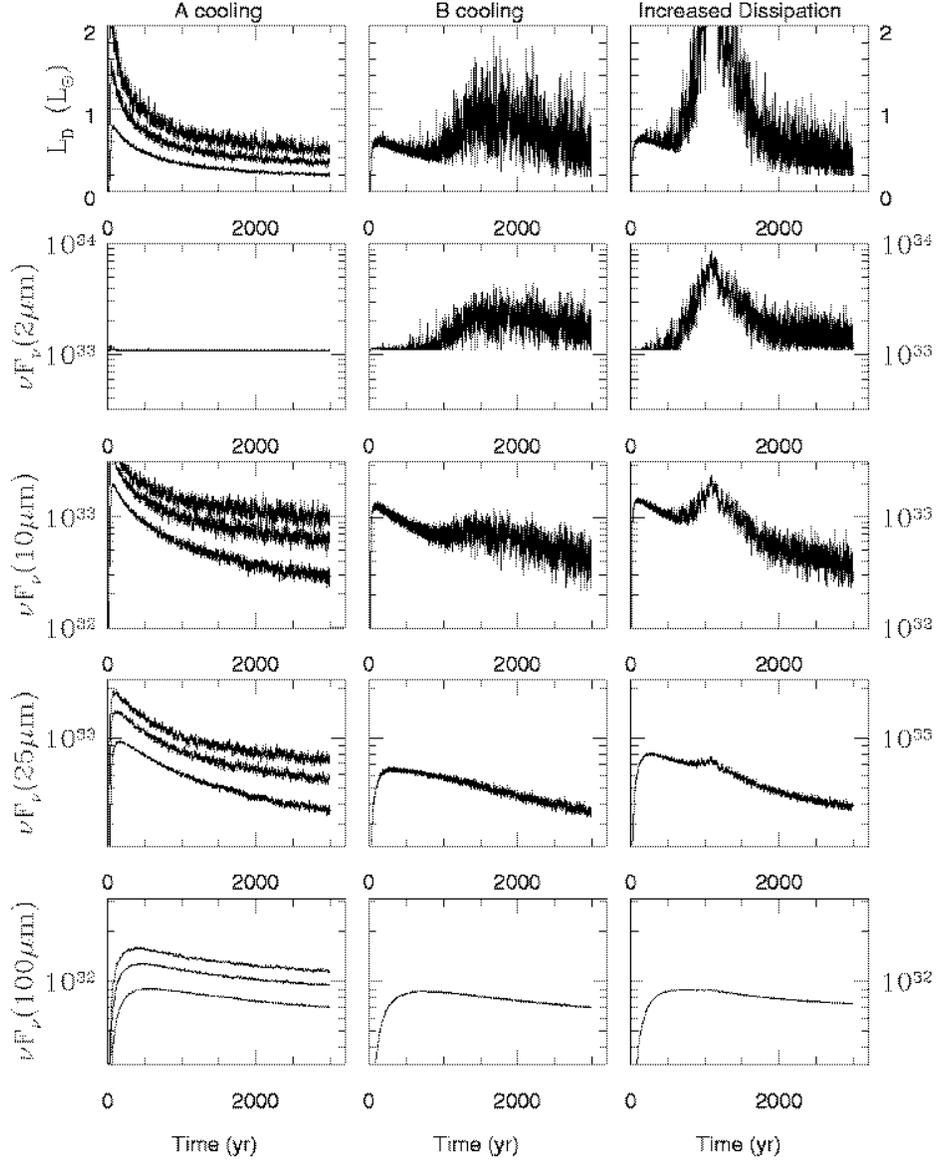}{6.75in}{0}{69}{69}{-210}{-35}
\caption[Disk flux in various wavelength bands and total luminosity
as a function of time.]
{\label{flux-var}
The luminosity and emitted power at four wavelengths: 2, 10, 25 and 100$\mu$m.
On the left are simulations {\it A2lo, A2me} and {\it A2hi} and in each panel
the top, middle and lower tracks originate from the low, middle and high 
resolution simulations respectively. The 2$\mu$m flux consists only of the
assumed constant contribution from the stellar photosphere, while the longer
wavelengths are dominated by the flux from the disk. The center panels show
only simulation {\it B2h3}. The lower resolution counterparts were suppressed
for clarity. The right panels show the results of simulation {\it H2h3}, for
which a higher effective thermal energy generation rate is present.}
\end{figure}

We concluded in section \ref{cool-improve} that the time averaged SED 
in the near and mid IR is strongly dependent upon the microphysics of 
the dust grain size distribution and its effect upon the opacity, but
that our model could only bracket the magnitude of the modification
required to accurately reproduce the time averaged spectrum. We now
study the time dependent behavior of our simulations.  In fig. 
\ref{flux-var} we plot the the total luminosity of the disk and the
emitted power, $\nu F_\nu$, at 2, 10, 25 and 100 $\mu$m as a function of
time for each of three resolutions for our `A' simulations and our
highest resolution `B' simulation (The right hand panels are
discussed in appendix \ref{app-numphys}). 

Within 100 yr of the beginning of the simulation, the total luminosity 
from the disk increases to a peak and then decreases over the next 
two \td to a more stable, long term evolution. The high luminosity
is a consequence of the the high mass accretion period seen above and 
as the mass accretion rate decreases the luminosity falls. Fluxes in
all wavelengths continue to decrease slowly even late times. We attribute
the continued decrease to the absence in our models of external sources
of thermal energy, such as stellar illumination and additional mass
accretion from the molecular cloud.

With the `A' cooling model only small flux variations in time are 
present: no short term variations larger than 10-20\%  are present at any
wavelength and short term variations at longer wavelengths are smaller, 
about 1\% at 100$\mu$m. At 2$\mu$m, no contribution from the disk 
is present; the flux is completely dominated by the assumed constant 
4000~K black body contribution of the star. In these simulations, at
short wavelengths the assumed 1~\lsun, 4000~K black body contribution 
from the star dominates the emission from the system. Further, the 
total luminosity of the disk is only a small fraction of the stellar
contribution. In our highest resolution simulation `A' simulation the 
disk luminosity is of order 0.2\lsun, while the increased near infrared 
emission in the `B' simulation increases the total to about 2/3~\lsun.

The `B' simulations show behavior similar to the `A' simulations in
the mid and far IR, but exhibit large variations in the near IR. At
2$\mu$m for example, the variation is about a factor of two or more
from peak to peak and the total disk luminosity shows a similar amount 
of variation. This simulation does not produce substantial 2$\mu$m flux 
until near the end of the initial fast mass accretion period and similar 
behavior occurs in the other `B' simulations. 

\begin{figure}
\plotfiddle{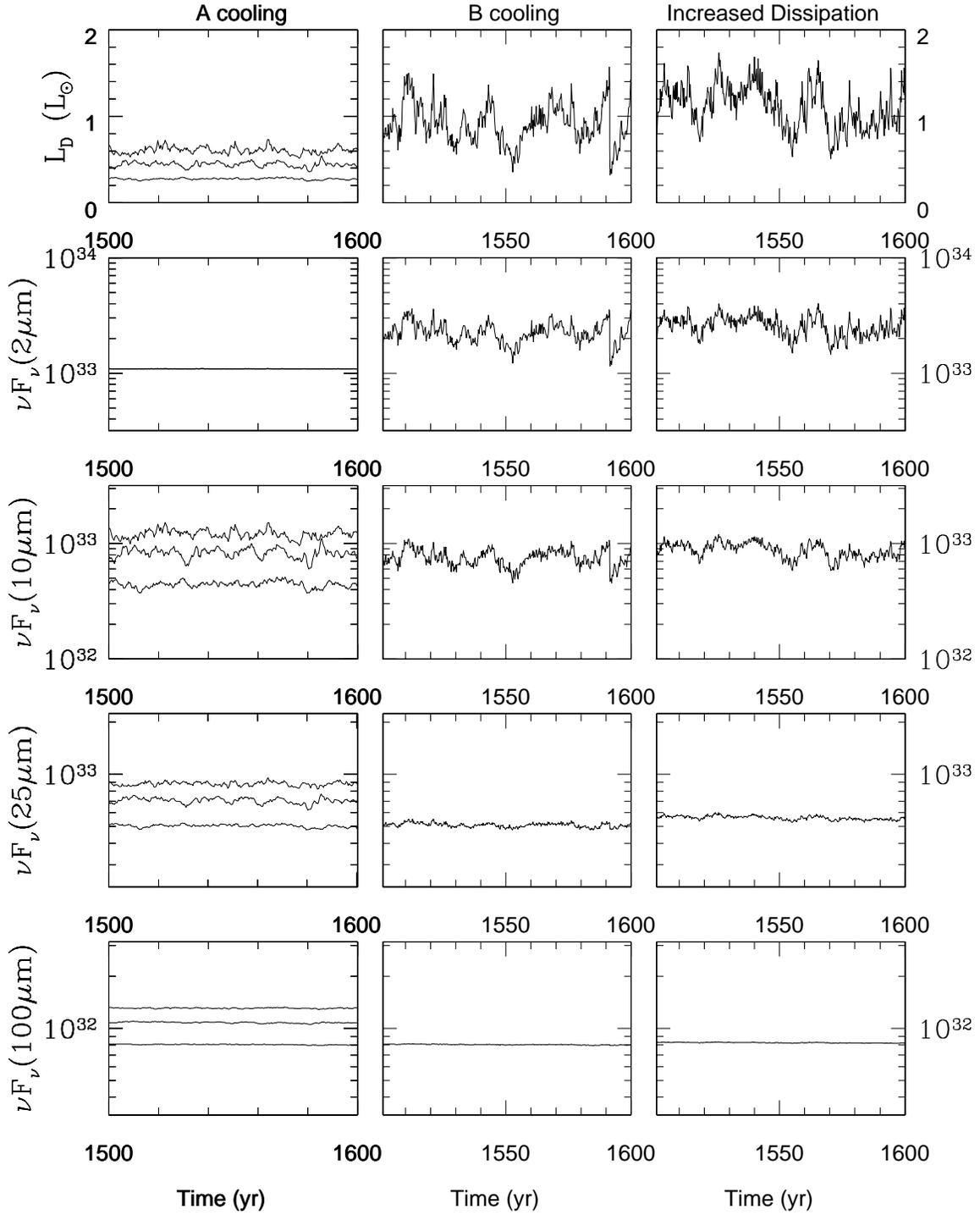}{6.75in}{0}{75}{75}{-230}{-25}
\caption[Time expanded disk flux in various wavelength bands and total 
luminosity as a function of time.]
{\label{flux-var-expand}
The same as figure \ref{flux-var} but expanded in time to show the 
details of the time dependence of the flux.}
\end{figure}

In order to see the shorter term structure in the flux variation, 
we show the same variables as in \ref{flux-var}, but expanded to 
show a small slice in time in fig. \ref{flux-var-expand}. The time
scale of the variation in the near and medium infrared is similar to
the orbital time scales of the inner disk, which is truncated at 
0.4~AU in our simulations. At some times, only the assumed stellar 
component of the flux contributes to the flux, while at others, the
flux is dominated by the disk contribution. The variations have no well
defined periodicity. Variations occur over periods of less than a year 
and over periods of as long as ten to twenty years. 

The variation which exists in the near infrared is due to the sensitive
temperature balance near the destruction temperature of dust grains.
When a column is heated so that the midplane temperature is above the
dust destruction temperature, the midplane is exposed and more near 
infrared radiation escapes. As it cools, dust reforms and the the hot 
midplane is again concealed by a cooler high altitude dust layer, which 
emits most efficiently at longer wavelengths. Because of this 
sensitivity, we conclude that a correct model of the dynamics and
spectral energy distributions of circumstellar disks must include
an accurate description of the full three dimensional spatial 
distribution of grains and of their size distribution. Qualitatively,
we can understand the dynamical origin of the variation by noting 
that heating processes in the inner disk do not occur at regular
intervals as the disk evolves, but rather occur sporadically as
spiral arm structures or other inhomogeneities interact and dissipate
orbital energy as heat. We discuss the significance of possible 
numerical origins of the variations and their amplitude in Appendix
\ref{app-numphys}.

Coupled with the sharp decrease in opacity required to reproduce
the near IR flux (sec \ref{cool-improve}), the time variability present 
in our simulations suggests the following interpretation. In the inner 
portion of the accretion disk, clouds of grains in small patches of the 
disk are destroyed and reform, intermittently obscuring the hottest parts 
of the disk midplane from view. Such an interpretation implies quite 
naturally the existence of intermittent variability in the near and mid 
infrared spectra of star/disk systems originating from within the disk
rather than from a stellar photosphere. Skrutskie \etal (1996) observe
such variation on time scales of a few days to a few weeks in the $J$,
$H$ and $K$ bands for several young stellar systems. They conclude that
such variations are due to processes in the accretion disk within a 
few tens of stellar radii from the star or less. Longer term mid-IR
variations observed by Wooden \etal (1999) show an apparent correlation
between a short lived 10~$\mu$m absorption feature and a sudden decrease
in the $V$ band luminosity in the DG Tau system. Such a correlation suggests 
that stellar or disk midplane photons escaping to space are later more 
and more strongly attenuated by dust as it recondenses and the medium
becomes optically thick.

Also, the class of higher mass pre-main sequence stars known as ``UXOR's''
show quasi-periodic optical variations on decade long time scales occurring 
(Herbst \& Shevchenko 1999). These variations have been attributed to dust 
clouds in a disk which intermittently obscure the stellar photosphere. While
UXOR stars are more massive than those modeled in our study, similar
mechanisms may apply to each. In our model, we do not require that the
system be observed edge on as discussed in Herbst \& Shevchenko and so
remove one of the more substantial objections to the model noted in
their discussion. Similar observational studies of longer term infrared
variations of other T-Tauri and UXOR stars would provide {\it in situ}
measurement grain evolution in the disk and dynamical activity within
a few AU of the star.

\subsection{The origin of thermal energy generation}\label{therm-gen}

The goal for this section is to understand which physical processes are 
responsible (and just as importantly which are not responsible) for the
luminosity produced by observed young star/disk systems in each portion 
of their spectrum.  Of particular interest will be to understand the 
origin of so called `flat' or shallow spectrum sources which may be 
representative of more massive disk systems like those in our study.
We have noted that the artificial viscosity incorporated into our
simulations approximately models the underlying physical dissipation 
of kinetic energy present in the disk. We now proceed to calculate the
magnitude and origin of the thermal energy generated in different 
portions of the disk. In appendix \ref{app-numphys}, we attempt to
quantify and understand properties of the artificial viscosity prescription
which are not of physical origin.

We first quantify the quantify the magnitude of turbulent dissipation 
in our simulations, as quantified by the Shakura \& Sunyaev $\alpha_{SS}$
parameter. Figure \ref{alpha-fig} shows its azimuth averaged value derived 
from our simulations. Its value in any $\delta r$ of the disk is of 
order a few $\times 10^{-3}$. Both the `A' and `B' simulations produce 
identical dissipation rates, so only the `A' results are shown. We 
also plot the ratio of the dissipation due to the turbulent and shock 
artificial viscosities and find that the ratio is near unity. The total 
budget of thermal energy generation from dissipation of large scale gas 
motions averaged over azimuth is therefore within a factor of a few of that
provided by the $\alpha_{SS}$ term alone. Two distinct regions exist in
the plot. In the inner disk ($\lesssim 10$~AU) shock dissipation dominates
the total, while further out turbulent dissipation dominates. The 
increases at large radii are due to the loss of resolution at the disk 
boundary and are not significant.

\subsubsection{The inner disk}

We have discussed the mass accretion onto the star from the inner 
disk in section \ref{rad-struc} above. Here we discuss the dissipation
responsible for it. The dominant source of of the near and mid 
infrared radiation is from radii $\lesssim 5-10$~AU from the star but
the exact allocation of radiated energy to specific wavelength bands
depends on the opacity assumptions. The internal dissipation processes
present at radii $\lesssim 5-10$~AU are sufficient to power a large 
fraction of the near/mid infrared SED. In this region, figure 
\ref{alpha-fig} shows that shock dissipation is as much as a factor of
2--3 larger than turbulent dissipation. Also, the turbulent dissipation
itself increases by a similar factor over its value at large radii.
A portion of this increase may be due to numerical and resolution
effects as discussed in appendix \ref{app-numphys} (higher resolution
implies lower numerical dissipation), but the exact proportion of
physical and numerical dissipation remains unclear.

The decrease in mass density of the central region later in the 
simulations poses a problem for our SPH code because of its reliance
upon particles to resolve the flow. As the density decreases so does
the particle density and therefore also the detailed resolution of
the flow. In order to resolve the physical conditions in the inner
disk, we expect that very high resolution simulations concentrating 
specifically on the dynamical behavior of the inner accretion disk must 
be performed in order to better define the physical conditions present 
over long periods.

\subsubsection{The outer disk}

The flux derived from the simulations at long wavelengths (figures
\ref{sed-1} and \ref{sed-2}) underestimates that
from observed systems (e.g. BSCG), and especially those thought to be
younger systems for which large disk masses are more likely (Adams
\etal 1990). Also, the long wavelength turnover in the SEDs typically
occurs near between 10 and 30$\mu$m ($\sim10^{13}$ Hz) rather than 
the $\sim$100--300$\mu$m (10$^{12}$ Hz) typical of observed systems
(BSCG, Adams \etal 1990). Is this deficit physical, or are the initial
conditions of our simulations responsible for its existence?

The dominant source of the long wavelength radiation is colder regions 
of the disk distant from the star. If the assumed disk radius is doubled, 
so that the total disk surface area radiating at, say 10--30~K, is 
increased by a factor of approximately four, will sufficient long
wavelength flux be produced? We examined the differences in the long
wavelength end of the SEDs generated from our 50~AU disks (simulations
{\it A2lo, A2me} and {\it A2hi}) with those generated from 100~AU disks
(simulations {\it a2lo, a2me} and {\it a2hi}). Time averaged SEDs for
each of the simulations are shown in fig \ref{sed-50-100cmp}. Looking
specifically at the long wavelength flux behavior, in which dust destruction
is unimportant, we find no significant differences between shape of the
SED seen in one or the other simulation. We conclude that the assumed
size of the accretion disk is not responsible for the shape of the SED at 
long wavelengths.

Is the initial temperature profile too cold, so that even with substantial
heating the disk never warms enough to produce appreciable long wavelength
flux? A test simulation identical to {\it A2me}, but with initial
\qmin=2.5 so that the initial disk temperatures everywhere are higher,
cools over the course of the first 1--3\td to resemble the conditions in 
simulation {\it A2me}. We conclude that the initial temperature
profile is not critical for the long term ($>2$\td) temperature of the
system.

In section \ref{rad-struc} we showed that the temperatures in the outer
part of the disks are lower than are determined from observations.
We find temperatures at the outer disk edge $\sim$10~K, while
Adams \etal 1990 for example determine temperatures at the outer disk
edge of order 15--30~K. In many parts of the disk, they are in fact
also lower than those observed for the molecular clouds in which the
disks reside (see e.g. Walker, Adams \& Lada 1990). We can conclude
from the low temperatures and the corresponding long wavelength flux
deficit that heating of the disk due to identifiable internal processes
(e.g. large scale shocks) in our simulations is insufficient to heat the 
outer disk to the `right' temperature, i.e. a temperature warm enough to 
produce SEDs from our models which are similar to observed systems. 

This conclusion implies an upper limit on the amount of mass and angular
momentum transport due to gravitational torques, since such torques
are ultimately responsible for the growth and evolution of the 
spiral structures in which the shocks occur. We obtain a conservative 
limit if we attribute all of the energy dissipation in our simulations 
in the outer accretion disk to gravitational torques. We conclude that 
they produce large scale shocks which dissipate kinetic energy at a 
rate no greater than an equivalent $\alpha_{SS}$ dissipation of
$\alpha_{SS}\sim2-5\times10^{-3}$. This limit is valid for marginally
self gravitating systems (\mrat=0.2) only, where spiral arms 
are both filamentary and of high order symmetry ($m>3$). In 
massive systems, where the dominant spiral patterns are of low 
order symmetry ($m\le 3$) and gravitational torques are likely to be
more effective, perhaps more efficient transport is possible.
Also in this case, gravitational torques may be more effective
in transporting matter non-dissipatively, so that their effects would
not be observable in the radiated energy spectrum. Then the transport
would be through direct and non-local exchange of gravitational 
potential energy of matter in one part of the disk for kinetic energy
of matter another part of the disk, rather than through wave 
amplification and dissipation.

\section{Discussion and comparisons to other work}\label{cmparother} 

In order to provide a context for further investigations, here we
compare the results and limitations of this study with other recent
work. The majority of previous multi-dimensional hydrodynamic 
simulations of circumstellar disks have been done completely or
partially within the assumption that the thermodynamic properties
of the gas are either locally isothermal or locally adiabatic. 
One subset of these (e.g. Pickett \etal 1998, \pone, Laughlin \& 
R\'o\.zyczka 1996) has investigated the linear and nonlinear 
development of spiral density patterns in massive systems. A second
subset (e.g. Boss 1998, Laughlin \& Bodenheimer 1994, Bonnell \& 
Bate 1997) has investigated physical outcomes of cloud collapse
and the early evolution of the disk which forms afterwards.

The results of first subset show that Fourier components of the 
spiral structures that form tend to saturate at amplitudes of 
$\delta\Sigma_m/\Sigma\sim10-20$\%, due to non-linear mode coupling
between spiral patterns of different symmetries (Laughlin,
Korchagin \& Adams 1997). This amplitude saturation is true of the
present work as well, however the saturation amplitude is lower
in the inner third of the disk relative to the growth found in 
isothermally evolved models with similar initial conditions (\pone).
In our simulations, thermal energy can be retained in the gas rather 
than being immediately radiated to space or preserved in the form 
of bulk kinetic energy. Similar to isothermally evolved runs, regions 
more distant from the star develop spiral structures which vary in time 
and do not persist as a single pattern over even a single orbit of the 
outer disk around the star.

The Pickett \etal (1998) models examine the evolution in a massive 
star/disk system (\mrat=0.4), including the star disk boundary and
the region within a few 10's of \rsun of the star. They investigate
the effects of the thermodynamic treatment of the fluid by evolving
systems in the limits of either isothermal and adiabatic (i.e. $K(r)$ 
rather than $T(r)$ fixed) evolution. Their isothermally evolved runs 
develop much more pronounced spiral structure than their adiabatic runs, 
similar to our own findings for isothermal vs. cooled simulations and
that their isothermally evolved runs break up into `arclets' similar
to the clumps in \pone and in Boss (1998). They also find that the 
star/disk boundary is subject to different instability morphologies
than is the disk itself. Coupled with our findings about the 
variability in the inner disk, such a finding makes even more clear
the need for a detailed study of the character of the disk within
a few AU of the star.

In \pone, Boss (1998) and Pickett \etal (1998), spiral structures 
collapse into clumps late in some of the simulations. In \pone and
Pickett \etal, the radial temperature structures were isothermal,
while in Boss (1998), the thermal structure of the disk was first
obtained from a synthesis of many axisymmetric calculations including
simply radiative transport, then three dimensional disk evolution 
was modeled assuming the structure was thereafter either locally adiabatic
or isothermal. Except in one unverified example (section \ref{morph}), 
spiral patterns in our current calculations do not collapse into 
clumps as they did in the isothermally evolved cases in \pone. It is
significant that only simulations of disks with predefined radial 
temperature or entropy structure also produce clumping in disks. As
we warned in \pone, the thermodynamic structure of disks is not yet
sufficiently well understood to make the interpretation of such 
clumping clear.

In both \pone and in the present work, dynamical activity (spiral 
structure) in the inner few AU develops and produces heating due to
shocks as various structures collide and reform. The dissipation
results in mass accretion onto the star and transport radially
outward to conserve angular momentum. Within $\sim$1000 yr, The assumed
$r^{-3/2}$ surface density power law becomes nearly flat as a function 
of radius inside 10~AU. From these results it seems improbable that such 
a steep density distribution can endure for a long time. Laughlin \& 
Bodenheimer (1994) showed that a massive disk forms during the collapse
of a cloud which is initially somewhat toroidal, and during later 
evolution, evolves toward a power law profile (Laughlin \& 
R\'o\.zyczka 1996). One could argue that the differences between their
calculation and ours are due to modeling different stages of an 
evolutionary sequence, though it is more likely that the true nature
of the density distribution as a function of radius is simply not yet 
sufficiently well constrained by the models. Neither set of simulations
evolve far enough in time to produce a steady state (if one exists) and 
do not include additional accretion of matter from the surroundings.

In a series of papers, Bell and her collaborators (BL94, Bell \etal 1995,
BCKH, Turner Bodenheimer \& Bell 1997) have developed an evolutionary 
model for accretion disks based on an $\alpha_{SS}$ model for radial
transport and a mixing length theory (MLT) based vertical structure 
model. In similar work D'Alessio \etal (1998, 1999) have also developed 
a `1$+$1' dimensional model. In an $\alpha_{SS}$ model, three input
parameters determine the amount of active and passive heating experienced
by the disk, as well as its time and radius dependent temperature and 
density structure. They are the accretion rate, $\dot M$, magnitude of
the viscosity, $\alpha_{SS}$ and the radiative flux impinging upon the 
disk surface

In the present work, we follow the evolution through a `2$+$1' dimensional 
model, including the evolution of matter in both radius and azimuth, but 
at the expense of simpler vertical structure and radiative transport 
(section \ref{energy-diss}). Our models assume that the disk exists 
in isolation (i.e. that radiative fluxes onto the disk are zero) in order 
to study the effect of dynamical processes internal to the disk itself.
We investigate the dynamical evolution (and consequent thermal energy
generation mechanisms) of the disk which can not be addressed in the
1$+$1D work due to the input assumptions of the $\alpha_{SS}$ disk 
formulation. The accretion rate is left unspecified and is determined 
only from dynamical conditions.

As the high pattern amplitudes noted above indicate, a full 
multi-dimensional evolution is important for a description of the
density structure. The energy dissipated by their variation
in time and space is also important for the specification the
temperature structure. The temperature structure does not develop 
such large variations in azimuth. Given an accurate approximation
of the energy input from dynamical heating, an azimuth averaged
temperature law will be accurate to a few percent. In the inner disk,
where grain destruction becomes important and the hot midplane 
intermittently becomes exposed to space, an azimuth average is 
not appropriate and a multi dimensional description must be 
invoked.

We assume in our vertical structure calculation that the structure
is adiabatic in $z$, which is equivalent to assuming that turbulent
processes smooth vertical entropy gradients. BCKH find that ordinarily
the structure is super adiabatic in $z$ when only convective turbulence
is active. Three dimensional calculations of a narrow annular section
of a minimum mass disk (Klahr, Henning \& Kley 1999), support this 
conclusion. The effect that a super adiabatic gradient would have 
on our simulations would be to systematically reduce the photosphere 
temperature of the disk for a given midplane temperature. 

Another possible source of turbulence which may be important
in dynamically active systems is turbulent mixing as large scale density
structures in the disk intermittently collide and reform. It is unclear 
however, how efficient this process might be in coupling large scale
($r,\phi$) motion into the small scale $z$ motion required to affect the 
vertical structure. Recently, Pickett \etal (1999) have extended their 
1998 calculations to include the effects of thermodynamic heating (but not 
cooling) on the disk matter and find that most thermal heating occurs away 
from the disk midplane, as waves generated at low altitudes are amplified 
upon entering lower density, high altitude regions. This result indicates
that the $z$ transport of energy due to structure in the $(x,y)$ plane
is at least non-negligible and can affect the thermal structure. It remains
unclear whether the effect will be limited only to a very high altitude 
`corona' or whether the structure throughout the vertical extent of the disk 
will be affected.

The magnitude of vertical turbulence will also have a critical affect
on the vertical distribution of the opacity. Across a grain destruction
boundary for example convective turbulence may be suppressed. If
the dynamical turbulence noted above is also inefficient then grains
would rarely be dragged into the hot midplane region and destroyed,
as assumed in our `B' models. In those models, a sensitive temperature
balance between dynamical heating radiative cooling occurs near the 
dust destruction temperature and can cause variation in the near 
infrared energy output of the disk. Because of this sensitivity, we
conclude that a correct model of the dynamics and spectral energy
distributions of circumstellar disks must include an accurate description
of the full three dimensional spatial distribution of grains and of 
their size distribution. With such a description more accurate opacity
models can be obtained and radiative transfer calculations which 
depends on them can be improved.

In spite of the differences between the modeling assumptions and 
procedures utilized in our work and the $\alpha_{SS}$ models above,
many of the results produced from each are quite similar. This should
perhaps not be too unexpected since one of our heating terms analogous
to the standard $\alpha_{SS}$ model, extended to include limited time 
and space dependence. The other term (shocks) is not a dominant source
of heating in most of the radial extent of the disk.

Both methods are able to produce SEDs which reproduce observed profiles
to varying degrees. The temperature and density profiles on which the
SEDs are based however are quite different. In our work, the inner 
disk provides nearly all of the flux and is characterized by an essentially
flat surface density profile and an $r^{-1}$ temperature law at the
photosphere. Only at distances $>10$~AU does the surface density begin to 
fall off steeply.  At these radii, the temperatures derived from our models 
are low enough not to contribute significantly to the flux, and indeed
our simulations are deficient in long wavelength radiation relative
to observed systems.

Both in our work and Bell \etal (1995), a mechanism by which the disk 
may vary its energy output (SED) in time is explored. We consider 
the variation of the opacity due to the destruction and reformation
of grains in the inner disk, while they consider variation in the
thermal ionization state of gas within a few stellar radii of the 
star. They are able to produce very large and long term temporal 
variations typical of FU~Orionis outbursts in accretion disk SEDs, 
while we find much smaller variations (factors of $\sim$2 or less) 
which occur over much shorter time scales.

The puffed up inner disk in our work is similar to the hump at
the same radii seen in BCKH, which they attribute to the temperature
dependence of the opacity (i.e. the temperature at which various 
component grains vaporize) and the our results are likely to be
the same effect. The smaller `volcano region' discussed by Turner
\etal (1997) in their simulations of the disks of FU~Orionis objects
(derived from the BCKH models) is much smaller in extent and is due
to reprocessed stellar photons. Neither model depends upon the 
validity of the thin disk assumption and both indicate that such an
approximation may be inappropriate in the inner regions of the disk.

We have linked the long wavelength flux deficit in our calculations to
a limit on the amount heating due to shocks derived from gravitational 
torques. Studies of shock dissipation (Spruit \etal 1987, Larson 1989)
have previously shown that shocks in disks are weak. Our results
gives an estimate of how weak, in terms of their effect on the observable
character of the system. If gravitational torque heating does not power
the long wavelength part of the SED, what does? A variety of mechanisms
are possible. The outer disk may be heated by internal means like three
dimensional turbulence originating in dynamical processes or magnetic
fields if there is enough ionization (e.g. Gammie 1996). The outer 
disk may also be heated by passively absorbing and re-radiating light
from the central star or the surrounding molecular cloud. As is shown
for the 1D case in Turner \etal (1997), absorption and reradiation of
infrared and microwave photons can heat the outer disk, flatten the
temperature profile and provide additional far infrared flux.  Chiang 
\& Goldreich (1997) show that at long wavelengths the flux may instead
be due to passively reradiated long wavelength flux from the hot disk 
interior, because at long wavelengths the disk becomes optically thin.
If the midplane regions can indeed supply enough flux `as is', then the
temperatures in the outer disk may not require modification at all and
disk systems may be colder than expected far from the star.

We have not attempted to model passive heating in this work, in part 
because of the radiative transport approximations we have implemented
(i.e. Rosseland mean opacities) preclude a reliable determination of
the magnitude of such radiative heating processes in optically thin 
regimes. Such a treatment in our model would require a multi-dimensional 
frequency dependent radiative transport code, which has not yet been
incorporated. The distinction between active and passive heating is
important because assuming the wrong proportions of each imply incorrect 
mass and momentum transport rates through the disk. Accurate models 
require that the correct balance between passive and active heating be
well understood in order both to reproduce observed SEDs {\it and} model
the correct dynamical evolution. 

\appendix

\section{Grain Growth and Destruction in a Hot Disk Midplane}\label{app-grains}

In this appendix we discuss the growth of grains in regions where
the disk midplane is above the dust destruction temperature. We 
base these calculations on the coagulation models of Weidenschilling 
\& Ruzmaikina (1994). For a complete description of the physical
model and the code, see Spaute \etal (1991); here we shall merely
summarize the model presented there.

In our calculations, the vertical density and temperature structure 
remain constant and are computed as described in section 
\ref{energy-diss}. We assume a midplane temperature of 1350~K and a
local mass surface density of $\Sigma=10^3$~gm/cm$^2$ at 1~AU. These
conditions correspond to the intermediate temperature curves in fig.
\ref{z-strucplot}. A vertical column in the disk is divided into 20 
vertical layers and particle aggregates are accounted for as a series 
of 84 bins spaced logarithmically in grain diameter with each bin
1.1 times larger than the previous bin. The smallest bin is assumed to 
contain grains of size 1$\times 10^{-2}\mu$m. Grains smaller than 
1$\times 10^{-2}\mu$m are not accounted for and nucleation of grains
from the gas phase is likewise neglected. Relative velocities of 
grains are associated with the turbulence, settling of dust aggregates
to the central plane, and radial drift due to gas drag. The turbulent 
velocity is set to $\sim255$~m/s, equivalent to $\alpha_{SS}\sim10^{-2}$
(assuming $v_T=\sqrt{\alpha_{SS}}c_s$, where $c_s$ is the sound speed
at the midplane), and the impact strength of the grains, $E_s$, is set
to $E_s=1\times 10^6$ erg/cm$^3$. 

Using the geometric cross section of each grain size, the number
density of grains of that size and the relative velocities between
grains in different size bins, we compute the number of collisions
between all possible pairs of size bins during one time step. The
result of these collisions may be coagulation, erosion or total
destruction depending upon the relative velocities of the particles
and their assumed strength. Collisions resulting in grain coalescence
remove aggregates from the smaller bins and change the mean mass of
aggregates in the larger bins. If a collision leads instead to
erosion or disruption, then the fragments are distributed into
appropriate smaller size bins. Vaporization of grains in hot
regions is modeled by lowering the grain strength so that any
collision causes fragmentation. In the initial state, all of the
grains are in the smallest size bin. Assuming that nucleation 
from the gas phase into 1$\times 10^{-2}\mu$m sized grains
occurs quickly, this initial condition is equivalent to the 
assumption that at some point in the evolution of a particular 
column of gas all of the dust has been destroyed and must now 
reform. 

Figure \ref{coagmos} shows snapshots of the grain size distribution
at one AU plotted as a function of altitude above the midplane.
Time step constraints within the coagulation model forbid a very
long time evolution of the size distribution, however such longer term 
evolution is of limited value because the state of the gas (its 
temperature and density) changes on these same time scales, making a
grain distribution derived from a single ($\rho, T$) configuration 
irrelevant physically.

At high altitudes, grains are unable to grow to large sizes in the
time shown because of the low densities (which imply low collision
cross sections) that are found there. The largest size to which grains
grow in the time shown is $\sim$0.02-0.05$\mu$m. At moderate altitudes, 
but still well above hot midplane where grains do not grow, grains ten 
times larger ($\sim0.2-0.3\mu$m) form in this same time interval. At low
altitudes near the midplane, grains are unable to grow and remain locked 
in the smallest size bin available. It seems likely that in this region
grain material is in fact locked in an atomic or molecular state rather
than in small grains, however this supposition cannot be investigated
in the context of our present computation.

The size distribution of grains is quite unlike that of the interstellar
medium (ISM), as characterized by MRN or KMH, whose work shows a 
distribution proportional to $a^{-3.5}$. Instead it is characterized a
quantity of grains in the smallest size bin, whose origin is in the 
partial erosion of larger particles, and an increasing or near flat 
spectrum near the upper edge of the size distribution with a sharp
cutoff. A flat size spectrum such as this is a characteristic feature
of the collisional coagulation of grains where little destruction
takes place. The flat spectrum forms because larger grains have longer
stopping times, encounter more grains and therefore grow faster than
their smaller neighbors. A declining power law distribution characterizes
destructive processes and is pronounced at lower altitudes. Near the
midplane only grains in the smallest two or three size bins exist
because the assumed vaporization of grains (modeled in our calculation
via grain destruction, which moves grains from larger bins to smaller)
is very efficient there.

How does the grain distribution vary as the turbulent velocity and
assumed grain strength vary? In figure \ref{coag-cmp} we show
the results of tests similar to that outlined above, but with varying
turbulent velocity and grain strength, $E_s$. We vary $E_s$, between
5$\times 10^4$ and 1$\times 10^6$, as discussed in Weidenschilling (1984).
Except for very weak particles, the largest grains grow to a size of a few
tenths of a micron within half an orbit around the star. The size distribution 
of grains is dependent upon the strength of the grains but not strongly on 
the turbulent velocity, except that higher velocities produce more
erosion and keep a larger population of smaller grains active. The
turbulent velocity plays a role in carrying grains into regions where
they otherwise would not grow, as is shown in the bottom three frames
of figure \ref{coag-cmp}. Strong grains in a highly turbulent flow were
able to penetrate much closer to the midplane before being completely 
disrupted. This penetration is a likely artifact of the way we 
implement grain destruction, since a grain much suffer a collision 
with another grain rather than be heated by collisions with gas 
molecules in order to be destroyed.

In all cases with impact strength, $E_s$, greater than 10$^5$ erg/cm$^3$, 
grain growth occurs more slowly at high altitudes than at lower altitudes.  
With the nominal grain strength of 10$^6$ erg/cm$^3$, no turbulent velocities 
produce power law like distributions, even at turbulent velocities of half
the sound speed. With weak turbulence grains form a relatively narrow size
distribution, since the growth times from the smallest grains are everywhere
the same and little destruction occurs. The small end of the size distribution
fills in with higher turbulence as more destructive or erosive collisions
occur. The distribution approaches power law form if grains are weaker than
our assumed impact strength of 1$\times 10^6$ erg/cm$^3$ and the turbulent
velocity is high, but even with the highest turbulent velocity shown, 
differences from power law form are present. With still weaker grains, very 
little growth is possible anywhwere, even in regions where the grains do not 
vaporize, because the thermal velocities of the grains themselves are 
large enough to cause fragmentation of the very small, weak grains we assume.
In this case, only high altitudes evolve grains of any size much larger than
our smallest assumed size bin. Under no conditions were we able to reproduce
the $a^{-3.5}$ distribution of interstellar grains obtained by MRN, and we 
conclude that opacity calculations based upon such size distributions may
produce misleading results, when used for radiative transport calculations 
in circumstellar disks.

\clearpage

\section{Sorting out numerical effects from physical processes}\label{app-numphys}

\subsection{Systematic errors in the dissipation}\label{visc-systematics}

A numerical question remains before we can make a physical interpretation
of figure \ref{alpha-fig}. We must certain that modeling the dissipation
of kinetic energy into heat using artificial viscosity gives a reasonably
accurate representation of the true thermal energy generation rate. For
example, we would draw incorrect conclusions if the thermal energy
generated by our artificial viscosity was systematically larger or smaller
than physically appropriate. Simply running a higher resolution
simulation does not address this question because the procedure for 
determining the dissipation remains the same and systematic errors
would produce the same effect.

We have performed a simulation similar to {\it B2h3} with the time
variation of the viscous coefficients discussed in section \ref{energy-gen}
turned off and the viscous coefficients set to $\bar\alpha=1$ and
$\beta=2$. This simulation is denoted {\it H2h3} in table
\ref{cool-params}. Effectively, this change will increase the global
rate of thermal energy generation because we find that the time
dependent viscous coefficients for each particle ($\bar\alpha_i(t)$)
ordinarily fall well below the value of unity assumed in simulation
{\it H2h3}.

The time dependent fluxes for this simulation are shown in the right hand 
panels of figure \ref{flux-var} and \ref{flux-var-expand}. The total 
luminosity is increased by the viscosity modification but the increase 
comes only from short wavelengths regime representative of the inner disk. 
The high luminosity and short wavelength flux over the first 2--3 \td 
again come from the initial fast mass accretion period seen in the `A'
and `B' runs. After this period the disk has adjusted from its initial 
condition to one which is stable over longer periods. The magnitude of
the initial burst in flux is larger because for a single heating event, 
the thermal energy transfered to column of matter is larger and leads to
higher temperatures than before. Higher temperatures over a larger 
region of the inner disk then produce a correspondingly increased flux
as the gas cools.

After the initial fast accretion period (i.e. after $\sim$2\td), the
near and mid IR flux are increased in magnitude above that of the `B'
simulation by only $\sim 5$\%--barely large enough to be detectable on
the plot. The long wavelength flux also increases by only $\sim 5$\%
relative to the `B' simulation. In both simulation {\it B2h3} and
{\it H2h3}, little thermal energy is produced in the outer disk. More
significantly, the bottom frame of figure \ref{sed-50-100cmp} shows
that the long wavelength turnoff does not shift further into the far
IR or submillimeter region. It remains instead near $10^{13}$~Hz 
($\sim$30$\mu$m). Therefore, we can be certain that the specific 
details of the artificial viscosity implementation to generate thermal
energy are not strongly affecting the outcome of our simulations.

\subsection{Variation of dissipation with resolution}\label{resol-diss}

\begin{figure}
\plotfiddle{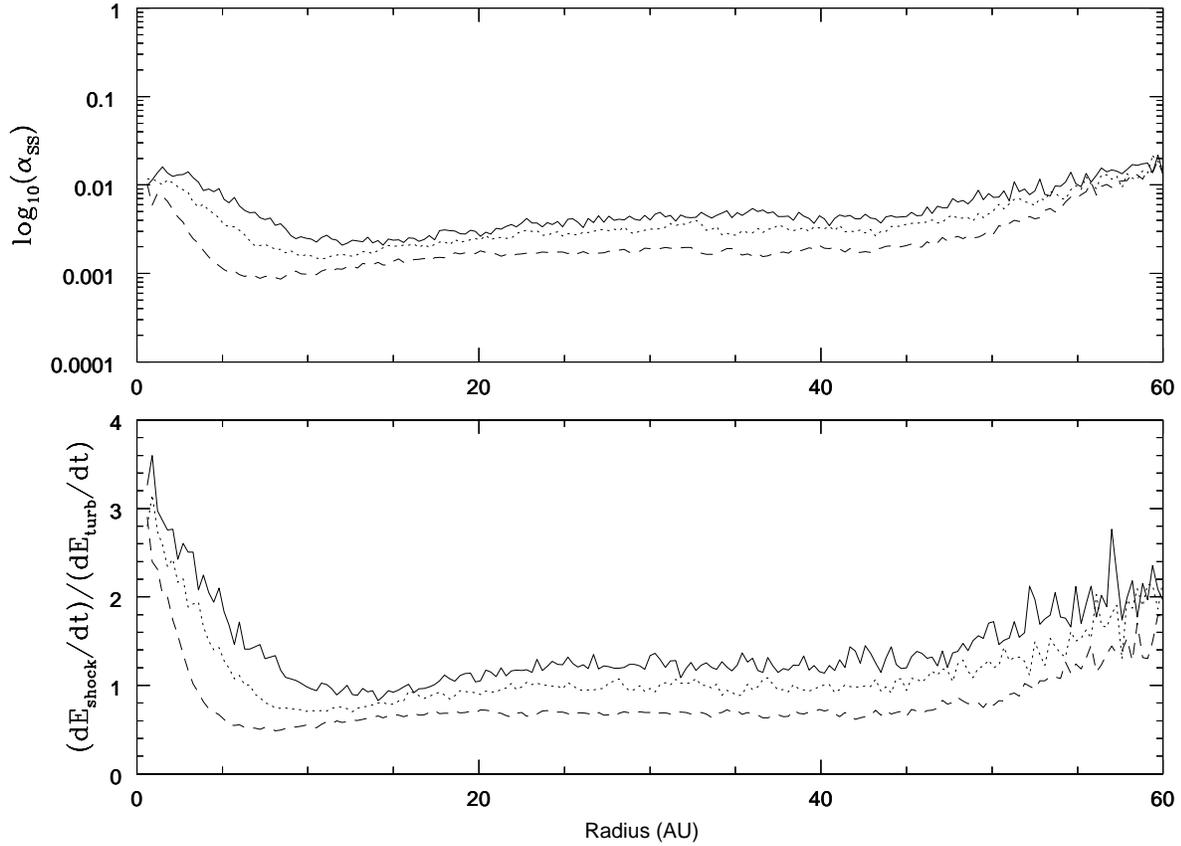}{5.5in}{-90}{60}{60}{-240}{500}
\caption[Approximate magnitude of the local dissipation in the disk
in terms of the well known `$\alpha$' parameter for viscous disks]
{\label{alpha-fig} 
Top panel: The azimuth averaged value of $\alpha_{\rm SS}$ near the end
of simulations {\it A2lo} (solid), {\it A2me} (dotted) and {\it A2hi} 
(dashed). Bottom panel: The azimuth averaged value of the ratio of the 
thermal energy generation rate due to shocks and turbulence. Values are
defined out to the edge of each panel ($>$60~AU) because a small fraction
of particles spread outward to this distance. The amount of this spreading
is contaminated by numerical diffusion of the boundary in the SPH code,
so no conclusions can be based on its magnitude.}
\end{figure}

\begin{figure}
\plotfiddle{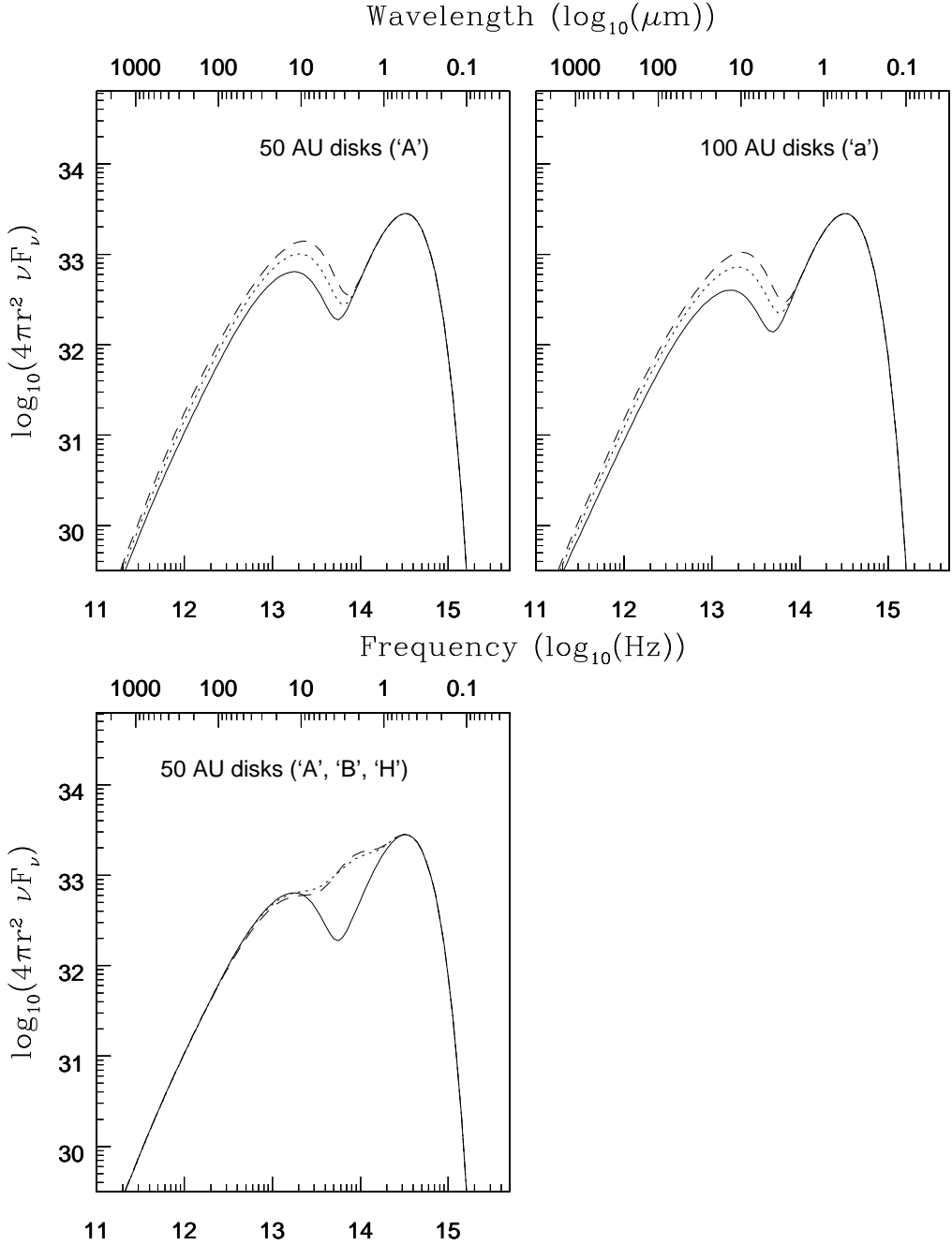}{6.4in}{0}{70}{70}{-230}{-50}
\caption[Disk flux in various wavelength bands and total luminosity
as a function of time.]
{\label{sed-50-100cmp}
Time averaged spectral energy distributions of simulations of
different resolution and with assumed disk radii of 50 (top left) or
100 AU (top right). Simulations {\it A2lo, A2me} and {\it A2hi} are
designated with dashed, dotted and solid lines respectively. Similar
designations define the curves for simulations {\it a2lo, a2me} and
{\it a2hi}. Little additional long wavelength radiation is
produced in the 100 AU disks in spite of the additional surface
area. The bottom left panel shows the differences between the SEDs
produced at high resolution but differing physical assumptions
(simulations {\it A2hi, B2h3} and {\it H2h3} with solid, dotted
and dashed curves respectively). The time averages are taken between
\td=1 and \td=5 except for simulation {\it H2h3}, which is taken
between \td=2.5 and \td=6 in order to reduce the effect of the
rapid mass accretion period (see fig. \ref{flux-var} above).}
\end{figure}

\begin{figure}
\plotfiddle{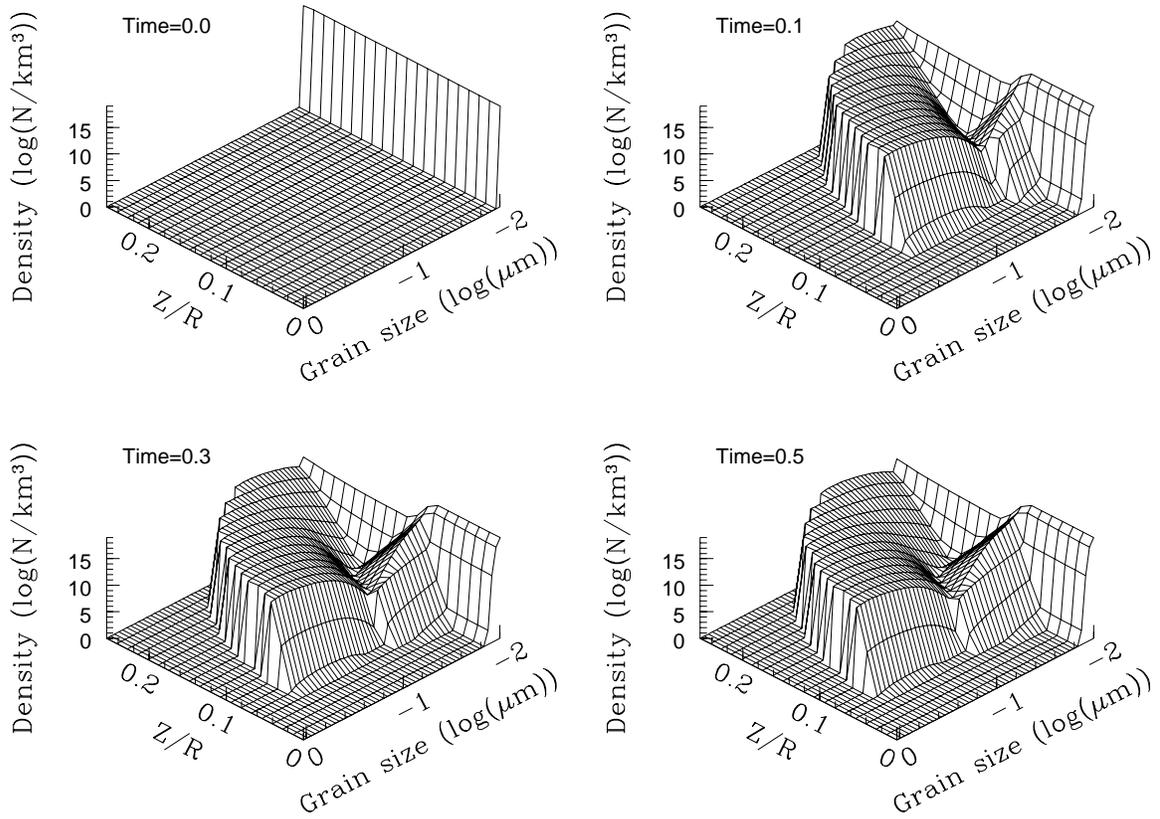}{6.0in}{-90}{60}{60}{-235}{500}
\caption[Grain size distribution at different altitudes above the midplane]
{\label{coagmos}
The grain size distribution as a function of altitude above the disk
midplane. The time units in the upper left of each frame are given in
orbit periods at the assumed 1~AU distance from the star
($T=1\approx 1.42$~yr). The vertical structure is identical to that
shown in figure \ref{z-strucplot} with a surface density of 
1000~gm/cm$^2$, midplane temperature of 1350~K at a distance of 1~AU 
from the central star. In the example shown, the midplane temperature 
is above the assumed grain destruction temperature of $T=1200$~K. }
\end{figure}

\begin{figure}
\plotfiddle{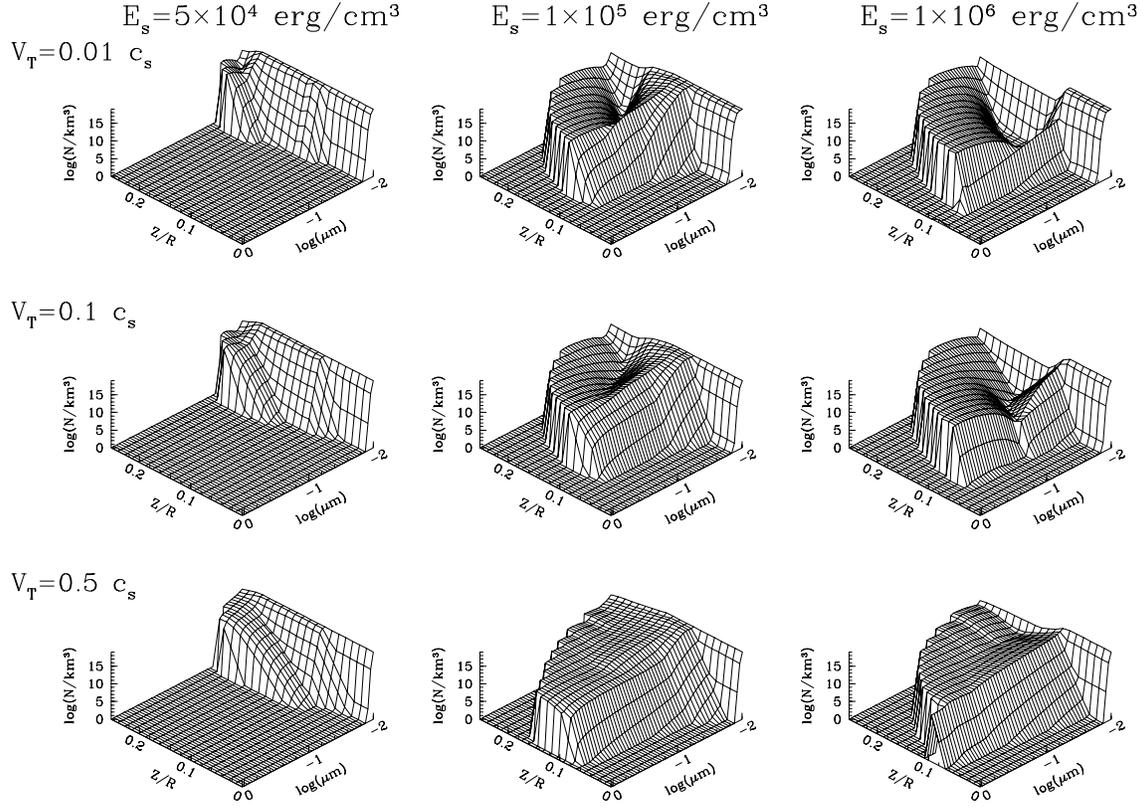}{6.0in}{-90}{60}{60}{-235}{500}
\caption[Grain size distribution compared for several grain strengths and
turbulent velocities]
{\label{coag-cmp}
The grain size distribution for a grid of models with the same
initial conditions but varying assumed grain strengths and gas turbulent
velocity. Each frame is a snapshot of the grain distribution at a time 
after the beginning of one half orbit, and are comparable to the last
frame of figure \ref{coagmos}. The middle frame in the right column here and
the last frame of figure \ref{coagmos} are identical. Values of assumed
grain strength and turbulent velocity are shown at the top of each column
and the left edge of each row, respectively. }
\end{figure}

In our higher resolution runs, the SEDs synthesized have a 
systematically smaller infrared excess and total luminosity than 
in lower resolution runs (fig. \ref{sed-50-100cmp}). This difference
is due to the correspondingly lower dissipation possible at high 
resolution, as shown in the top panel of fig \ref{alpha-fig}. Because 
the purely compressional dissipation (i.e. shocks) is better resolved 
at higher resolution, as the resolution increases the shock dissipation
term will more closely model the physical dissipation present in 
shocks and purely numerical dissipation will decrease elsewhere
in the flow. Therefore until a set of simulations converges to a
well defined amount of dissipation which is not a function of 
resolution, the dissipation which is present will represent an upper 
limit on that present in a real system.

Ideally, the contribution of unidentified sources of dissipation
to the energy output of the system would be negligible. In such a
case, specification of known dissipation mechanisms and the known
passive heating mechanisms in the model assumptions would specify
the observable appearance of the system. In our models this ideal
can only be approached, rather than definitely specified. We have
identified the bulk viscosity term in eq.~\ref{Piij}
with the Shakura \& Sunyaev $\alpha_{SS}$, which attempts to model
the effects of turbulence with a single parameter. Since our resolution
is finite, this source of dissipation is non-zero. Further, at 
progressively higher resolution, the magnitude of the dissipation 
decreases and we conclude that we have not fully resolved the 
hydrodynamics important for energy generation in the system.
We estimate the unknown, `turbulent' contribution to the dissipation 
(via eq. \ref{alpha-eq}) to be of order $2\times 10^{-3}$ between
10 and 50~AU for the highest resolution simulations we have run
(see fig. \ref{alpha-fig}).

We cannot assign a definite physical origin to the bulk 
viscosity term, other than to say that it models turbulence.
However, we still can constrain the magnitude of other, known 
sources of thermal energy generation by comparing their computed 
contribution to those of our bulk term. The comparison is useful 
because of the correspondence between this bulk term and the well 
known and widespread use of $\alpha_{SS}$ models to describe
accretion disk evolution.

We have identified the von~Neumann-Richtmyer dissipation term in
the artificial viscosity with shock dissipation. In the lower panel
of fig \ref{alpha-fig} we show the ratio between the dissipation
rates from turbulence and shocks. Higher resolution produce less shock
dissipation relative to lower resolution runs, and we conclude that
the shock dissipation is not fully resolved. By convervatively
ascribing all dissipation as being due to shocks, we conclude that
the effects of shocks on the disk can be no greater than an equivalent
$\alpha_{SS}$ dissipation of a few $\times 10^{-3}$.  

At radii close to the star this conclusion is reversed. Shock dissipation
apparently dominates the total dissipation rate. Further, the dissipation
attributed to turbulence also increases. There are two numerical reasons 
for the turbulent dissipation increase. First, the value of the
viscous coefficient $\bar\alpha$ itself is never able to relax to its
lowest value in this region. (Recall that we have implemented a time
dependence in the viscous coefficients, which causes them to grow
during compressions and decay to a small, steady state value in a
smooth flow). This indicates that strongly compressive events repeat
on very short temporal and spatial scales not well resolved by the code, 
and leads (by eq. \ref{alpha-eq}) to an artificial increase in the 
derived Shakura \& Sunyaev viscous coefficient, $\alpha_{SS}$, which 
should be attributed instead to shock dissipation. This effect would 
tend to make the dominance of shock dissipation even stronger.

Second, within only a few ten's of orbits, mass accretion onto the 
central star begins to reduce the density (see section \ref{rad-struc}). 
Since SPH resolves the flow using particles of finite mass, lower mass
density in a given region is equivalent to lower resolution (fewer 
particles) and higher numerical dissipation. Notably, higher resolution
simulations produce progressively more centrally peaked dissipations 
(fig. \ref{alpha-fig}), which means that the region in which shock 
dissipation may be important is limited to a smaller portion of the
disk near the star. The derived value of the turbulent dissipation at
all three resolutions reaches $\alpha_{SS}\approx 10^{-2}$ at the 
inner edge, suggesting that this value is fairly well resolved. At
early times, for which little mass transport has yet occurred and
the structures developing in the inner disk are best resolved, the
conclusion that shock dissipation is a strong contributor to the 
thermal energy generation remains.

\subsection{Dependence of the near infrared variation on resolution 
and the opacity modification}\label{var-var}

The variations in the near infrared flux of our `B' simulations
are a function of the resolution of the simulations, decreasing as
resolution increases. We cannot be certain of the amplitude at which
variations become independent of resolution or the opacity 
approximations we have made. In order to accurately quantify the true
amplitude of variations, the balance between dynamical heating and
radiative cooling must be better understood. Heating due to shock 
dissipation in the inner disk must be well resolved and the effects of 
grain vaporization, reformation and size evolution on the opacity must 
be determined to accurately determine the magnitude of the radiative 
cooling.

On the other hand, the variation time scales are dominated by the
dynamical (orbital) times of the inner disk and so will be relatively
insensitive to resolution. Our simulations are insensitive to shorter
term variations since our disks were truncated at a relatively large
distance from the star. If dynamical activity continues to modify the
opacities closer to the star in a similar manner to our simulations,
signatures of shorter time scale variations could also exist in 
the emitted flux.

\acknowledgments
We wish to thank Robbins Bell for an excellent referee report,
We thank Phil Pinto, Sarah Maddison, Mike Meyer, Brian Pickett
and Jim Stone for useful conversations and correspondence.
We thank David Alexander for providing the `grain free' opacities
noted in section \ref{cool-improve}. This work was supported under
the NASA Origins of the Solar System grant NAG5-4380.

\newpage

\singlespace
\begin{deluxetable}{lcccccc}
\tablecaption{\label{cool-params} Initial Parameters For Simulations}
\tablehead{
\colhead{Name}  & \colhead{Number of} & \colhead{ Disk}  &
\colhead{Opacity} & \colhead{End Time} 
\\
\colhead{}  & \colhead{Particles} & \colhead{Radius (AU)} &
\colhead{Factor $R$} & \colhead{(T$_{\rm D}$=1)} }

\startdata
I2lo & \phn 16520 & \phn 50 &  \nodata & 1.6 \nl
A2lo & \phn 16520 & \phn 50 &  \nodata & 6.0 \nl
I2me & \phn 33399 & \phn 50 &  \nodata & 1.8 \nl
A2me & \phn 33399 & \phn 50 &  \nodata & 6.0 \nl
B2m1 & \phn 33399 & \phn 50 &   0.001  & 6.0 \nl
B2m2 & \phn 33399 & \phn 50 &   0.010  & 6.0 \nl
B2m3 & \phn 33399 & \phn 50 &   0.050  & 6.0 \nl
B2m4 & \phn 33399 & \phn 50 &   0.075  & 6.0 \nl
B2m5 & \phn 33399 & \phn 50 &   0.025  & 6.0 \nl
I2hi &     101016 & \phn 50 &  \nodata & 1.8 \nl
A2hi &     101016 & \phn 50 &  \nodata & 6.0 \nl
B2h3 &     101016 & \phn 50 &   0.050  & 6.0 \nl
H2h3 &     101016 & \phn 50 &   0.050  & 6.0 \nl
a2lo & \phn 16182 &     100 &  \nodata & 3.0 \nl
a2me & \phn 33134 &     100 &  \nodata & 2.3 \nl
a2hi &     100971 &     100 &  \nodata & 3.0 \nl
\enddata
\end{deluxetable}
\doublespace

\newpage

\singlespace

\clearpage

\clearpage

\clearpage

\clearpage

\clearpage

\clearpage

\clearpage

\clearpage

\clearpage

\clearpage

\clearpage

\clearpage

\clearpage

\clearpage

\clearpage

\clearpage

\clearpage

\clearpage

\clearpage

\clearpage

\clearpage

\clearpage

\clearpage

\clearpage

\doublespace

\end{document}